\documentclass[preprint,review,10pt]{elsarticle}
\usepackage[english]{babel} 
\usepackage{amssymb}                  
\usepackage{times}			
 \usepackage[latin1]{inputenc}   
\usepackage[T1]{fontenc} 

\usepackage[T1]{fontenc}
\usepackage{epsfig}
\usepackage{subfigure}
\usepackage{calc}
\usepackage{amssymb}
\usepackage{amstext}
\usepackage{amsmath}
\usepackage{multicol}
\usepackage{pslatex}
\usepackage[small]{caption}
\usepackage{amsmath,epsfig}
\usepackage{amstext,amssymb,amsbsy,amsmath,amscd}
\usepackage{graphics,graphicx,epsfig,enumerate,color,pstricks,cancel,ulem, rotating,theorem,array,longtable,multirow,multicol}
\def\x{{\mathbf x}}
\def\V{{\mathbf V}}
\def\N{{\mathbf N}}

\newcommand{\be}{\begin{eqnarray}}
\newcommand{\ee}{\end{eqnarray}}

\newtheorem{theorem}{Theorem}
 
 \newtheorem{proposition}{Proposition}

\DeclareMathOperator{\mes}{mes}

\usepackage{amsmath,epsfig}

\journal{ArXiv}

\begin{document}

\begin{frontmatter}
\title{A Mutual Reference Shape for Segmentation Fusion and Evaluation}

\author{\small
  S. Jehan-Besson {(a)},
  R. Clouard {(b)},
  C. Tilmant {(c)},
  A. de Cesare {(d)}, 
  A. Lalande {(e)},
  J. Lebenberg {(f)}, 
  P. Clarysse {(g)},
  L. Sarry {(h)},
  F. Frouin {(i)}, 
  M. Garreau {(j)}
 }

 \address[1]{\footnotesize CNRS now with LITO U1288 INSERM Institut Curie Paris,
   \\ stephanie.jehan-besson@cnrs.fr }
 \address[2]{\footnotesize Normandie Univ, UNICAEN, ENSICAEN, CNRS, GREYC, 14000 Caen, France }
 \address[3]{\footnotesize Institut Pascal, UMR 6602 UBP CNRS, Clermont-Ferrand, France }
 \address[4]{\footnotesize Sorbonne Universit\'es, UPMC Univ. Paris 06, CNRS, INSERM, Laboratoire d'Imagerie Biom\'edicale (LIB), 75013, Paris, France }
 \address[5]{\footnotesize ImViA EA 7535 laboratoire, UniversitÈ de Bourgogne, France }
 \address[6]{\footnotesize CEA NeuroSpin UNATI, France}
 \address[7]{\footnotesize Univ Lyon, INSA-Lyon, Universit\'e Lyon 1, UJM-Saint Etienne, CNRS, Inserm, CREATIS UMR 5220, U1206, F-69621, LYON, France. }
 \address[8]{\footnotesize Clermont Auvergne UniversitÈ, CNRS, Institut Pascal, Clermont-Ferrand, France}
 \address[9]{\footnotesize LITO INSERM U1288 Institut Curie, Paris, France}
  \address[10]{\footnotesize  INSERM U1099, Universit\'e de Rennes 1, LTSI, Rennes, F-35000 France }
 

\date{Written on March 2017}

\maketitle

\begin{abstract}
This paper proposes the estimation of a mutual shape from a set of different segmentation results using both active contours and information theory. The mutual shape is here defined as a consensus shape estimated from a set of different segmentations of the same object. In an original manner, such a shape is defined as the minimum of a criterion that benefits from both the mutual information and the joint entropy of the input segmentations. This energy criterion is justified using similarities between information theory quantities and area measures, and presented in a continuous variational framework. In order to solve this shape optimization problem, shape derivatives are computed for each term of the criterion and interpreted as an evolution equation of an active contour. A mutual shape is then estimated together with the sensitivity and specificity of each segmentation. Some synthetic examples allow us to cast the light on the difference between the mutual shape and an average shape. The applicability of our framework has also been tested for segmentation evaluation and fusion of different types of real images (natural color images, old manuscripts, medical images).
\end{abstract}

\begin{keyword}
Mutual shape \sep segmentation \sep variational approaches \sep segmentation fusion \sep segmentation evaluation \sep active contours \sep shape gradients \sep shape optimization \sep average shape \sep information theory
\end{keyword}
\end{frontmatter}

\section{Introduction}
\label{sec:intro}

Constructing a ``consensus'' shape from a set of different segmentation results is an important point when dealing with image segmentation evaluation when no expert reference is available (evaluation without gold standard). It is also a key point for an appropriate fusion of several segmentation results in a single shape. Such a shape must ideally take advantage of the information provided by each input shape while being robust to outliers. We propose to tackle the estimation of such a reference shape using information theory (mutual information and joint entropy) through the definition of a shape optimization problem. The consensus shape will then be defined as the minimum of an original criterion based on information theory and area measures and computed within the framework of active contours and shape gradients. This shape is called ``a mutual shape'' and its applicability is tested for image segmentation evaluation and fusion.

In the context of segmentation evaluation, the estimation of such a reference shape may be important especially when dealing with large databases of medical images when the manual delineation of all the frames by an expert becomes a tedious and time consuming task. The obtained contour is also subject to inter- and intra-variability and being expert-dependent, it can then not be considered as an absolute reference. As far as the evaluation without gold standard is concerned, the STAPLE algorithm (Simultaneous Truth and Performance Level Estimation) proposed by Warfield et al. \cite{Warfield_TMI_04} is now classically used in this difficult context. Their algorithm consists in one instance of the EM (Expectation Maximization) algorithm where the true segmentation is estimated by maximizing the likelihood of the complete data. Their pixel-wise approach leads to the estimation of a reference shape simultaneously with the sensitivity and specificity of each input segmentation. From these measures, the performance level of each input segmentation can be estimated and a classification of all the segmentation entries can be performed. The most recent MAP-STAPLE approach \cite{Warfield_TMI_12} is semi-local and takes benefit of a small window or patch around the pixel. In this paper, we rather propose to estimate the reference shape within a continuous optimization setting by considering such a shape estimation under the umbrella of shape optimization tools \cite{Zolesio} and deformable models \cite{Kass88}. Indeed, the computation of a reference shape can be advantageously modeled as the optimum of a well chosen energy criterion and estimated by a shape gradient descent that corresponds to the deformation of an active shape. Moreover, we propose a new theoretical criterion based on information theory that allows to well understand the behaviour of our reference shape.

Let us also note that shape optimization algorithms have already been proposed in order to compute shape averages \cite{Charpiat_FCM04,Soatto_IJCV02} or more recently median shapes \cite{Berkels_JMIV2010} by minimizing different shape metrics like the Hausdorff distance in \cite{Charpiat_FCM04} or the symmetric area difference between shapes in \cite{Soatto_IJCV02}. Some other approaches also take advantage of well-appropriate distances between level-set shapes (see for example \cite{Leventon_CVPR00}). Comparing with these previous approaches, our goal is quite different since our aim is to compute a consensus shape from $N$ input segmentations. 

The main contribution of this paper is then to propose a new theoretical model to carry out the estimation of a consensus or reference shape from several segmentation entries using active contours and shape gradients. This theoretical model is based on information theory and justified using the analogies between information theory and area measures. In order to estimate what we call a ``mutual shape'', we then propose to maximize the mutual information between the $N$ input segmentations while minimizing the joint entropy. Such a statistical criterion can be interpreted as a robust measure of the symmetric area difference. The minimization is performed through the computation of the evolution equation of an active contour. This evolution equation is computed using advanced shape derivation tools. In order to perform such a derivation, the criterion must be expressed in a continuous settings and non parametric probability density functions are estimated using Kernel methods \cite{duda-hart:73}. In this variational setting, we also propose to add a classical regularization term based on the curvature of the deformable contour. Such a term is weighted using a regularization parameter that controls the smoothness of the obtained contour. The advantage of this formalism is to make explicitly appear, in the criterion to minimize, the domain and the associated contour. The criterion is then easier to understand and interpret and some geometrical and photometric priors could be directly added in the criterion to minimize. The derivation is directly performed according to the domain using shape derivation tools.

The proposed algorithm is first experimented on a synthetic example that allows to understand the differences between a classic average shape based on a symmetric area minimization \cite{Soatto_IJCV02}, a simple majority voting shape and the proposed mutual shape. It is also evaluated for segmentation fusion and evaluation on different images : color real natural images, old manuscripts or medical images, in order to show the genericity of this framework.  The first application concerns segmentation evaluation and fusion on a real color natural image using segmentations from the Berkeley database \cite{MartinFTM01}. The second application is dedicated to text segmentation in old manuscripts and we propose two main examples of segmentation. One of them takes benefit of the DIBCO database \cite{DIBCO13}. The last application is devoted to segmentation fusion and evaluation of different delineation methods of the left ventricular cavity in Magnetic Resonance Imaging (MRI). For this application, we propose to compare the mutual shape to the reference algorithm STAPLE \cite{Warfield_TMI_04} classically used for segmentation fusion and evaluation in medical images.

In section \ref{sec:format}, we present the problem statement and in section \ref{sec:criterion}, the proposed criterion for the estimation of the mutual shape. The criterion is then estimated in a continuous framework and expressed using domain or contour integrals in section \ref{sec:stat_criterion}. Such a continuous criterion can then be derived using shape optimization tools in order to compute the mutual shape (see section \ref{sec:optimization}). Experimental results on synthetic examples are detailed in section \ref{sec:res} and the different applications in section \ref{sec:applis}.

\section{Problem statement}
\label{sec:format}
\begin{figure}[htb]
\centering
\includegraphics[width=12cm]{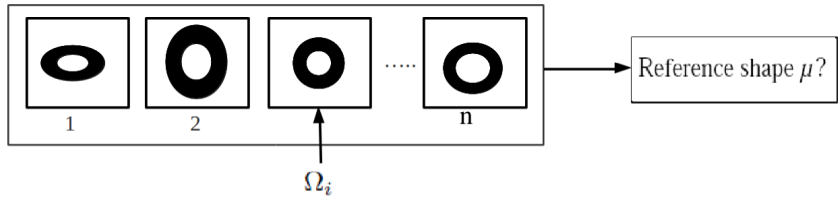} 
\caption{Diagram of the problem statement: evaluation of a reference shape $\mu$ from a set of $n$ segmented shapes of the same object.}
\label{fig:problem_statement}
\end{figure}

Let $\mathcal{U}$ be a class of domains (open regular bounded sets, i.e. $\mathcal{C}^2$) of $\mathbb{R}^{d}$ (with $d=2$ or $3$). In this paper theoretical results are stated for $d=2$ or $d=3$ but the experimental results are conducted on 2D-images. We denote by $\Omega_i$ an element of $\mathcal{U}$ of boundary $\partial \Omega_i$. We consider $\{\Omega_1,...,\Omega_n\}$ a family of $n$ shapes where each shape corresponds to the segmentation of the same unknown object $O$ in a given image. The image domain is denoted by $\Omega \in \mathbb{R}^{d}$. Our aim is to compute a reference shape $\mu$ that can closely represent the true object $O$ (Fig.\ref{fig:problem_statement}). We propose to define the problem through a statistical representation of shapes embedded in an information theory criterion. Let us first recall the main shape representation models and criteria proposed in the literature. 

\subsection{Shape representation}
\label{sec:shape_representation}

The computation of a reference shape is closely linked to the choice of a representation. An analytical representation may be used as in \cite{Kendall_84} where the authors propose a statistical study of shapes by representing them as a finite number of points. Some authors prefer to choose an implicit representation of shapes which avoids the parametrization step. For example, in \cite{Berkels_JMIV2010,Charpiat_ICIP03} shapes are represented using their characteristic function as follows:
\begin{eqnarray}
\label{eq:di}
d_i(\x)=
\left\{
\begin{array}{c c c}
1 &\text{if} &\x \in \Omega_i \\ 
0 &\text{if}  &\x \, \not\in \Omega_i
\end{array}
\right.
\end{eqnarray}
where $\x \in \Omega$ is the location of the pixel within the image. We denote by $\overline{\Omega_i}$ the complementary shape of $\Omega_i$ in $\Omega$ with $\Omega_i \cup \overline{\Omega_i} =\Omega$.
One may also take advantage of the distance function associated to each shape. In \cite{Leventon_CVPR00} the authors propose to perform a principal component analysis on shapes in order to provide a statistical shape prior. In the same vein, some statistical shape priors have been proposed by \cite{Cremers03b,Paragios_03} using this implicit representation.

More recently shapes have been represented using Legendre moments in order to define shape priors for segmentation using active contours \cite{Foulonneau_PAMI06}. This representation can also be easily included in a variational setting \cite{Foulonneau_PAMI06,Lec_icip06}.

We may also consider that each shape is a realization of a random variable. Such a representation has been introduced in \cite{Warfield_TMI_04} in order to evaluate a reference shape in a statistical framework, in \cite{Angulo_ICPR2010} for the morphological exploration of shape spaces and statistics, and also in \cite{Herbulot_JMIV06,Kim_ICIP02} for image segmentation using information theory. In this paper, we take advantage of this statistical representation that appears to be well adapted to the definition of a statistical criterion. The shape is represented through a random variable $D_i$ whose observation is the characteristic function $d_i$ defined in \eqref{eq:di}. The reference shape $\mu$ is also represented through an unknown random variable $T$ with the associated characteristic function $t(\x)=1$ if $\x \in \mu$ and $t(\x)=0$ if $\x \in \overline{\mu}$.

\subsection{Definition of average shapes}

In the literature, average shapes are defined through the minimization of the sum of the distances of the unknown shape $\mu$ to each shape $\Omega_i$ as follows:
\begin{eqnarray}
\mu=\arg\min_{\mu^*} \sum_{i=1}^{n} d(\Omega_i,\mu^*)
\end{eqnarray}
Of course, the definition of the distance $d$ is crucial and may lead to different results and average shapes.
For example, an average shape can be computed by minimizing the area of the symmetric differences \cite{Soatto_IJCV02} using $d(\Omega_i,\mu):=|\Omega_i \triangle \mu|$  where $|.|$ stands for the cardinal of the considered domain. In a continuous optimization framework, the criterion to minimize can be expressed as follows:
\begin{eqnarray}
\label{eq:criterion_SD}
SD(\mu)=  \sum_{i=1}^{n} |\Omega_i \triangle \mu|= \sum_{i=1}^{n}\left( \int_{\mu}(1-d_i(\x)) d\x + \int_{\overline{\mu}} d_i(\x) d\x \right)
\end{eqnarray}
In \cite{Charpiat_ICIP03,Charpiat_FCM04}, the authors prefer to introduce the Hausdorff distance to perform shape warping while in \cite{Berkels_JMIV2010}, the authors modify the previous criterion in order to compute \textit{a median shape}. In addition to the previous works, we can also cite \cite{Angulo_ICPR2010} where the authors propose to explore shape spaces using mathematical morphology. The optimal shape is computed using a watershed performed on the squared sum of the distance functions or using a morphological computation of a median set. Another class of algorithms was proposed for the estimation of an unknown shape from multiple channels (color or multimodal segmentation). We can cite the work of Chan et al. \cite{Chan00} or the multimodal segmentation approaches proposed in \cite{Herbulot_JMIV06,Kim_ICIP02}. These works were not designed at first for segmentation evaluation or fusion but they are worth mentioning because they propose to treat the different channels in a single criterion (may also be useful for information fusion). Moreover in \cite{Herbulot_JMIV06,Kim_ICIP02}, some information theory quantities are used. Our work is different especially due to the fact that we consider both the maximization of mutual information coupled with the minimization of joint entropies and the joint estimation of evaluation quantities (sensitivity and specificity measures).

\section{Proposition of a criterion for the estimation of a mutual shape}
\label{sec:criterion}

Our goal is here to mutualize the information given by each segmentation to define a consensus or reference shape. Such a shape cannot be considered as a simple average shape. 

In this context, we propose to take advantage of the analogies between information measures (mutual information, joint entropy) and area measures. As previously mentioned, $D_i$ represents the random variable associated with the characteristic function $d_i$ of the shape $\Omega_i$ and $T$ the random variable associated with the characteristic function $t$ of the reference shape $\mu$. Using these notations, $H(D_i,T)$ represents the joint entropy between the variables $D_i$ and $T$, and $I(D_i/T)$ the mutual information. We then propose to minimize the following criterion :
\begin{eqnarray}
E(T)=\sum_{i=1}^{n} \left(H(D_i,T) - I(D_i,T) \right)= JH(T) + MI(T)
\label{eq:crit_mutualshape}
\end{eqnarray}
where the sum of joint entropies is denoted by $JH(T)= \sum_{i=1}^{n} H(D_i,T)$ and the sum of mutual information by $MI(T)=-\sum_{i=1}^{n} I(D_i,T)$.

Introducing this criterion can be justified by the fact that $\varphi(D_i,T)=(H(D_i,T)-I(D_i,T))$ is a metric which satisfies the following properties : 
\begin{enumerate}
\item $\varphi(X,Y) \geq 0$ 
\item $\varphi(X,Y)=\varphi(Y,X)$
\item $\varphi(X,Y)=0$ if and only if $X=Y$
\item $\varphi(X,Y)+ \varphi(Y,Z) \geq \varphi(X,Z)$
\end{enumerate}
Indeed, we can show easily that $\varphi(D_i,T)=H(T/D_i)+H(D_i/T)$ using the following classical relations between the joint entropy and the conditional entropy and the mutual information and the conditional entropy :
\begin{eqnarray*}
H(D_i,T)=H(D_i)+H(T/D_i) \\
I(D_i,T)=H(D_i)-H(D_i/T)
\end{eqnarray*}
Finally, $H(T/D_i)+H(D_i/T)$ is shown to be a metric that satisfies the four properties above \cite{cover-thomas:91}. We then minimize a sum of distances between $D_i$ and $T$ expressed using information theory quantities.

\begin{figure}
\centering
\includegraphics[width=10cm]{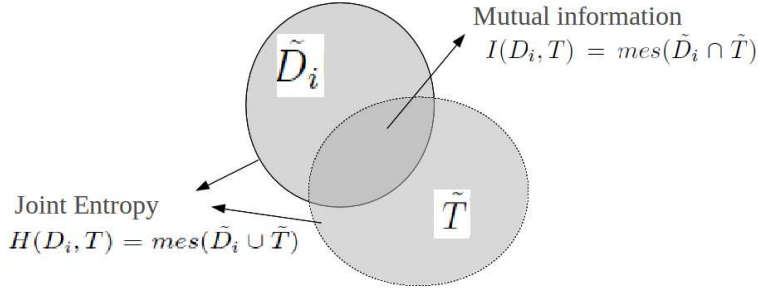} 
\caption{Mutual information and joint entropy as area measures}
\label{fig:mesure_aires}
\end{figure}

Moreover, we can give a further interesting geometrical interpretation of the proposed criterion. We propose to take advantage of the analogies between information measures (mutual information, joint entropy) and area measures. In \cite{Reza_book94,Yeung_IEEEIT_91}, it is shown that Shannon's information
measures can be interpreted in terms of area measures as follows:
\begin{eqnarray}
H(D_i,T)=\mes(\tilde{D_i} \cup \tilde{T}) \quad \text{and} \quad I(D_i,T)=\mes(\tilde{D_i}\cap \tilde{T}),
\end{eqnarray}
 with $\tilde{X}$ the abstract set associated with the random variable $X$ and the term $\mes$ corresponds to a signed measure defined on an algebra of sets with values in $]-\infty,+\infty[$. The signed measure must satisfy $\mes(\emptyset)=0$ and $\mes(\bigcup_{k=1}^{n} A_k)=\sum_{k=1}^{n} \mes(A_k)$ for any sequence $\{A_k\}_{k=1}^n$ of disjoint sets. Each quantity can then be viewed as an operation on the sets (Fig.\ref{fig:mesure_aires}). These properties help us to better understand the role of each term in the criterion to optimize. 

Indeed, when estimating a classic average shape using the criterion (\ref{eq:criterion_SD}), one performs the minimization of the sum of the union of the shapes $\Omega_i$ with $\mu$ while maximizing the sum of the intersection between the same shapes. By analogy with this criterion, we minimize a measure of the union while maximizing a measure of the intersection through the use of information quantities. In other words, the sum of the joint entropies (union of sets) will be minimized while the sum of the mutual information quantities (intersection) will be maximized. The proposed criterion can then be interpreted as a statistical measure of the area of the symmetric difference which is really interesting for the estimation of a consensus shape.

\section{Expression of the criterion in a continuous framework}
\label{sec:stat_criterion}

In order to take advantage of the previous statistical criterion \eqref{eq:crit_mutualshape} within a continuous shape optimization framework, we propose to express the joint and conditional probability density functions according to the reference shape $\mu$. This step is  detailed in this section for both the mutual information and the joint entropy. 

\subsection{Maximization of mutual information (MI)}

Here we try to express $MI(T)=-\sum_{i=1}^{n} I(D_i,T)$ in a continuous setting according to the unknown shape $\mu$. In order to simplify the criterion, we use the classic relation between mutual information and conditional entropy: $I(D_i,T)=H(D_i)-H(D_i/T)$. Since $H(D_i/T) \geq 0$ and $H(D_i)$ is independent of $T$, we will rather minimize $\sum_{i=1}^{n} H(D_i/T)$. Denoting by $t$ and $d_i$ the observations of the random variables $T$ and $D_i$, the conditional entropy of $D_i$ knowing $T$ can be written as follows:
\begin{eqnarray}
H(D_i/T)=-\sum_{t \in \{0,1\}}  \left[ p(t)\sum_{d_i \in \{0,1\}} p(d_i/t) \log (p(d_i/t)) \right],
\end{eqnarray}
with $p(T=t)=p(t)$ and $p(D_i=d_i/T=t)=p(d_i/t)$. \\
The conditional probability $p(d_i=1/t=1)$ corresponds to the sensitivity parameter $p_i$ (true positive fraction): 
\begin{eqnarray}
 p_i(\mu)= p(d_i=1/t=1)=\frac{1}{|\mu|}\int_{\mu}K(d_i(\x)-1)d\x.
\end{eqnarray}
where the function $K$ represents a Gaussian Kernel of $0$-mean and variance $\sigma$. This function allows a rigorous application of the shape derivation tools due to the fact that the function under the integral is differentiable. In this paper, we choose a very small $\sigma=0.1$ for all the experimental studies. \\
The conditional probability $p(d_i=0/t=0)$ corresponds to the specificity parameter $q_i$ (true negative fraction) :
\begin{eqnarray}
 q_i(\mu)= p(d_i=0/t=0)=\frac{1}{|\overline{\mu}|}\int_{\overline{\mu}}K(d_i(\x))d\x.
\end{eqnarray}
In the rest of the paper, for the sake of simplicity, $p_i(\mu)$ is replaced by $p_i$ and $q_i(\mu)$ by $q_i$.
The random variable $T$ takes the value $1$ with a probability $p(t=1)=|\mu|/|\Omega|$ and $0$ with a probability $p(t=0)=|\overline{\mu}|/|\Omega|$.  The $MI$ criterion can then be expressed according to $\mu$:
\begin{eqnarray}
\label{eq:criterion_IM}
MI(\mu)=& - &\sum_{i=1}^{n} \Big[ \frac{|\mu|}{|\Omega|} \left( (1-p_i) \log\left(1- p_i\right) + p_i \log p_i \right) \nonumber \\
&+&\frac{|\overline{\mu}|}{|\Omega|} \left(  q_i \log q_i + \left(1-q_i\right) \log \left(1-q_i\right)  \right) \Big]
\end{eqnarray}
The parameters $p_i$ and $q_i$ depend explicitly on $\mu$, which must be taken into account in the optimization process. Indeed if $\mu$ is updated in an iterative process, the parameters $p_i$ and $q_i$ must also be updated which implies a joint estimation of these quantities with the unknown mutual shape.

\subsection{Minimization of joint entropy}

Let us now express, according to $\mu$ and in a continuous setting, the sum of the joint entropies $JH(T)=\sum_{i=1}^{n}H(D_i,T)$. The following expression of the joint entropy is considered:
\begin{eqnarray}
H(D_i,T)=- \sum_{t \in \{0,1\}} \sum_{d_i \in \{0,1\}}  p(d_i,t) \log \left(p(d_i,t)\right),
\end{eqnarray}
with $p(D_i=d_i,T=t)=p(d_i,t)$. \\
The following estimates for the joint probabilities are then used ($a=0$ or $a=1$):
\begin{eqnarray}
 p(d_i=a,t=1)=\frac{1}{|\Omega|}  \int_{\mu} \left( K(d_i(\x)-a) \right) d\x, \nonumber \\
 p(d_i=a,t=0)=\frac{1}{|\Omega|}  \int_{\overline{\mu}} \left( K(d_i(\x)-a) \right) d\x. 
\label{eq:pdi}
\end{eqnarray}
where the function $K$ represents a Gaussian Kernel of $0$-mean and variance $\sigma$.
The criterion to minimize is now denoted by $JH(\mu)$ and can be written as follows:
\begin{eqnarray}
\label{eq:criterion_JH}
JH(\mu)&=& -\sum_{i=1}^{n} \Big[ p(d_i=1,t=1) \log(p(d_i=1,t=1))\\ \nonumber
 &-& p(d_i=1,t=0) \log(p(d_i=1,t=0)) \\ \nonumber
&-& p(d_i=0,t=1) \log(p(d_i=0,t=1)) \\ \nonumber
&-& p(d_i=0,t=0) \log(p(d_i=0,t=0)) \Big]
\end{eqnarray}
where $p(d_i=a,t=1)$ and $p(d_i=a,t=0)$ depends on $\mu$ as expressed in equations \eqref{eq:pdi}.
\subsection{Continuous expression of the criterion}

Using the two previous sections, we can express the global criterion to minimize according to $\mu$ as follows:
\begin{eqnarray}
\label{eq:continuousE}
E(\mu)&=&JH(\mu)+MI(\mu) \\ \nonumber
&=&\sum_{i=1}^{n} \Big[ -p(d_i=1,t=1) \log(p(d_i=1,t=1))\\ \nonumber
 &-& p(d_i=1,t=0) \log(p(d_i=1,t=0)) \\ \nonumber
&-& p(d_i=0,t=1) \log(p(d_i=0,t=1)) \\ \nonumber
&-& p(d_i=0,t=0) \log(p(d_i=0,t=0)) \\ \nonumber
& - &\sum_{i=1}^{n} \Big[ \frac{|\mu|}{|\Omega|} \left( (1-p_i) \log\left(1- p_i\right) + p_i \log p_i \right) \\ \nonumber
&+&\frac{|\overline{\mu}|}{|\Omega|} \left(  q_i \log q_i + \left(1-q_i\right) \log \left(1-q_i\right)  \right) \Big]\nonumber
\end{eqnarray}
In this given form, the minimization of such a criterion can be considered using active contours and shape gradients as detailed in the following section.

\section{Optimization using shape gradients}
\label{sec:optimization}

In order to compute a local minimum of the criterion $E$ defined in \eqref{eq:continuousE}, we propose to take advantage of the framework developed in \cite{Aubert03} which is based on the shape optimization tools proposed in \cite[Chap.8]{Zolesio}. The main idea is to deform an initial curve (or surface) towards the boundaries of the region of interest.

Formally, the contour then evolves according to the following Partial Differential Equation (PDE):
\begin{eqnarray}
\frac{\partial\Gamma(z,\tau)}{\partial \tau}=v(\x,\mu) \N(\x)
\end{eqnarray}
where $\Gamma(z,\tau)$ is the evolving curve, $z$ a parameter of the curve, $\tau$ the evolution parameter, $v(\x,\mu)$ the amplitude of the velocity in $\x=\Gamma(z,\tau)$ directed along the normal of the curve $\N(\x)$. The evolution equation and more particularly the velocity $v$ must be computed in order to make the contour evolve towards an optimum of the energy criterion. From an initial curve $\Gamma_0$ defined by the user, we will have $\lim\limits_{\tau \rightarrow \infty}\Gamma(\tau)=\mu$ at convergence of the process. 

The main issue lies in the computation of the velocity $v$ in order to find the unknown shape $\mu$ at convergence. This term is deduced from the derivative of the criterion according to the shape. The method of derivation is explained in details in \cite{Aubert03} and is based on shape derivation principles developed formally in \cite{Zolesio,Sokolowski_book_92}. 
For completeness, we recall some useful definitions and theorems.

\subsection{Main mathematical tools}
The following theorem is the central theorem for derivation of integral domains of the form $\int_\mu k(\x,\mu)\,d\x$. It gives a general relation between the Eulerian derivative and the shape derivative for region-based terms.
\begin{theorem}
\label{th:der}
Let $\Omega$ be a $C^1$ domain in $\mathbb{R}^n$ and $\V$ a $C^1$ vector field. Let $k$ be a $C^1$ function. The functional $J(\mu)=\int_\mu k(\x,\mu)\,d\x$ is differentiable and its Eulerian derivative in the direction of $\V$ is the following:
\begin{eqnarray}
<J'(\mu),\V>= \int_{\mu} k_s(\x,\mu)\, d\x-\int_{\partial \mu} k(\x,\mu) ( \V \cdot \N)\, d {\mathbf{a}}
\end{eqnarray}
where $k_s$ is the shape derivative of $k$ defined by $k_s(\x,\mu)=\lim_{\tau \rightarrow 0}\frac{k(\x,\mu(\tau))-k(\x,\mu)}{\tau}$. The term $\N$ denotes the unit inward normal to $\partial\mu$ and
$d{\mathbf{a}}$ its area element (in $\mathbb{R}^2$, we have $d{\mathbf{a}}=ds$ where $s$ stands for the arc length). 
\end{theorem}
The Eulerian derivative of $J$ in the direction $\V$ is defined as $$<J'(\mu),\V>= \lim_{\tau \rightarrow 0} \frac{J(\mu(\tau))-J(\mu)}{\tau}$$ if the limit exists, with $\mu(\tau)=T_{\tau}(\V)(\mu)$ the transformation of $\mu$ through the vector field $\V$.
The proof of the theorem can be found in \cite{Zolesio}.

\subsection{Methodology for the computation of the evolution equation}
The following proposition gives us a way to compute the evolution equation of the active contour when the Eulerian derivative can be expressed as an integral over the boundary of the domain.
\begin{proposition}
\label{th:velocity}
Let us consider that the shape derivative of the criterion $J(\mu)$ in the direction $\V$ may be written in the following way: 
\begin{equation}
\label{eq:Eul_der}
<J'(\mu),\V>=-\int_{\partial \mu} v(\x,\mu) ( \V \cdot \N )  d\mathbf{a}
\end{equation}
Interpreting this equation as the $L^2$ inner product on the space of velocities, the straightforward choice in order to minimize $J(\mu)$ consists in choosing $\V=v\N$ for the deformation. We can then deduce that, from an initial contour $\Gamma_0$, the boundary $\partial \mu$ can be found at convergence of the following evolution equation:
\begin{eqnarray}
\label{eq:EDP}
\frac{\partial\Gamma}{\partial\tau}=v(\x,\mu) \,\N
\end{eqnarray}
where $v$ is the velocity of the curve and $\tau$ the evolution parameter.
\end{proposition}
The shape derivatives of the criteria $SD(\mu)$ \eqref{eq:criterion_SD}, $MI(\mu)$ \eqref{eq:criterion_IM} and $JH(\mu)$ \eqref{eq:criterion_JH}, can be written in the form \eqref{eq:Eul_der} which allows us to find some geometrical PDEs of the form \eqref{eq:EDP} for each criterion. The derivation is developed thereafter.

\subsection{Shape derivatives}

This paragraph details the shape derivatives of the criteria $SD(\mu)$ \eqref{eq:criterion_SD}, $MI(\mu)$ \eqref{eq:criterion_IM} and $JH(\mu)$ \eqref{eq:criterion_JH}. Proofs of the two main new theorems \ref{theo:der_IM} and \ref{theo:der_JH} are given in the appendix of the paper.

\begin{theorem}
\label{theo:histogram-gen} 
The shape derivative in the direction $\V$ of the functional $SD(\mu)$ given in (\ref{eq:criterion_SD}) is:
$$
<SD'(\mu),\V>= -\int\limits_{\Gamma} v_{SD} \, ( \V \cdot \N )  d\mathbf{a} 
$$
with the velocity :
\begin{eqnarray}
\label{eq:vSD}
 \quad v_{SD}= \sum_{i=1}^{n}(1-2 \, d_i(\x)).
\end{eqnarray}
\end{theorem}
The computation of the shape derivative of $MI(\mu)$ is more complex because the functions inside the integrals depend on $\mu$. 
\begin{theorem}
\label{theo:der_IM} 
The shape derivative in the direction $\V$ of the functional $MI(\mu)$ defined in
(\ref{eq:criterion_IM}) is:
$$
<MI'(\mu),\V>= -\int_{\Gamma} v_{MI} \, (\V \cdot \N) d\mathbf{a}
$$
with the velocity 
\begin{eqnarray}
\label{eq:vMI}
v_{MI}=& & \frac{1}{|\Omega|} \sum_{i=1}^{n} \Big[(p_i -K(d_i-1)) \log \left(\frac{pi}{1-p_i} \right) \\
&-& (q_i -K(d_i)) \log \left(\frac{qi}{1-q_i} \right) \nonumber \\
&+& q_i \log q_i + (1-q_i) \log (1-q_i) \nonumber \\ 
&+& p_i \log p_i + (1-p_i) \log (1-p_i) \Big] \nonumber
\end{eqnarray}
\end{theorem}
The computation of the shape derivative of $JH(\mu)$ is also complex and leads to the following theorem:
\begin{theorem}
\label{theo:der_JH} 
The shape derivative in the direction $\V$ of the functional $JH(\mu)$ defined in
(\ref{eq:criterion_JH}) is:
$$
<JH'(\mu),\V>= -\int_{\Gamma} v_{JH} \, (\V \cdot \N)  d\mathbf{a} \nonumber 
$$
 The velocity $v_{JH}$ is given by the following equation:
\begin{eqnarray}
\label{eq:vJH}
v_{JH}&=&\frac{-1}{|\Omega|} \sum_{i=1}^{n} \Big[ K(d_i-1) \log \left(\frac{p(d_i=1,t=1)}{p(di=1,t=0)}\right) \nonumber \\
&+&K(d_i) \log \left(\frac{p(d_i=0,t=1)}{p(di=0,t=0)}\right) \Big].
\end{eqnarray}
where $v_{JH}$ is directed along $\N$.
\end{theorem}

\subsection{Global evolution equations for the different criteria}

A standard regularization term is added in the criterion to minimize in order to favor smooth shapes.
This term corresponds to the minimization of the curve length and is defined by $Reg(\mu)=\int_{\partial \mu} ds$. It is balanced with a positive coefficient $\lambda$ in the criterion and leads to the following velocity in the evolution equation:
\begin{eqnarray}
\label{eq:vReg}
v_{Reg}=\kappa
\end{eqnarray}
where $\kappa$ is the curvature of the contour $\Gamma(\tau)$.

Finally, we propose to define our mutual reference shape through the minimization of a global criterion called $J_{IT}$ (Information Theoretic criterion):
\begin{eqnarray}
\label{eq:crit-IT}
J_{IT}(\mu)=JH(\mu)+MI(\mu)+ \lambda Reg(\mu).
\end{eqnarray}
In order to minimize this criterion, the following evolution equation is used:
\begin{eqnarray}
\label{eq:evol_MI_JH}
\left(\frac{\partial \Gamma}{\partial \tau}\right)_{IT}=\left( v_{JH} + v_{MI}+ \lambda v_{Reg} \right) \bf{N} 
\end{eqnarray}
where $v_{MI}$, $v_{JH}$ and $v_{Reg}$ are defined respectively in equations \eqref{eq:vMI}, \eqref{eq:vJH} and  \eqref{eq:vReg}. The term $\bf{N}$ designates the inward unit normal of the active contour.
In the experimental results, the mutual reference shape is also compared to the average shape (SD) that corresponds to the minimization of the following criterion:
\begin{eqnarray}
J_{SD}(\mu)=SD(\mu)+ \lambda Reg(\mu).
\end{eqnarray}
In order to minimize this criterion, the following evolution equation is applied:
\begin{eqnarray}
\label{eq:evol_SD}
\left(\frac{\partial \Gamma}{\partial \tau}\right)_{SD}=\left( v_{SD} + \lambda v_{Reg} \right) \bf{N} 
\end{eqnarray}
where $v_{SD}$ and $v_{Reg}$ are defined respectively in equations \eqref{eq:vSD} and  \eqref{eq:vReg}. These velocities are directed along the unit inward normal $\bf{N}$ of the active contour.

Note also that using this formalism, some other prior information (photometric or geometric) can be inserted by adding some additional velocities in the PDE. For example, we may take advantage of the tools developed in \cite{Chan01,Foulonneau_PAMI06,Lec_icip06,Paragios_03}. 

\subsection{Implementation of the active contour}
\label{ssec:implementation}
As far as the numerical implementation is concerned, we use the level set method  \cite{Osher_1988}. The key idea is to introduce an auxiliary function \(U(\x,\tau)\) such that \(\Gamma(\tau)\) is the zero level set of \(U\). The function \(U\) is often chosen to be the signed distance function of
\(\Gamma(\tau)\). The evolution equation then becomes:
\begin{eqnarray}
\label{eq:EDP_U}
\frac{\partial U}{\partial \tau}=F |\nabla U|.
\end{eqnarray}
The velocity is chosen as $F=v_{JH} + v_{MI}+ \lambda v_{Reg}$ for the estimation of the mutual shape and $F=v_{SD} + \lambda v_{Reg}$ for the estimation of the SD shape. 
This method is accurate and allows to automatically handle the topological changes of the initial curve. However, the same evolution equations could be implemented using faster implementation algorithms such as B-splines \cite{Precioso05}. Convex optimization methods \cite{Bresson2007} may perhaps be interesting but the criterion is not convex and some assumptions are needed before a direct application of these methods.

\section{Experimental results on a synthetic example}
\label{sec:res}

The behavior of our mutual shape estimation is first tested on a synthetic example. The mutual shape, the classic average shape and a simple majority voting approach are compared. We also study the joint evolution of the sensitivity and specificity parameters.

\subsection{Difference between a mutual shape and a classic average shape}

\begin{figure}[htb]
\center
\begin{tabular}{cc}
\fbox{\includegraphics[height=1.7cm]{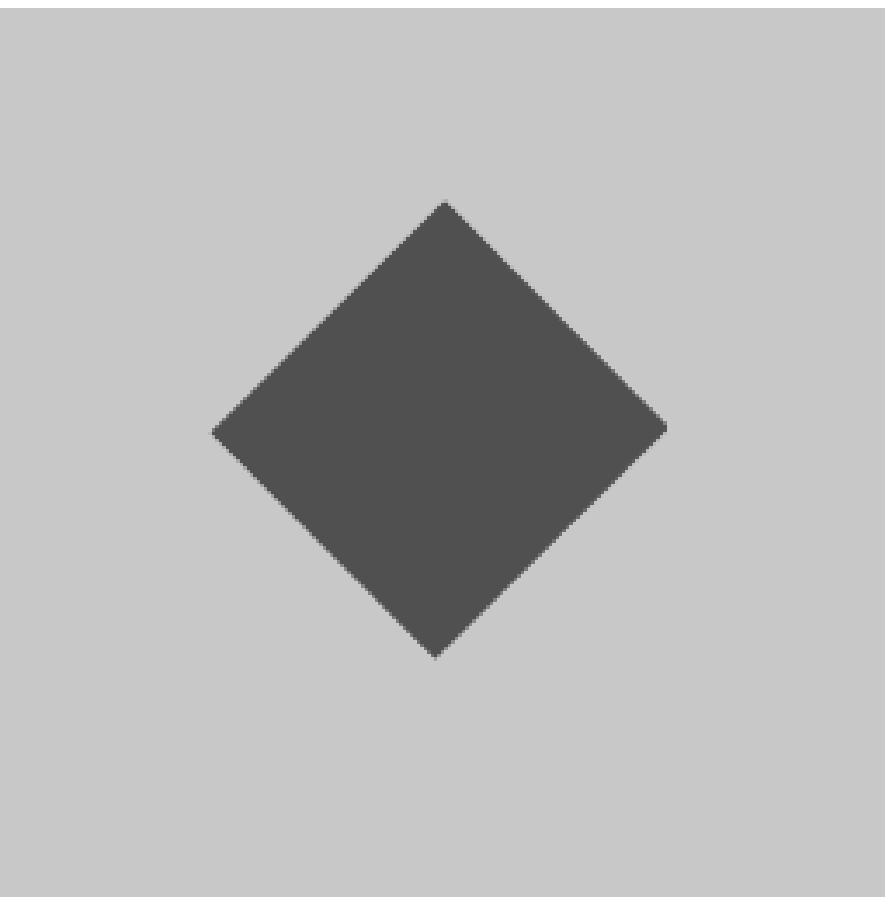}} & \fbox{\includegraphics[height=1.7cm]{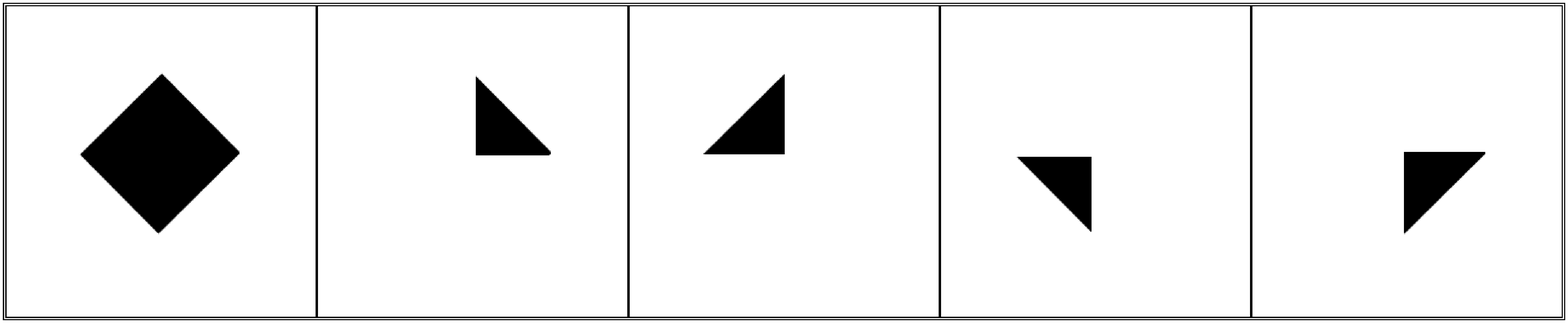} }\\
(a) & (b)
\end{tabular}
\caption{\footnotesize The image to segment is given in (a) and the different segmentation entries (masks) for this image are given in (b).}
\label{Fig:mask_lozenge}
\end{figure}

In this section, a test sequence consisting of different segmentations of a lozenge (Fig.\ref{Fig:mask_lozenge}) was built. The first entry is the true segmentation mask, the other entries represent the segmentation of $1/4$ of the true lozenge (Fig.\ref{Fig:mask_lozenge}.b). 

\begin{figure}[h]
\center
\begin{tabular}{cc}
\fbox{\includegraphics[height=1.7cm]{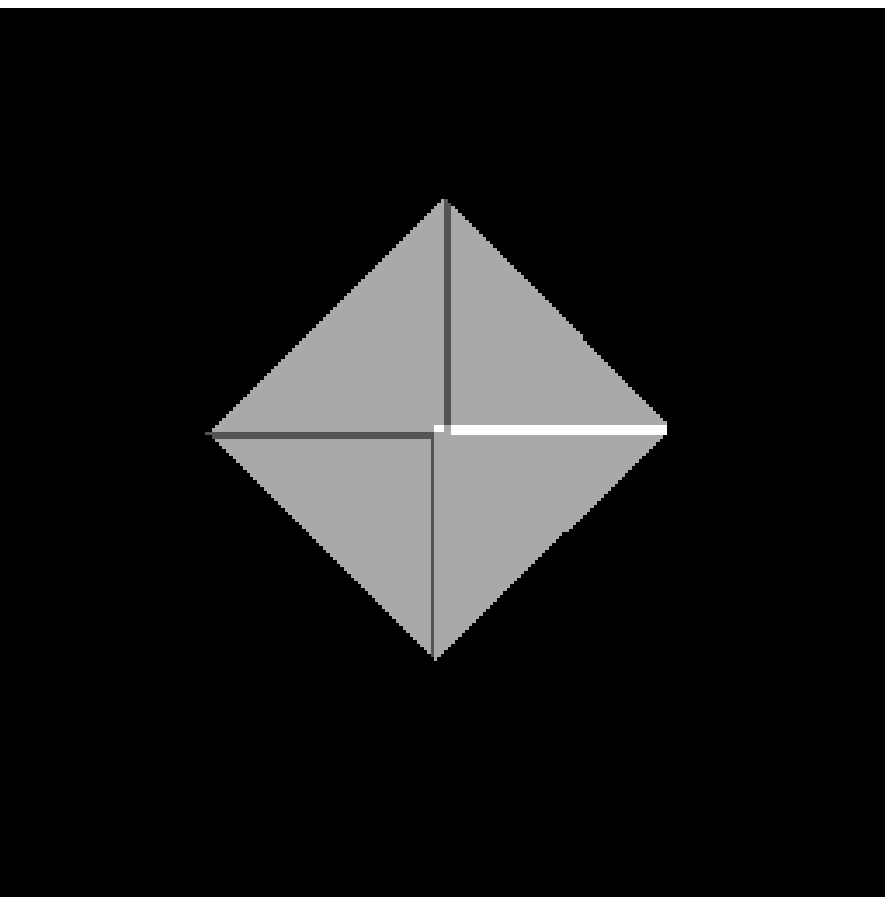} }&  \fbox{\includegraphics[height=1.7cm]{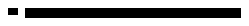}} \\
(a)  & (b)  
\end{tabular}
\caption{\footnotesize The average image $I_A$ (a) and the corresponding binarized average image $I_{AT}$ (b) of the masks of the Fig.\ref{Fig:mask_lozenge}(b) (simple majority voting procedure).}
\label{Fig:mean_mask_lozenge}
\end{figure}

When computing the average of the different characteristic functions using the formula : $\sum_{i=1}^{n} d_i / n$, we remark (Fig.\ref{Fig:mean_mask_lozenge}.b) that some masks share an intersection. Indeed the values of the average image belong to the interval $[0,0.6]$. The value $0$ corresponds to black points in Fig.\ref{Fig:mean_mask_lozenge}.a and the value $0.6$ corresponds to white points. We then binarize this average image $I_A$ in an image named $I_{AT}$ displayed in (Fig.\ref{Fig:mean_mask_lozenge}.b). If $I_A(\x) \geq 0.5$ then $I_{AT}=0$ (black points) and if $I_A(\x) < 0.5$ then $I_{AT}=255$ (white points). This procedure gives us a simple majority voting procedure. The result is the black line inside the lozenge. 

\begin{figure}[h]
\center
\begin{tabular}{c c c c}
\includegraphics[height=2.3cm]{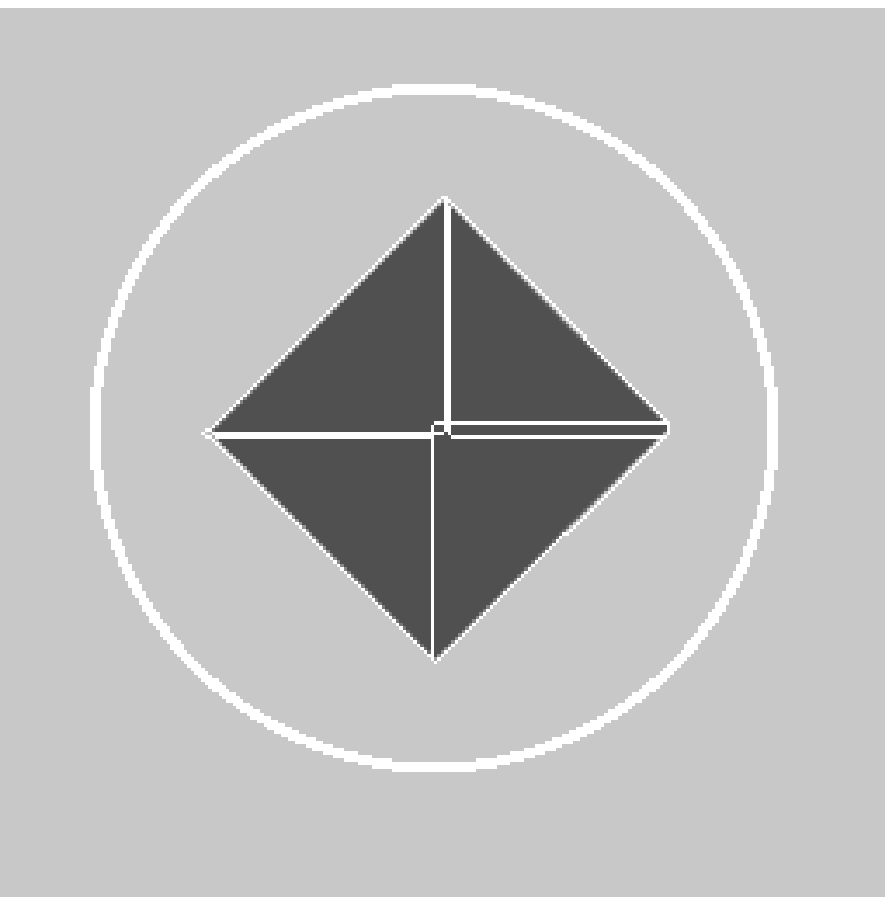} &
 \includegraphics[height=2.3cm]{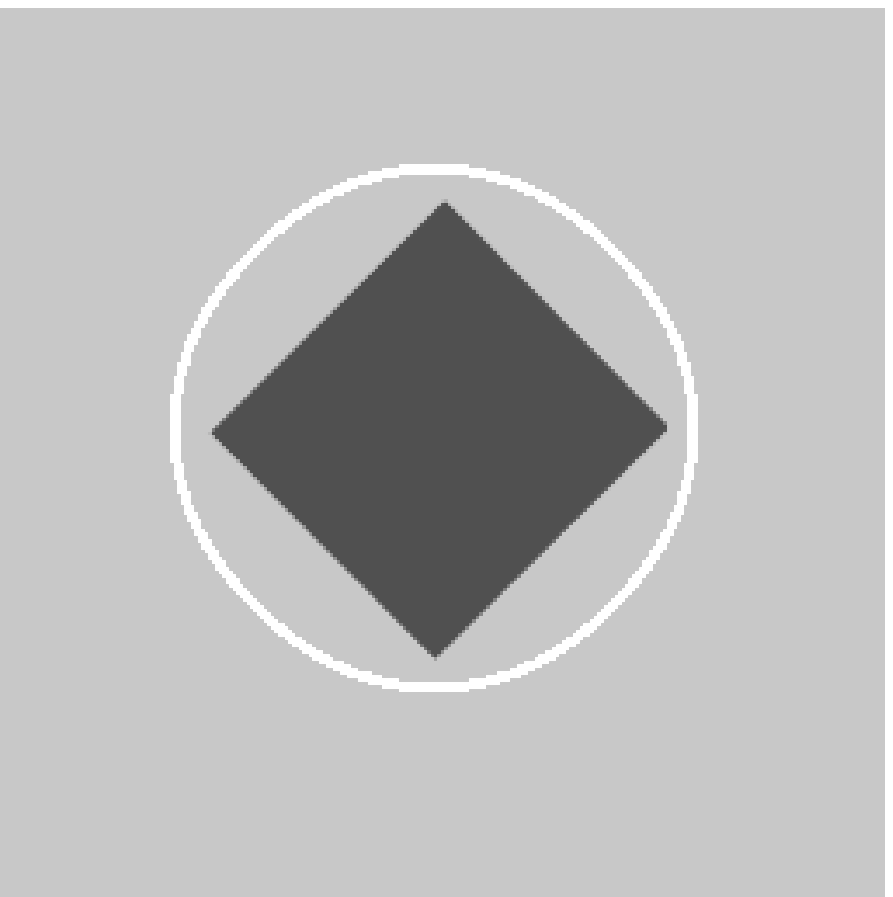} &
 \includegraphics[height=2.3cm]{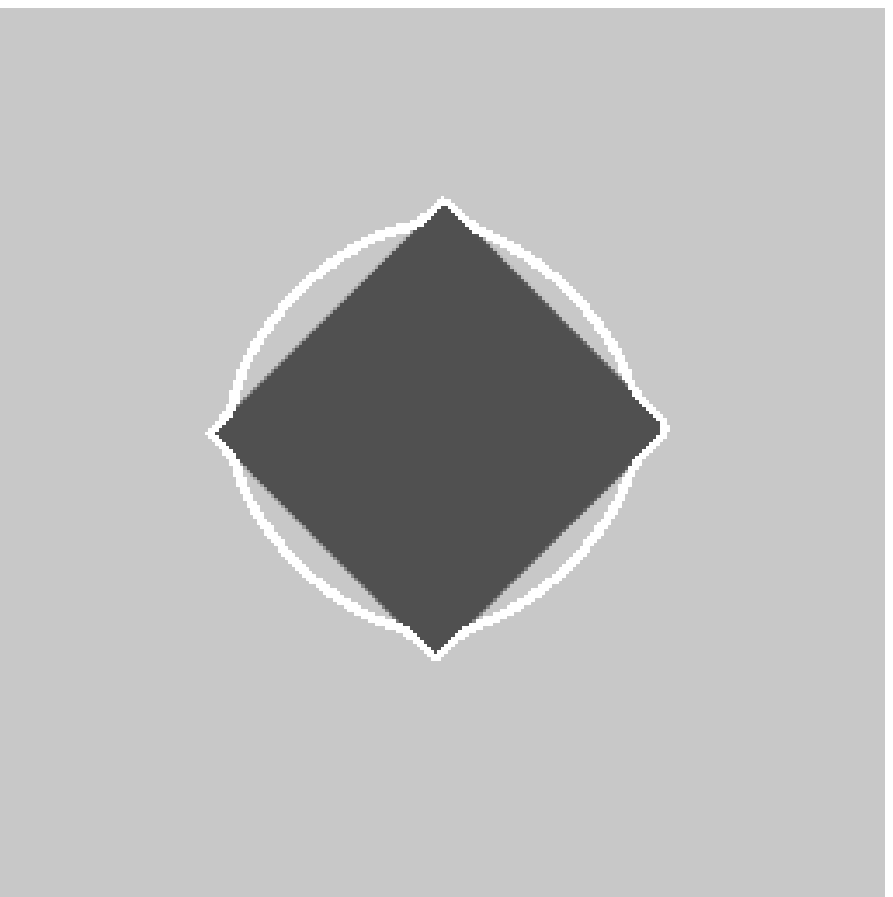} &
 \includegraphics[height=2.3cm]{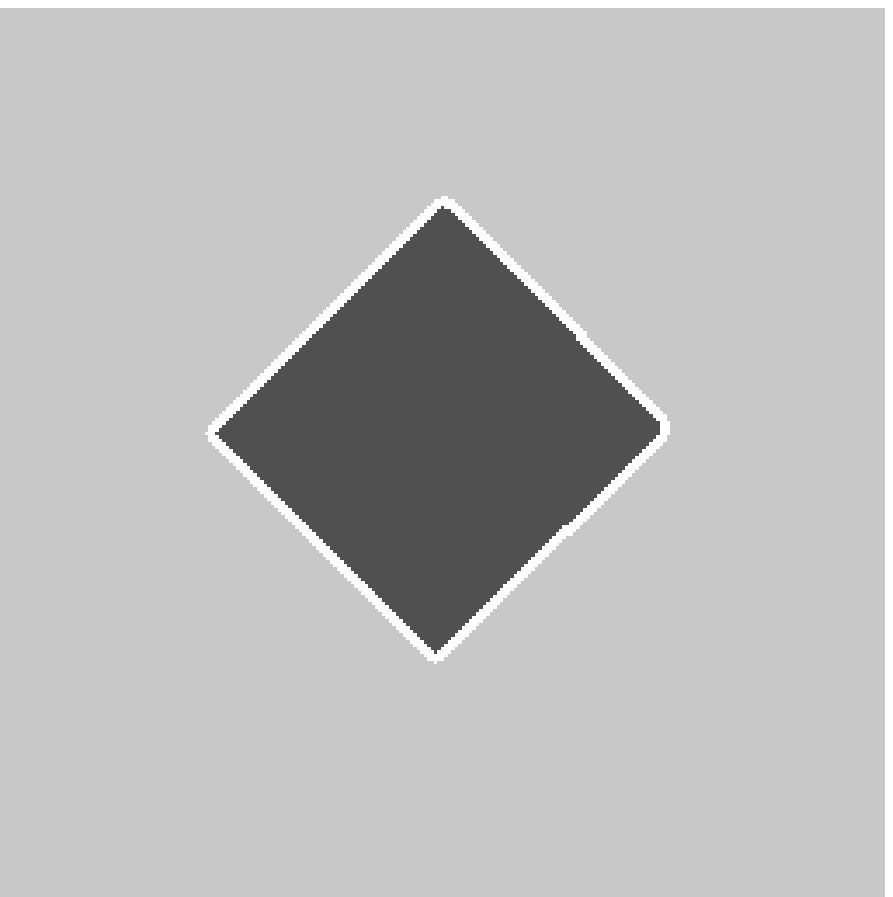} \\
(a) Initial contour & (b) It. 80  & (c) It. 140 & (d) Mutual shape
\end{tabular}
\caption{\footnotesize Evolution using the mutual shape (evolution equation \eqref{eq:evol_MI_JH} with $\lambda=10$). In the first image (a), the initial contour is in white (circle) and the other white lines represent the boundaries of the different segmentation entries. Intermediate results obtained from $80$ and $140$ iterations are displayed in images (b) and (c) and the final estimated mutual shape in (d) (240 iterations).}
\label{Fig:evol_lozenge_mutual}
\end{figure}

\begin{figure}[h]
\center
\begin{tabular}{c c c c}
\includegraphics[height=2cm]{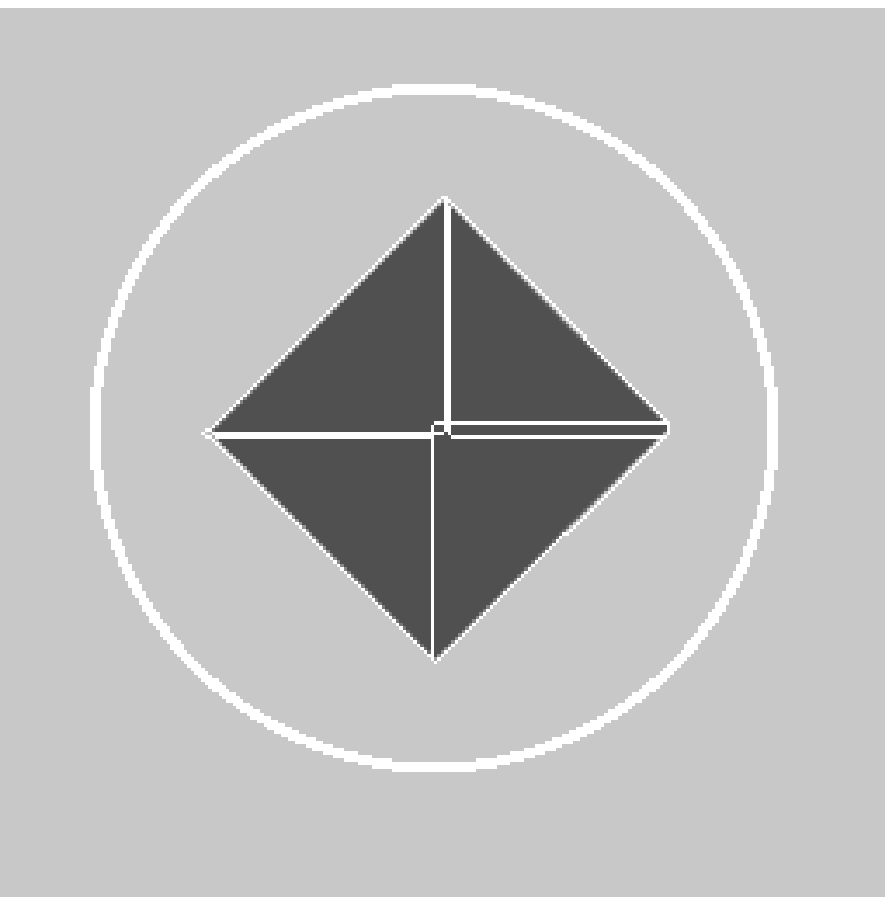} &
 \includegraphics[height=2cm]{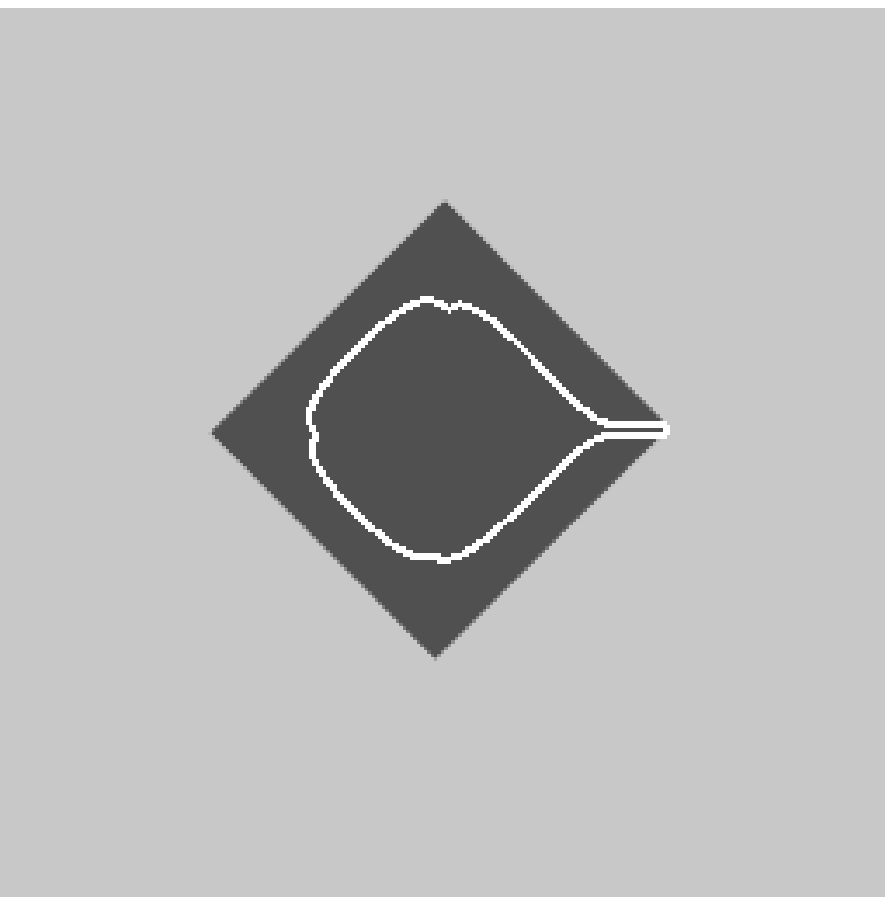} &
 \includegraphics[height=2cm]{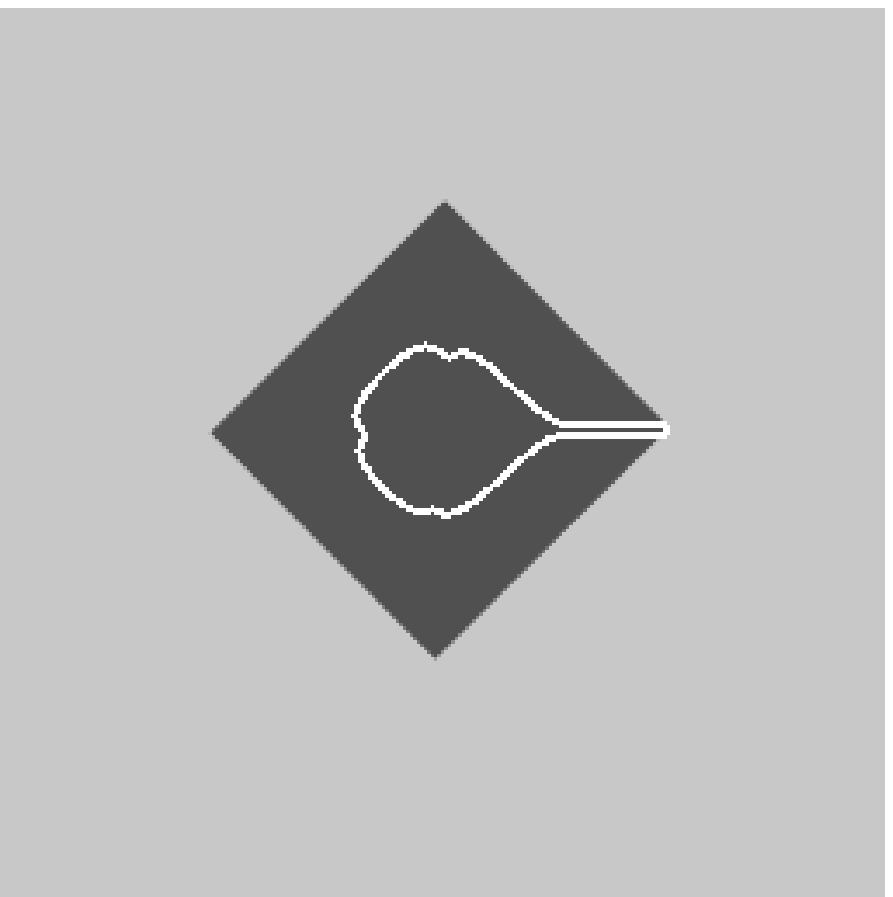} &
 \includegraphics[height=2cm]{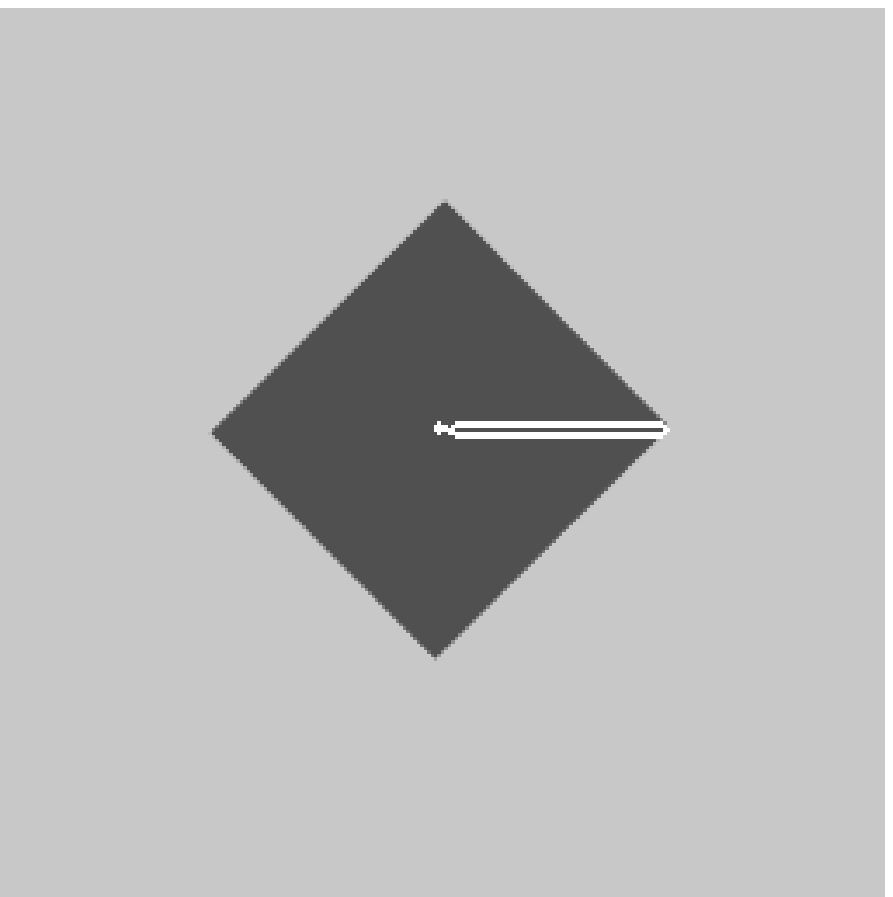} \\
(a) Initial contour & (b) It. 300  & (c) It. 400 & (d) SD shape 
\end{tabular}
\caption{\footnotesize Evolution using the SD shape (evolution equation \eqref{eq:evol_SD} with $\lambda=10$). In the first image, the initial contour is in white (circle)  and the other white lines represent the boundaries of the different segmentation entries. Intermediate results obtained from $300$ and $400$ iterations are displayed in images (b) and (c) and the final estimated SD shape in (d) (600 iterations).}

\label{Fig:evol_lozenge_SD}
\end{figure}

We then use the evolution equations of both the mutual shape \eqref{eq:evol_MI_JH} and of the SD shape \eqref{eq:evol_SD}. The initial contour is chosen as a circle including the lozenge (Fig.\ref{Fig:evol_lozenge_mutual}.a and Fig.\ref{Fig:evol_lozenge_SD}.a). The mutual shape algorithm is able to recover the whole lozenge and is then different from a classic average shape (see Fig.\ref{Fig:evol_lozenge_mutual} and Fig.\ref{Fig:evol_lozenge_SD}). The curve evolves and segments the whole lozenge by an iterative process (images resulted from different iterations in Fig.\ref{Fig:evol_lozenge_mutual}.b and Fig.\ref{Fig:evol_lozenge_mutual}.c). The final contour is given in Fig.\ref{Fig:evol_lozenge_mutual}.d. The mutual shape is compared to a shape average computed using the minimization of the classic symmetrical difference (criterion $J_{SD}$ with evolution equation \eqref{eq:evol_SD}). The evolution is given in Fig.\ref{Fig:evol_lozenge_SD}. In this case, the final contour is similar to the result obtained by computing a binarized mean $I_{AT}$  (Fig.\ref{Fig:mean_mask_lozenge}.b) since it corresponds to a line due to the small overlap between masks 2 and 5 (Fig.\ref{Fig:mask_lozenge}.b). The same small value is taken for the regularization parameter $\lambda$ in order to give a higher importance to the data term. 

\subsection{Difference between the mutual shape and the union of the masks}

\begin{figure}[h]
\center
\begin{tabular}{c c c c c}
\fbox{\includegraphics[height=1.9cm]{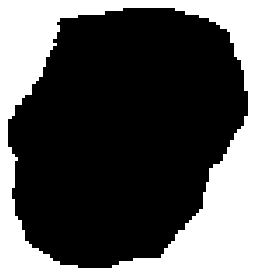}} & 
\includegraphics[height=2cm]{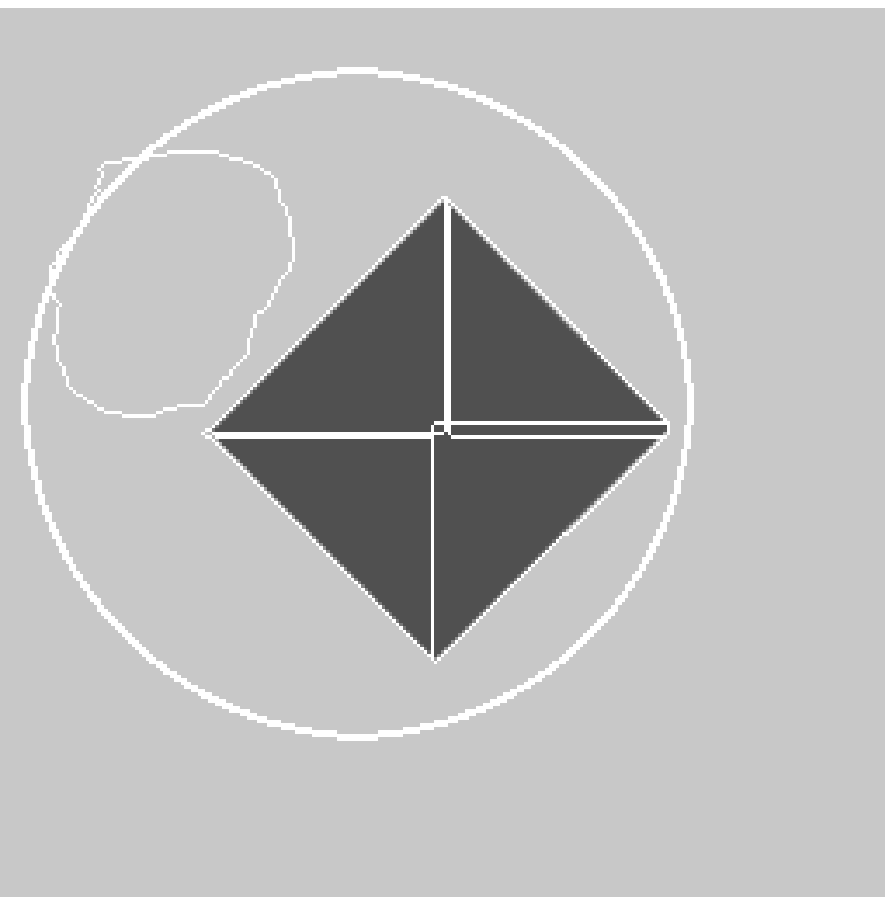} 
& \includegraphics[height=2cm]{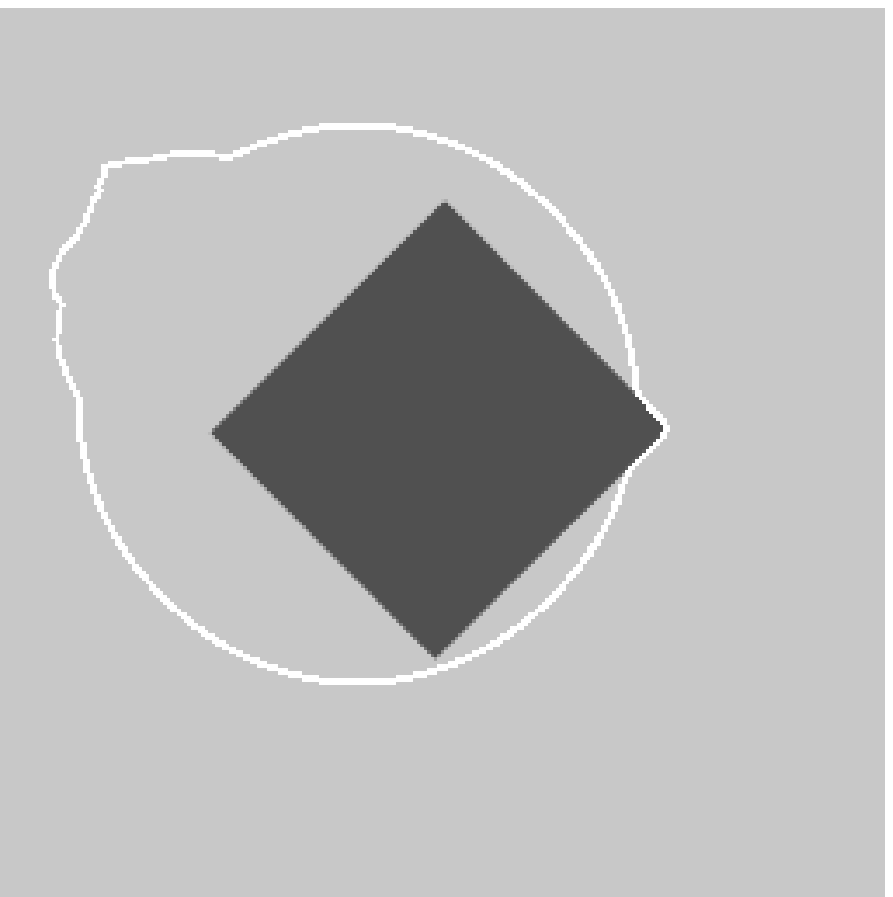}
& \includegraphics[height=2cm]{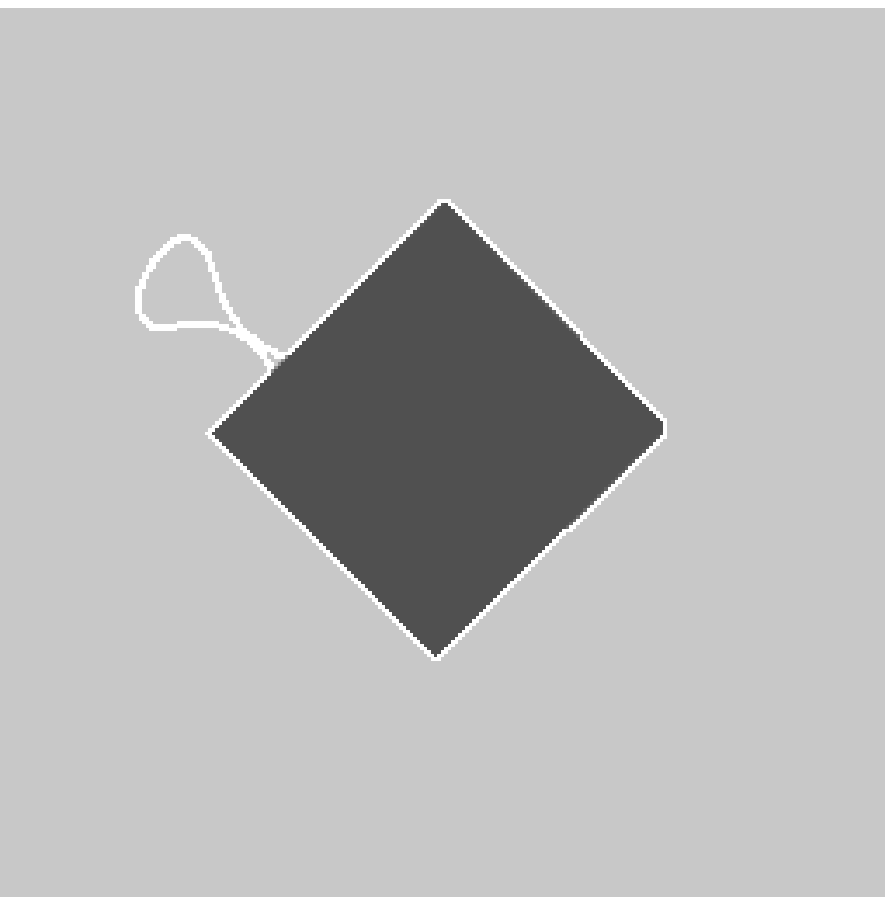}
& \includegraphics[height=2cm]{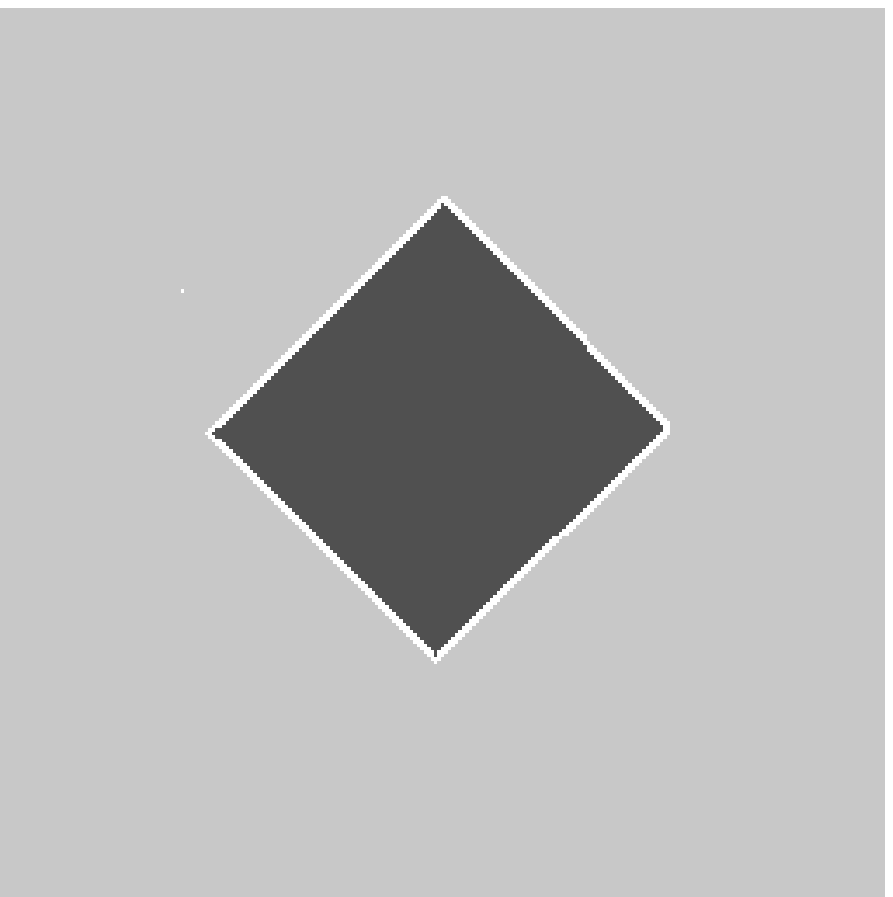} \\
(a) Input outlier & (b) Initial & (c) It. 100 & (d) It. 380 & (e) Mutual shape
\end{tabular}
\caption{\footnotesize Introduction of an outlier (a) in the initial sequence of masks (Fig.\ref{Fig:mask_lozenge}.b) and estimation of the mutual shape (evolution equation \eqref{eq:evol_MI_JH} with $\lambda=10$). In the image (b), the initial contour is in white and the other white contours and lines represent the different boundaries of the initial masks (the segmentation entries and the outlier).}
\label{Fig:lozenge_outlier_init}
\end{figure}

An outlier (Fig.\ref{Fig:lozenge_outlier_init}.a) was introduced in the initial sequence of masks in order to test the robustness of the mutual shape estimation. Indeed, our goal is to test that the mutual shape is also different to a simple union of the different masks. In Fig.\ref{Fig:lozenge_outlier_init}, the different steps of the evolution of the contour are displayed. The final contour (Fig.\ref{Fig:lozenge_outlier_init}.d) fits the lozenge and excludes the outlier from the final contour.

\subsection{Joint evolution of the sensitivity and specificity parameters}

When the active contour evolves using the evolution equation \eqref{eq:evol_MI_JH}, the parameters $p_i$ and $q_i$ are estimated jointly with the mutual shape as proposed in STAPLE \cite{Warfield_TMI_04}. The different values of these parameters along the evolution of the curve are given in Table \ref{tab:pi_qi_lozenge}. These results are obtained using masks displayed in the first row of this Table.  According to the final values reported in Table \ref{tab:pi_qi_lozenge}, we can conclude that the best segmentation corresponds to the shape $1$ with $p_1=1$ and $q_1=1$ and that the shape $6$ is an outlier since the sensitivity coefficient is equal to $0$. The other segmentations correspond to one quarter of the lozenge which leads to a sensitivity parameter around the value of $0.25$. Note that the initial values of $p_i$ and $q_i$ are computed directly using the initial contour.

We can notice that the specificity parameter $q_i$ is less relevant. Indeed this parameter is estimated using the external domain ($\bar{\mu}$) and is then estimated using a higher number of pixels. It should be normalized in order to be comparable to the $p_i$ value. One solution consists in the selection of a smaller working area (a mask that includes the union of masks chosen in order to get two regions with a comparable size). 

\begin{table}[h]
\begin{small}
\begin{tabular}{| c | c | c | c | c |c | c |}
\hline 
& & & & & & \\
& \includegraphics[height=0.7cm]{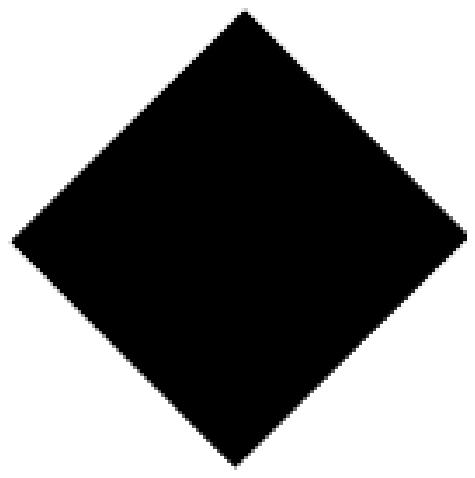} &  \includegraphics[height=0.7cm]{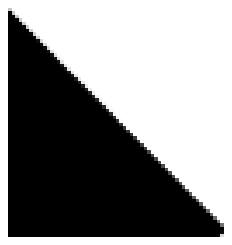} &  \includegraphics[height=0.7cm]{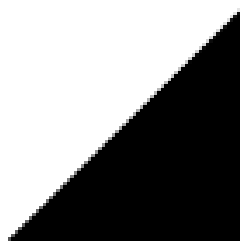} &  \includegraphics[height=0.7cm]{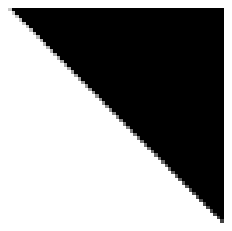} &  \includegraphics[height=0.7cm]{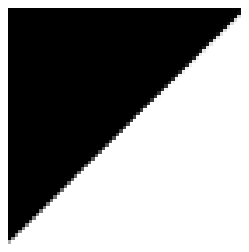} & \includegraphics[height=0.7cm]{img/losange_gd_outlier.eps}  \\
Iterations & mask 1 & mask 2 & mask 3 & mask 4 & mask 5 & mask 6 \\
\hline
It. 0 & $p_1=0.35$  & $p_2=0.09$  & $p_3=0.09 $ & $p_4=0.08$ & $p_5=0.09$ & $p_6=0.15$ \\ 
(Fig. \ref{Fig:lozenge_outlier_init}.b) & $q_1=1$ & $q_2=1$  & $q_3=1$ & $q_4=1$ & $q_5=1$ & $q_6=1$ \\
\hline
It. 100 & $p_1=0.60 $  & $p_2=0.15$  & $p_3=0.15 $ & $p_4=0.13$ & $p_5=0.16$ & $p_6=0.27$ \\ 
 (Fig. \ref{Fig:lozenge_outlier_init}.c) & $q_1=1$ & $q_2=1$  & $q_3=1$ & $q_4=1$ & $q_5=1$ & $q_6=1$ \\
\hline
Final  & $p_1=1$  & $p_2=0.24$  & $p_3=0.26 $ & $p_4=0.22$ & $p_5=0.27$ & $p_6=0$ \\ 
(Fig. \ref{Fig:lozenge_outlier_init}.e) &  $q_1=1$ & $q_2=1$  & $q_3=1$ & $q_4=1$ & $q_5=1$ & $q_6=0.93$ \\
\hline
\end{tabular}
\end{small}
\caption{\footnotesize Joint evolution of the contour and of the sensitivity and specificity parameters $p_i$ and $q_i$ for the masks 1 to 6. The values correspond to the evolution of the contour displayed in Fig.\ref{Fig:lozenge_outlier_init} (initial contour, iteration 100  and final contour).}
\label{tab:pi_qi_lozenge}
\end{table}

\section{Experimental results on real images}
\label{sec:applis}

In this section, our aim is to provide a variety of examples where the proposed mutual shape can be valuable. Indeed the theoretical framework proposed above is generic and can be applied to different images, modalities, shapes and applications. First of all, in subsection \ref{ssec:color}, we provide a simple example on a real color image from the Berkeley database \cite{MartinFTM01} to show the robustness of our estimation to an outlier, the accuracy of the obtained contour and the relevance of the classification performed using $p_i$ and $q_i$. As already mentioned, the implementation is performed using the level set method which automatically handles topological changes. Therefore, we then apply the estimation of the mutual shape for complicated shapes composed of multiple separated components such as the text in old documents. In the subsection (\ref{ssec:text}), we give two examples : the first one is dedicated to the fusion of very simple binarization techniques while the second one performs fusion and evaluation of real automatic binarization methods from the DIBCO challenge \cite{DIBCO13}. In the subsection \ref{ssec:medical}, we propose to test the mutual shape for the fusion and the evaluation without gold standard of different segmentation methods or expert delineations of the left ventricle in cardiac magnetic resonance images (cardiac MRI) and notably expert segmentations. This estimated mutual shape is compared to the classical STAPLE estimation \cite{Warfield_TMI_04} and evaluation results are analysed on the basis of some previous works on evaluation without gold standard \cite{Lebenberg_PLOS15}. 

Let us note that the parameter $\lambda$ is chosen small ($1$ to $10$) for non convex shapes and may be chosen higher for convex shapes ($10$ to $100$). In this last case, it can help to get a more regularized contour.

\subsection{Application on a real natural color image}
\label{ssec:color}
\begin{figure}[htb]
\center
\begin{tabular}{c c c}
 \includegraphics[width=3cm]{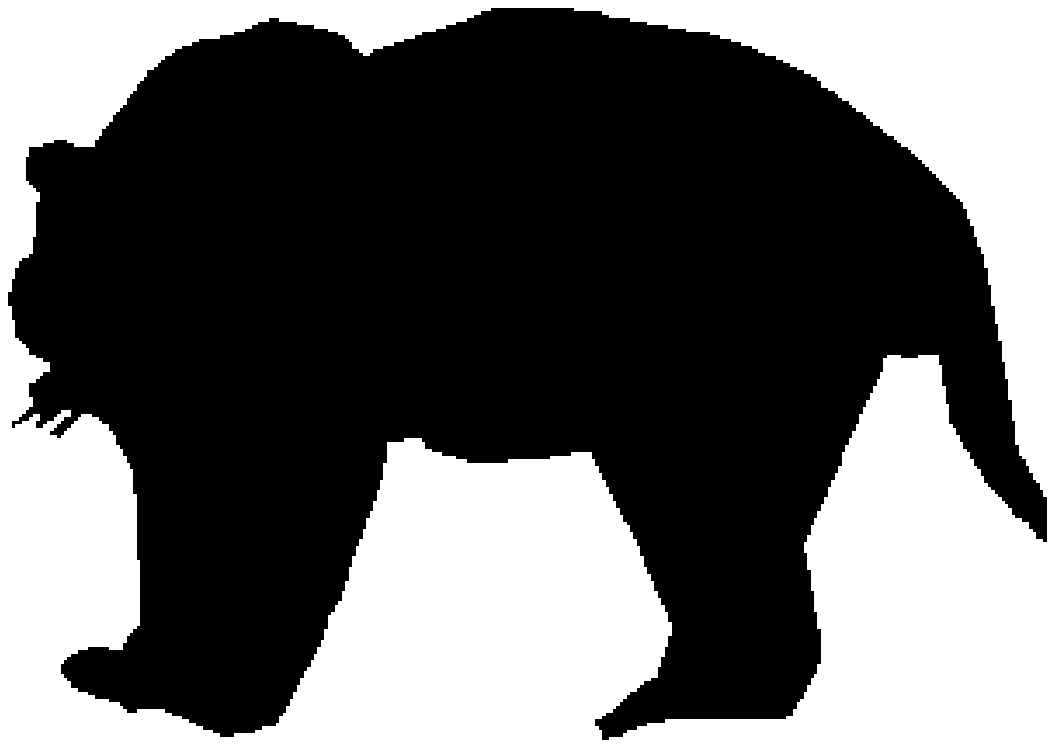} & 
 \includegraphics[width=3cm]{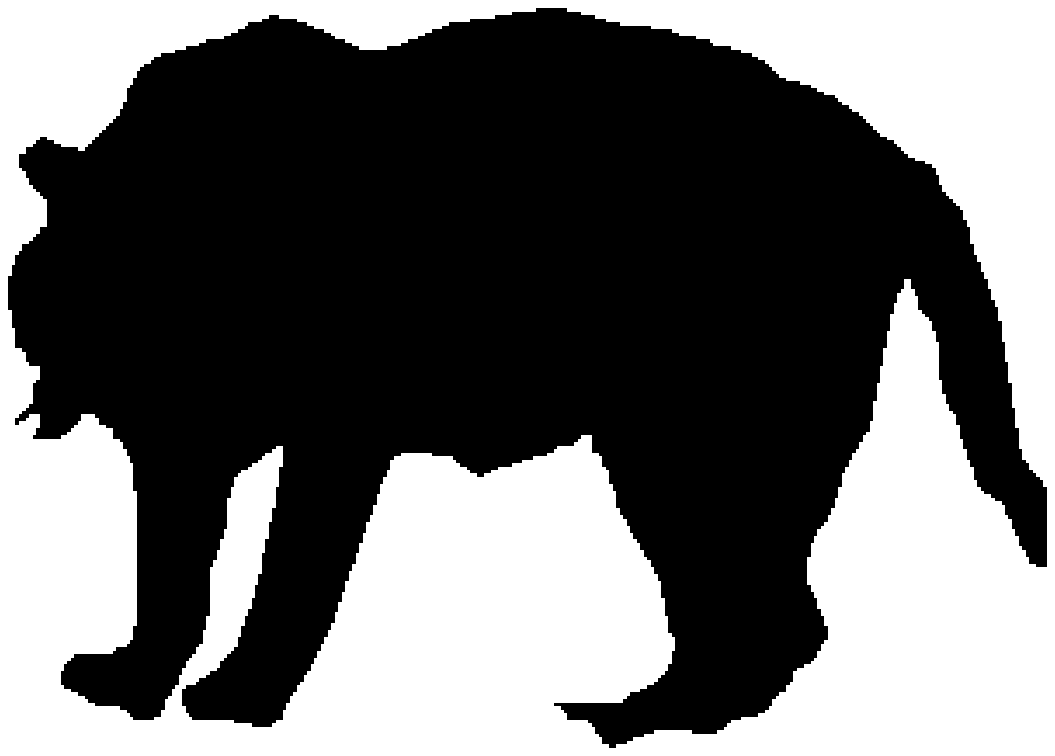} &
 \includegraphics[width=3cm]{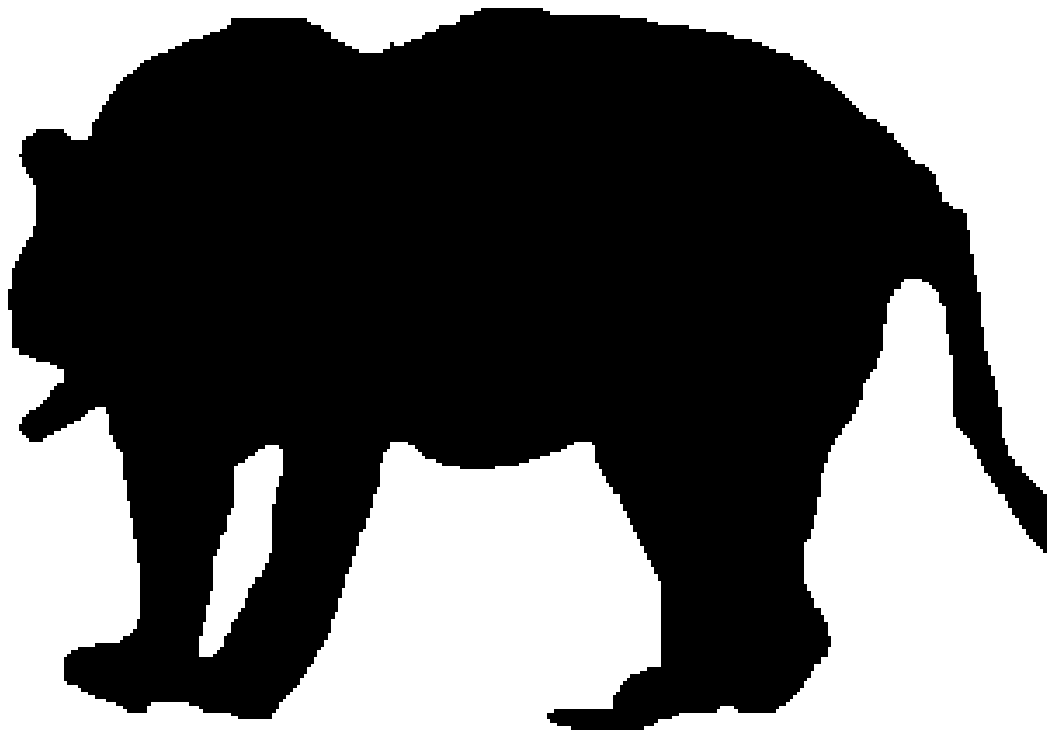} \\
 $m_1$ & $m_2$ & $m_3$ \\
\includegraphics[width=3cm]{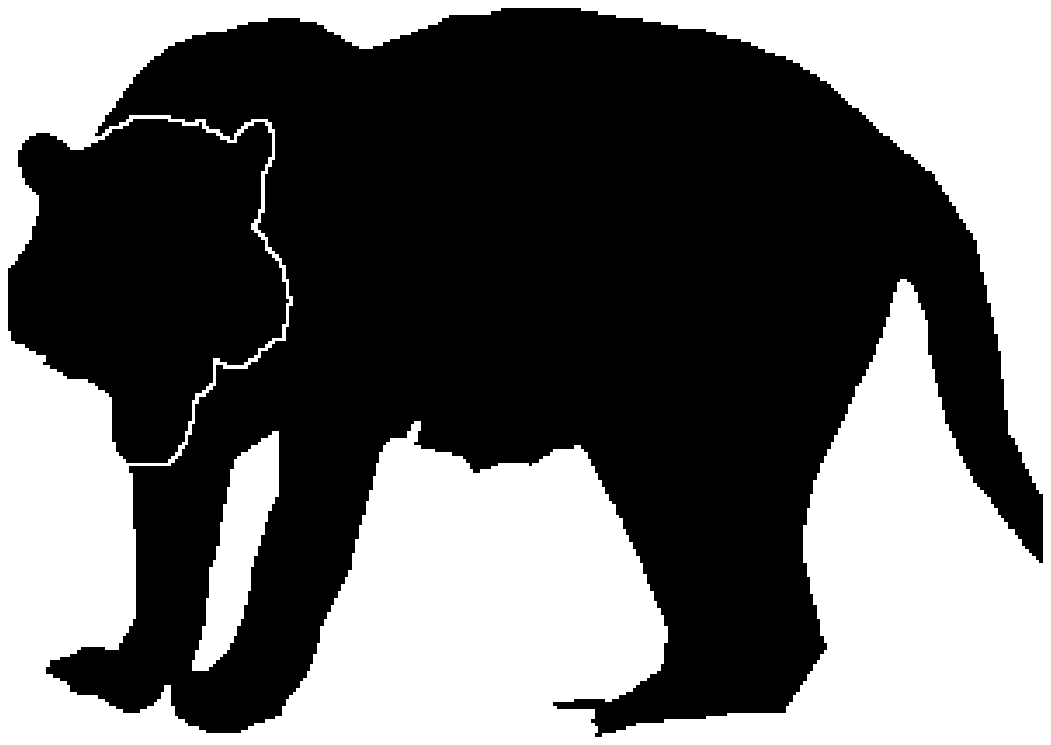} & 
\includegraphics[width=3cm]{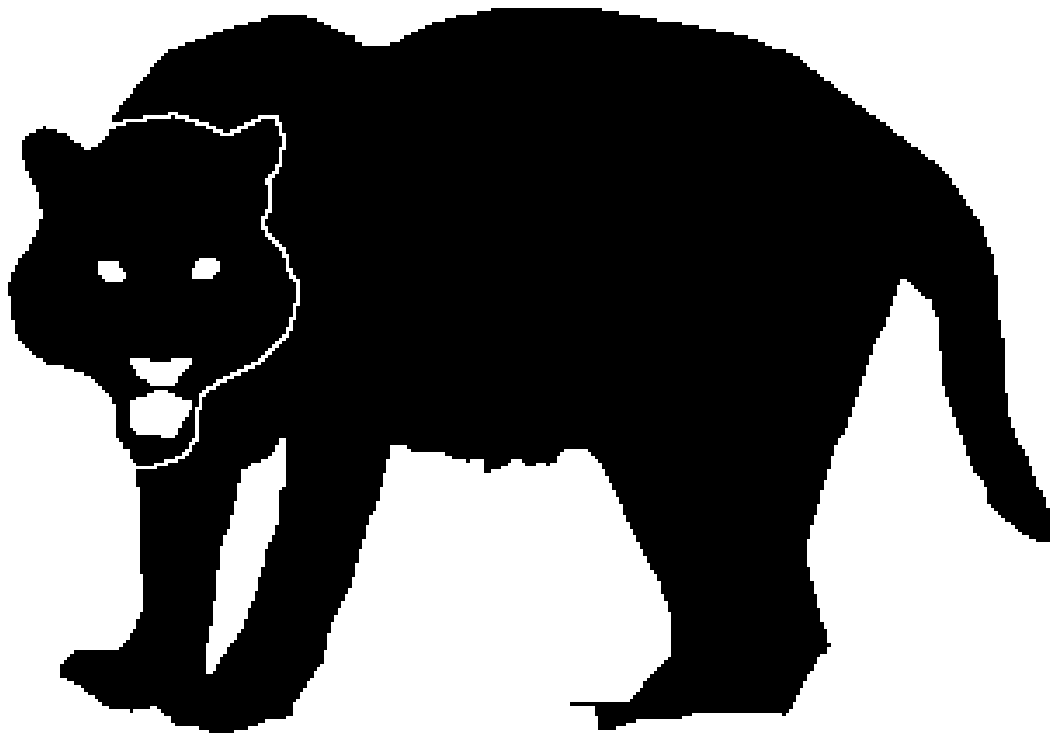} & 
\includegraphics[width=3cm]{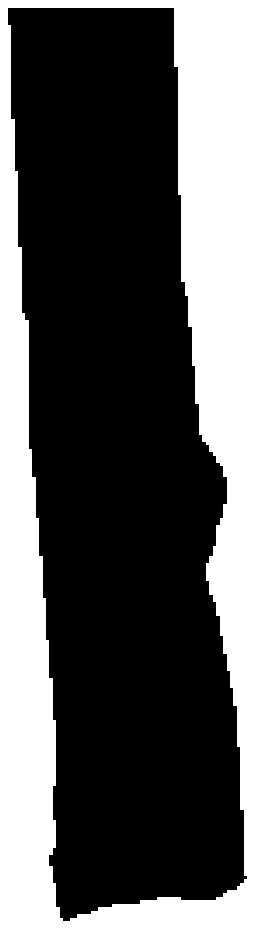} \\
$m_4$ & $m_5$ & $m_6$ \\
\end{tabular}
\caption{The different segmentation masks $m_1$ to $m_5$ and an outlier $m_6$ are taken as segmentation entries for the mutual shape estimation. }
\label{fig:tigre_seg}
\end{figure}

\begin{figure}[htb]
\center
\begin{tabular}{c c}
\includegraphics[width=4.5cm]{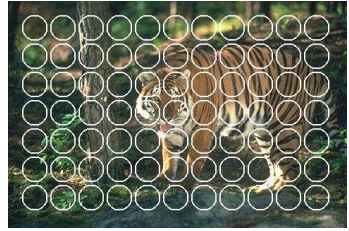} &
 \includegraphics[width=4.5cm]{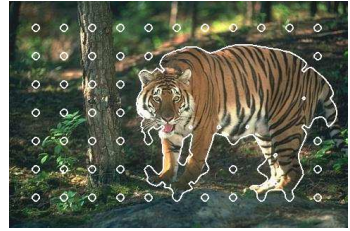} \\
(a) Initial contour & (b) Iteration 200 \\
 & \\
\includegraphics[width=4.5cm]{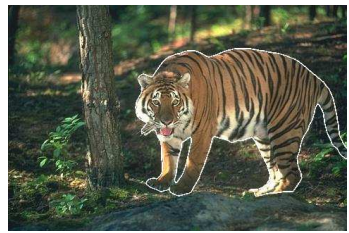} &
\includegraphics[width=4cm]{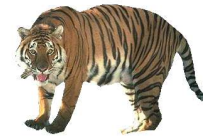} \\
(c) Estimated mutual shape (final contour)  &  (d) Final segmentation
\end{tabular}
\caption{Evolution of the active contour (in white, with $\lambda=10$), from the initial contour (a), for the estimation of the mutual shape using the masks $m_1$ to $m_6$ (Fig.\ref{fig:tigre_seg}). The final mutual shape is given in (c) and the segmented tiger in (d).}
\label{fig:res_tigre}
\end{figure}

The estimation of such a mutual shape is first tested for the unsupervised evaluation of segmentation methods of real images. The object of interest is the tiger of the image displayed in Fig.\ref{fig:res_tigre}. We then extract the object of interest from the different segmented images proposed in the Berkeley database\cite{MartinFTM01}. The different segmentation entries $m_1$ to $m_5$ are given in Fig.\ref{fig:tigre_seg} and we add an outlier $m_6$, which corresponds to the segmentation of the tree behind the tiger, to the five main segmentation entries.

\begin{table}[htb]
 \begin{center}
\begin{tabular} {|c | c | c | c | c | c | c|}
  \hline
  i  & $m_1$ & $m_2$ & $m_3$ & $m_4$ & $m_5$ & $m_6$\\
 \hline
$p_i$  & 0.988 & 0.980 & 0.975 & 0.979 & 0.971 & 0 \\
$q_i$ & 0.988 & 0.993 & 0.994 & 0.998 & 0.996 & 0.89  \\
  \hline
\end{tabular}
\caption{ Sensitivity and specificity parameters $p_i$  and $q_i$ for the segmentations $m_1$ to $m_6$ displayed in Fig.\ref{fig:tigre_seg}.}
\label{tab:tigre}
\end{center}
\end{table}

In Fig.\ref{fig:res_tigre}, we show the evolution of the active contour from the initial contours (bubbles) given in Fig.\ref{fig:res_tigre}.a. One intermediate contour is given Fig.\ref{fig:res_tigre}.b, and the final mutual shape is shown in Fig.\ref{fig:res_tigre}.c. The mutual shape provided in Fig.\ref{fig:res_tigre}.d provides an interesting result for segmentation fusion that takes benefit of the different segmentation entries while being robust to the outlier shape. The evolution of the active contour displayed in Fig.\ref{fig:res_tigre} shows that the initial shape evolves correctly towards the boundaries of the object of interest. The sensitivity and specificity parameters are computed together with this reference shape and provided in Table \ref{tab:tigre}. These parameters provide an interesting classification of the different segmentation without any given reference. The mask $m_2$ seems to be the best segmentation regarding with the estimated consensus and $m_6$ clearly appears as an outlier. However, such a classification is dependent on the choice of the different segmentation entries. We can however conclude that the mutual shape is robust to the introduction of an outlier segmentation in the entry sequence and that the outlier is clearly detected at the end of the process through the low values of its corresponding $p_i$ ($p_i=0$ in Table \ref{tab:tigre}).

Concerning the computational cost, for this example, the estimation of the mutual shape takes $37s$ using Intel-based CPU \symbol{64} 2.70GHz. The size of the image is $321*481$. 
\subsection{Application to text segmentation in old parchments}
\label{ssec:text}
A second real application of our algorithm is dedicated to the fusion of different segmentations of the text in old parchments. 

\begin{figure}[h]
\center
 \includegraphics[width=5cm]{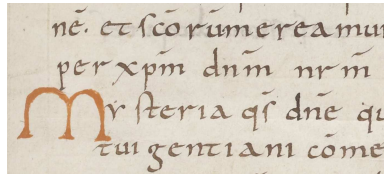} 
\caption{An original image from an old manuscript from Gallica (Gallica is the online numerical library of the BNF (National French Library)).}
\label{fig:parchment_ori}
\end{figure}

\begin{figure}[h]
\center
\begin{tabular}{c c c }
\includegraphics[width=3.2cm]{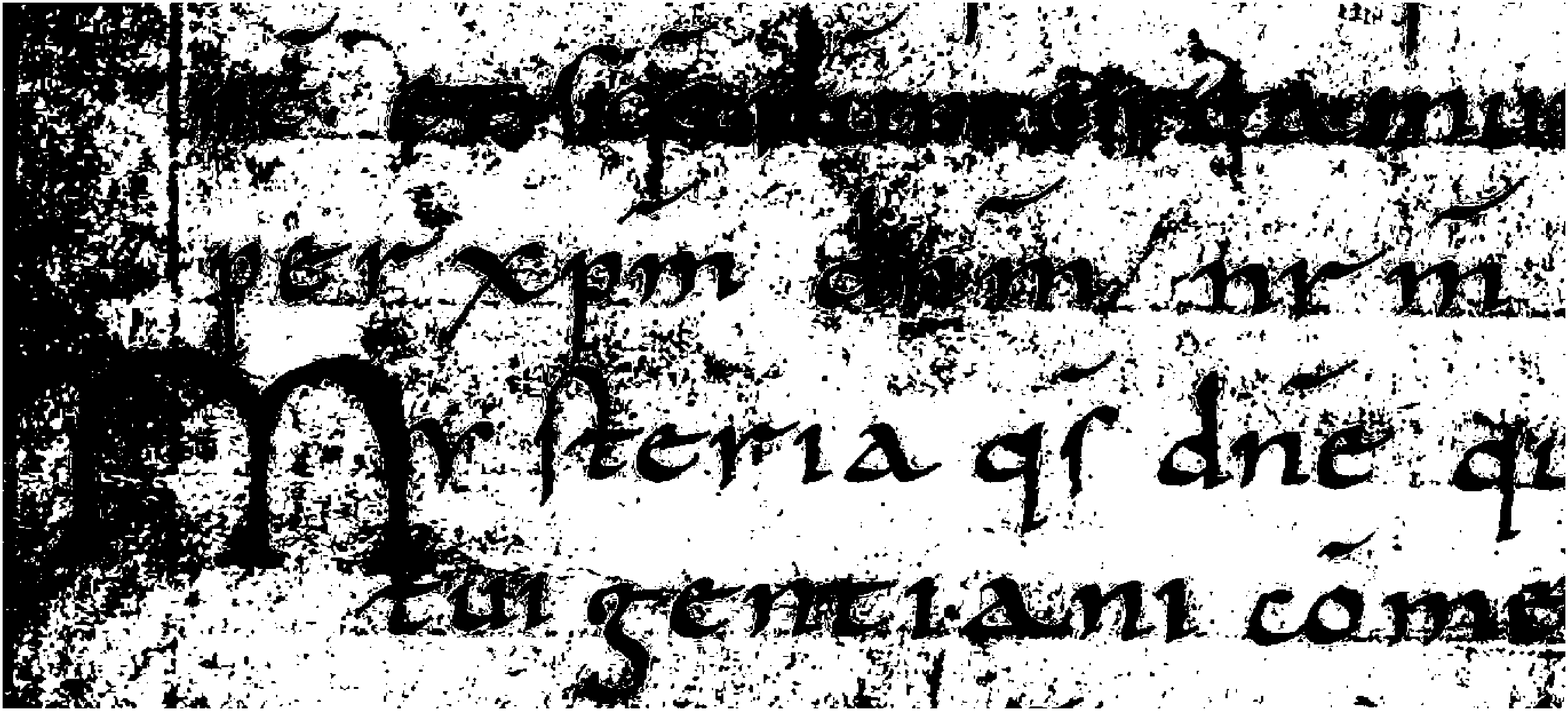} & \includegraphics[width=3.2cm]{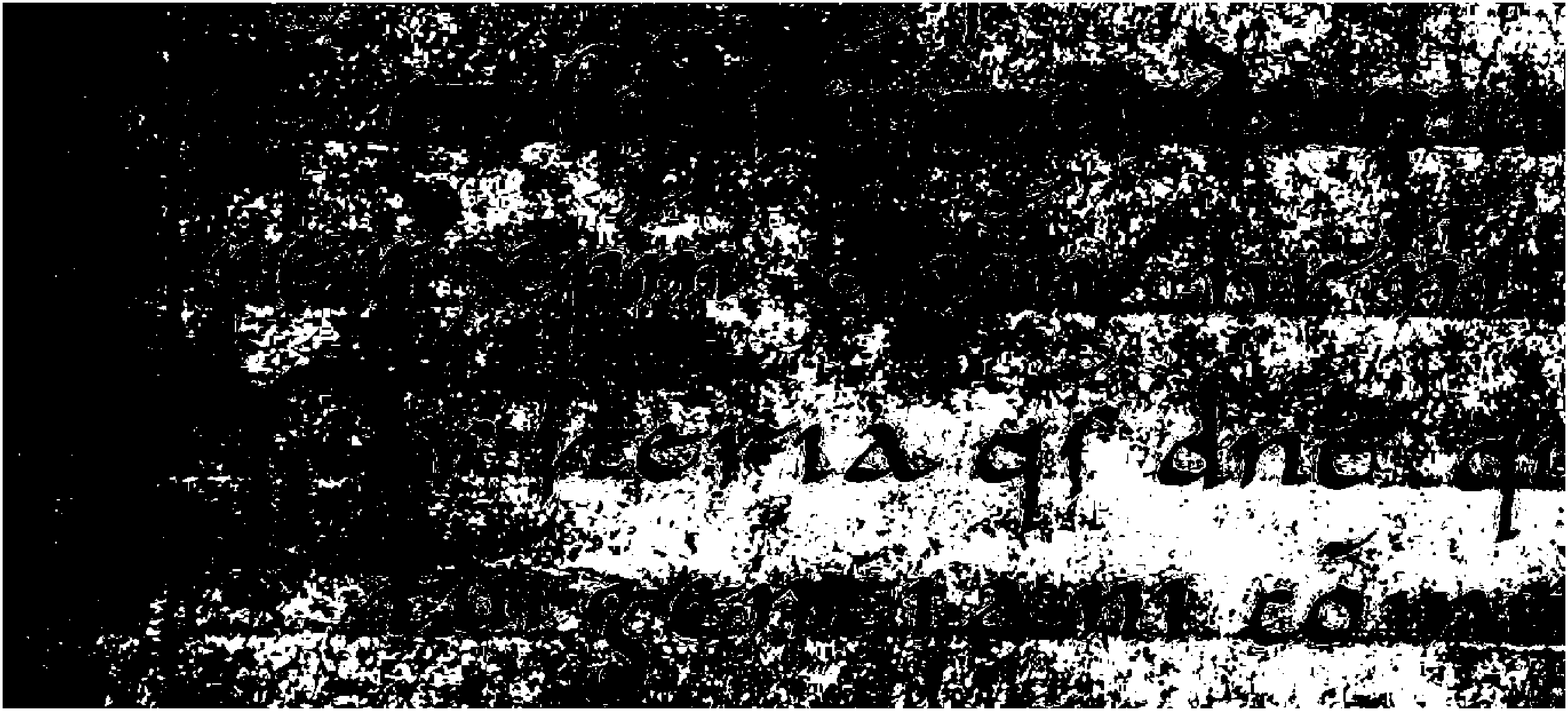} & \includegraphics[width=3.2cm]{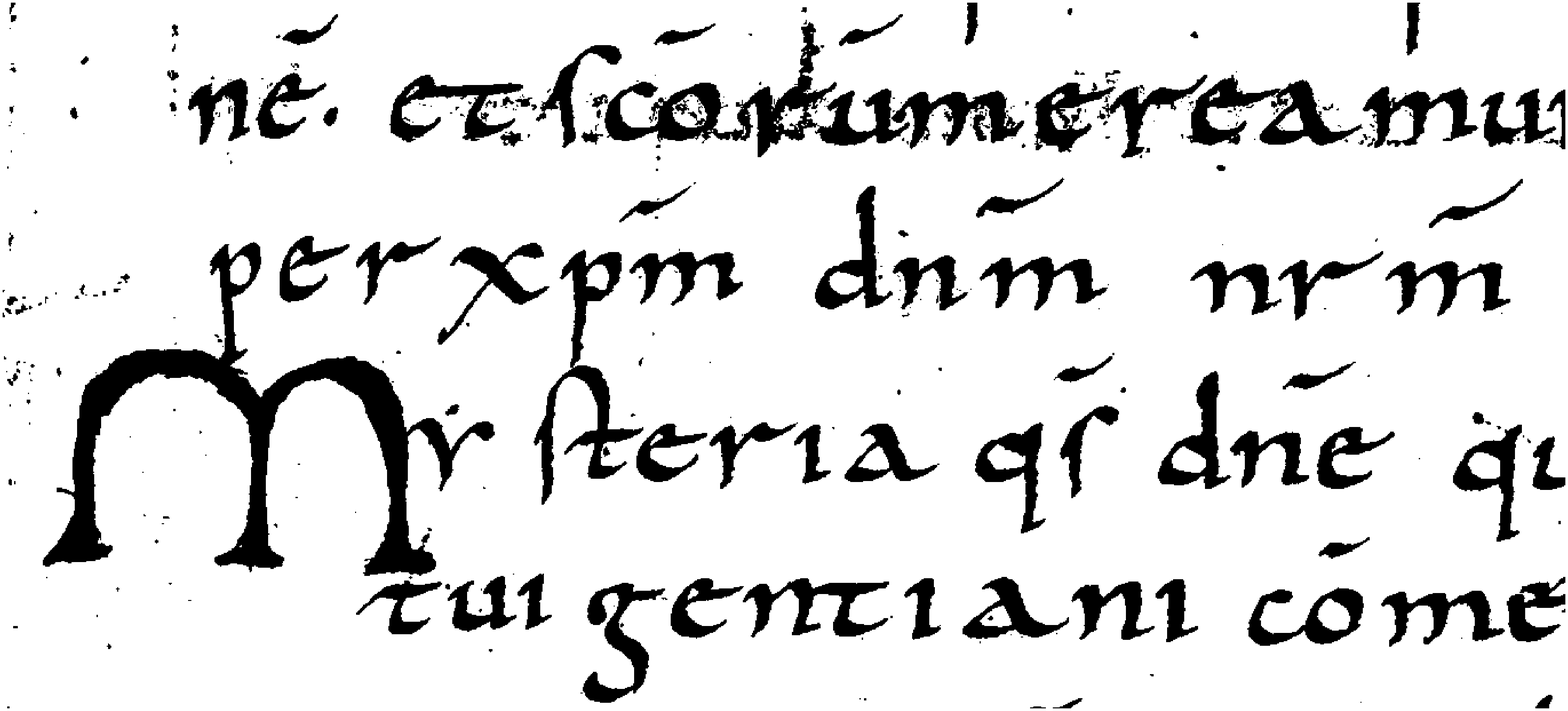} \\
(a) mask 1 & (b) mask 2 & (c) mask 3 \\ \\
 \includegraphics[width=3.2cm]{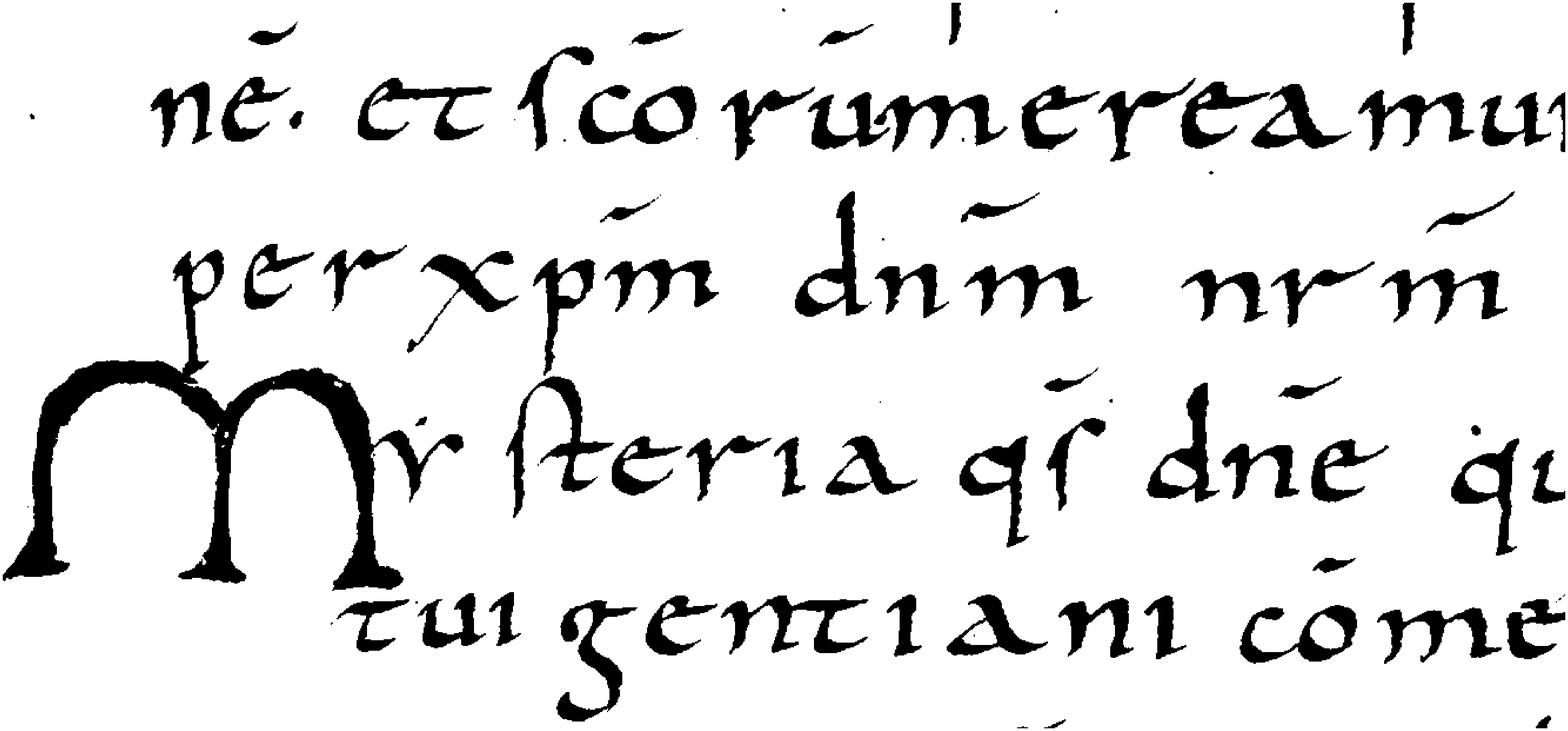} &
 \includegraphics[width=3.2cm]{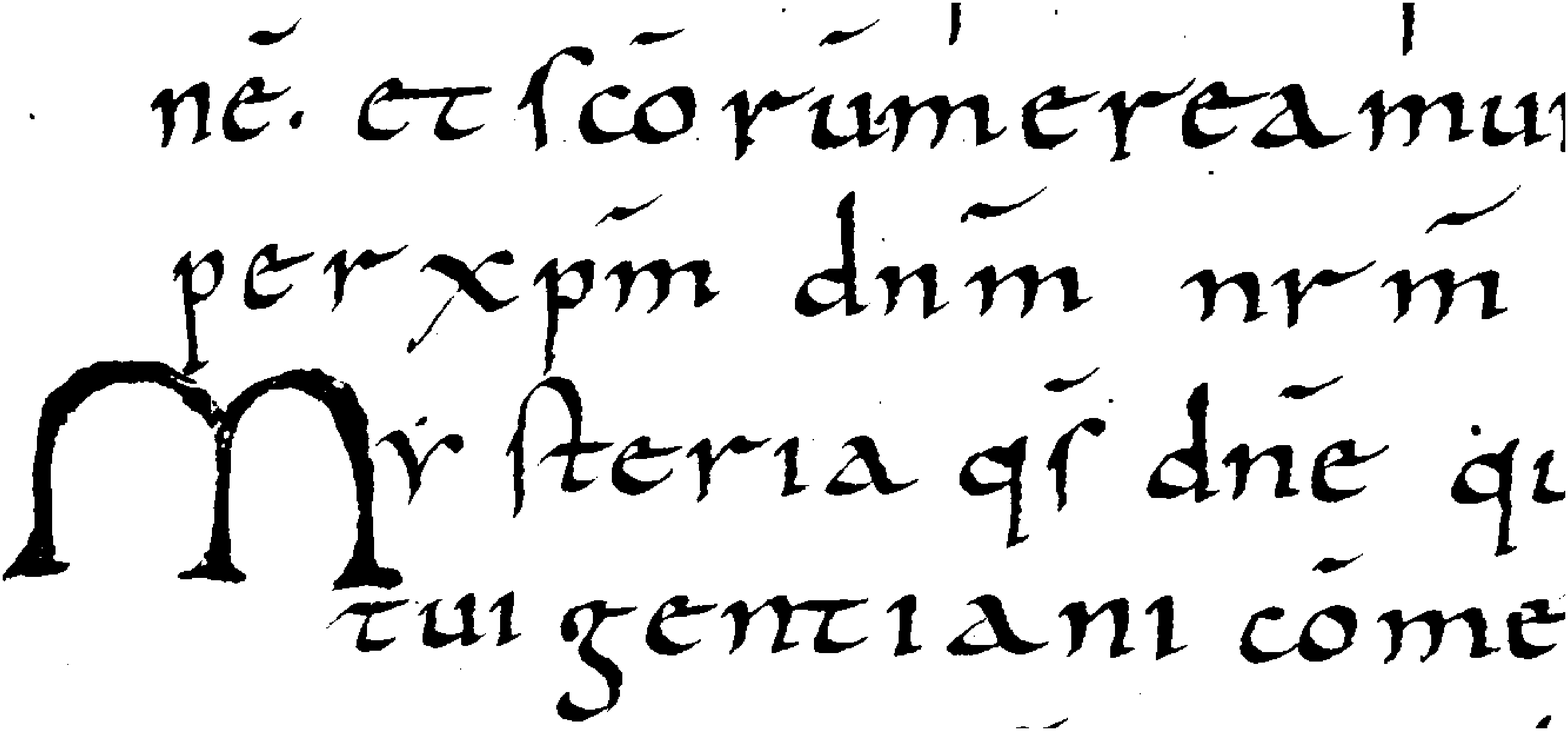} & \includegraphics[width=3.2cm]{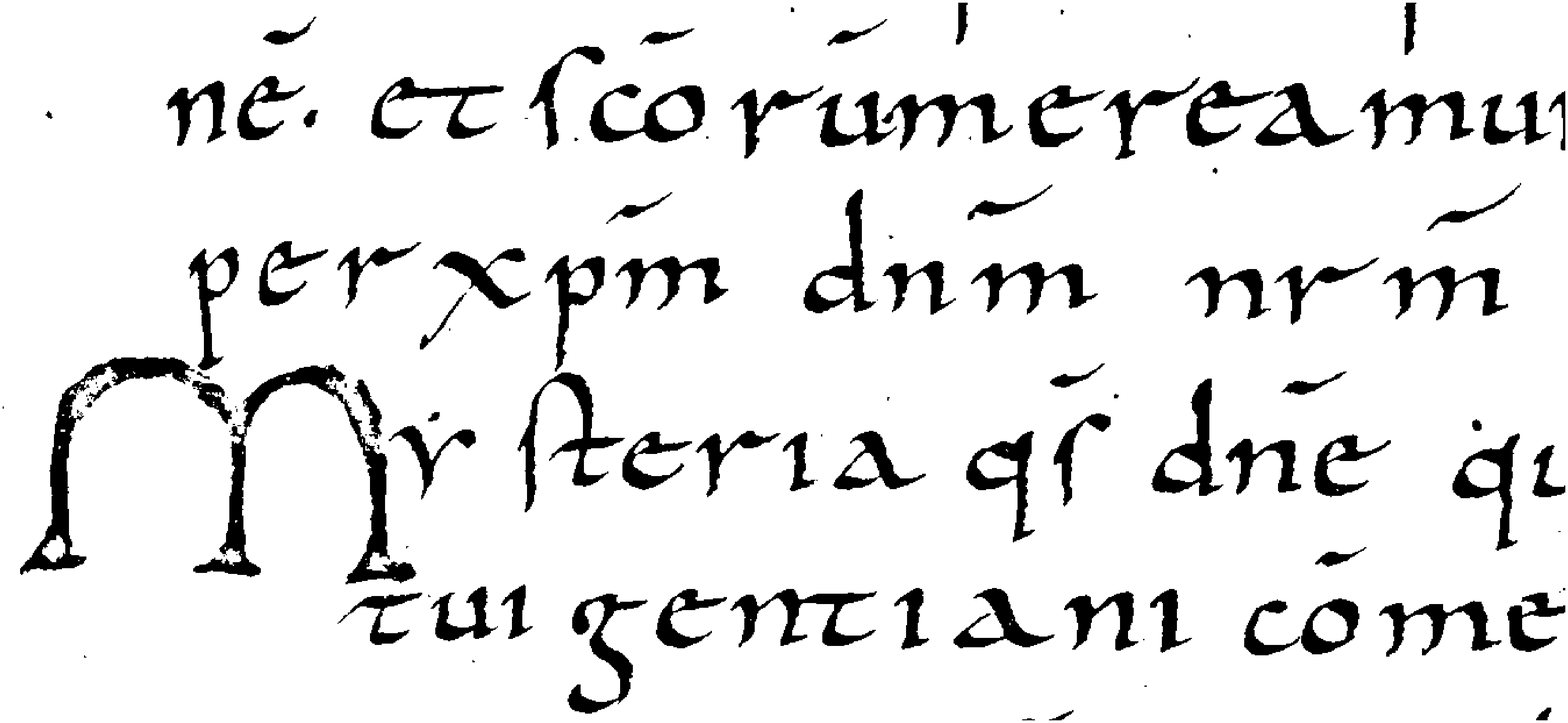} \\
 (d) mask 4 & (e) mask 5 & (f) mask 6  \\
\end{tabular}
\caption{ The different segmentation masks of the text (a,b,c,d,e,f).}
\label{fig:parchment}
\end{figure}

As a first example, we propose to combine different basic binarization methods using the mutual shape in order to construct a better segmentation. Let us consider for example the original image given in Fig.\ref{fig:parchment_ori} where the object of interest is the whole text. The input masks are obtained using classical binarization techniques provided by the library of image processing Pandore \cite{Pandore}. The techniques used are namely \textit{"pmassbinarization''} (based on a percentage of pixels, mask 1 and 2), \textit{``pcorrelationbinarization''} (maximization of the correlation between two classes, mask 3), \textit{``pvariancebinarization''}  (maximization of the interclass and intraclass distance, mask 4), \textit{``pniblackbinarization''}  (based on an adaptive binarization technique described in \cite{Sauvola_PR2000} mask 5) and \textit{``padaptativemeanbinarization''}  (based on the analysis of the mean value of the intensities on a sliding window mask 6) . The corresponding masks (shown in Fig.\ref{fig:parchment}) are used as segmentation inputs of our mutual shape algorithm.

\begin{figure}[h]
\center
\begin{tabular}{c c}
\includegraphics[width=4.5cm]{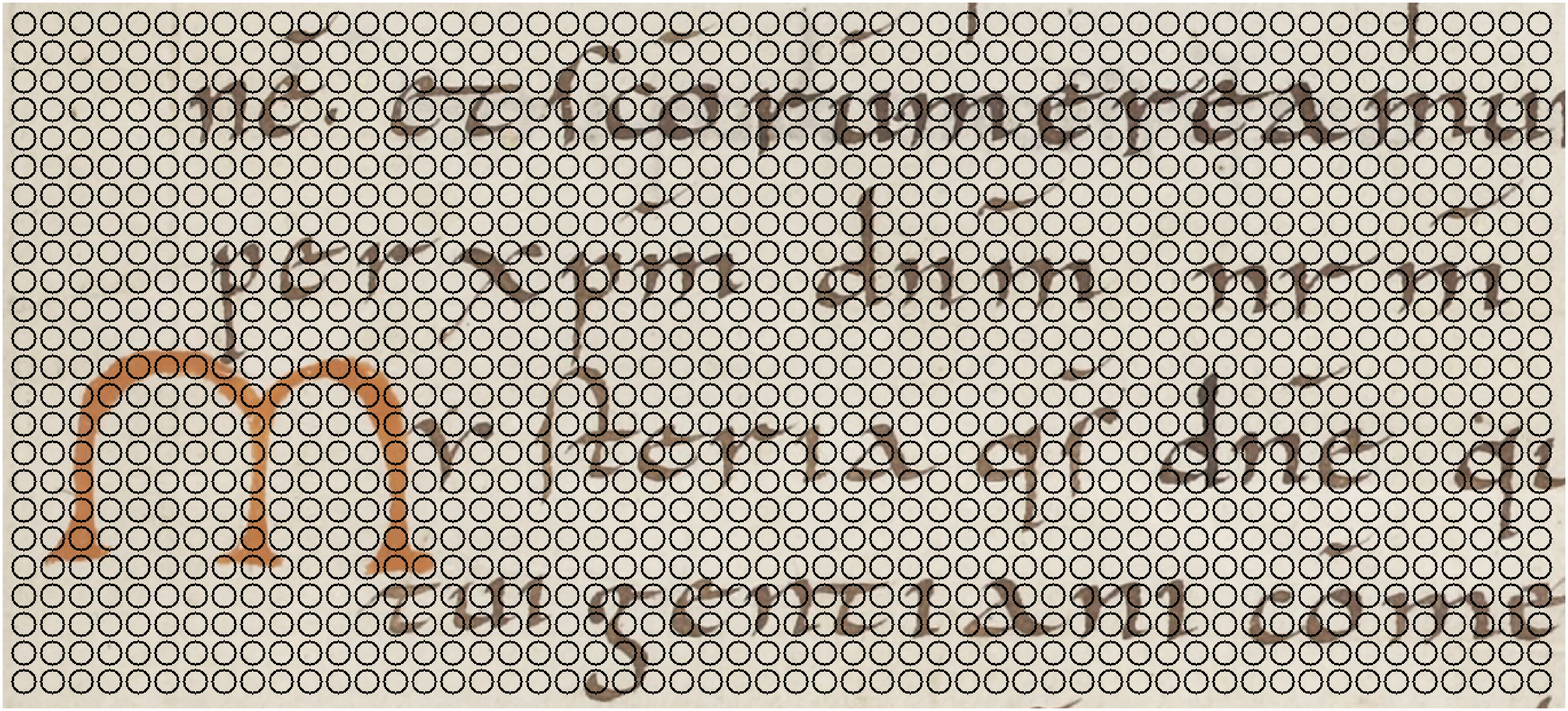} &
 \includegraphics[width=4.5 cm]{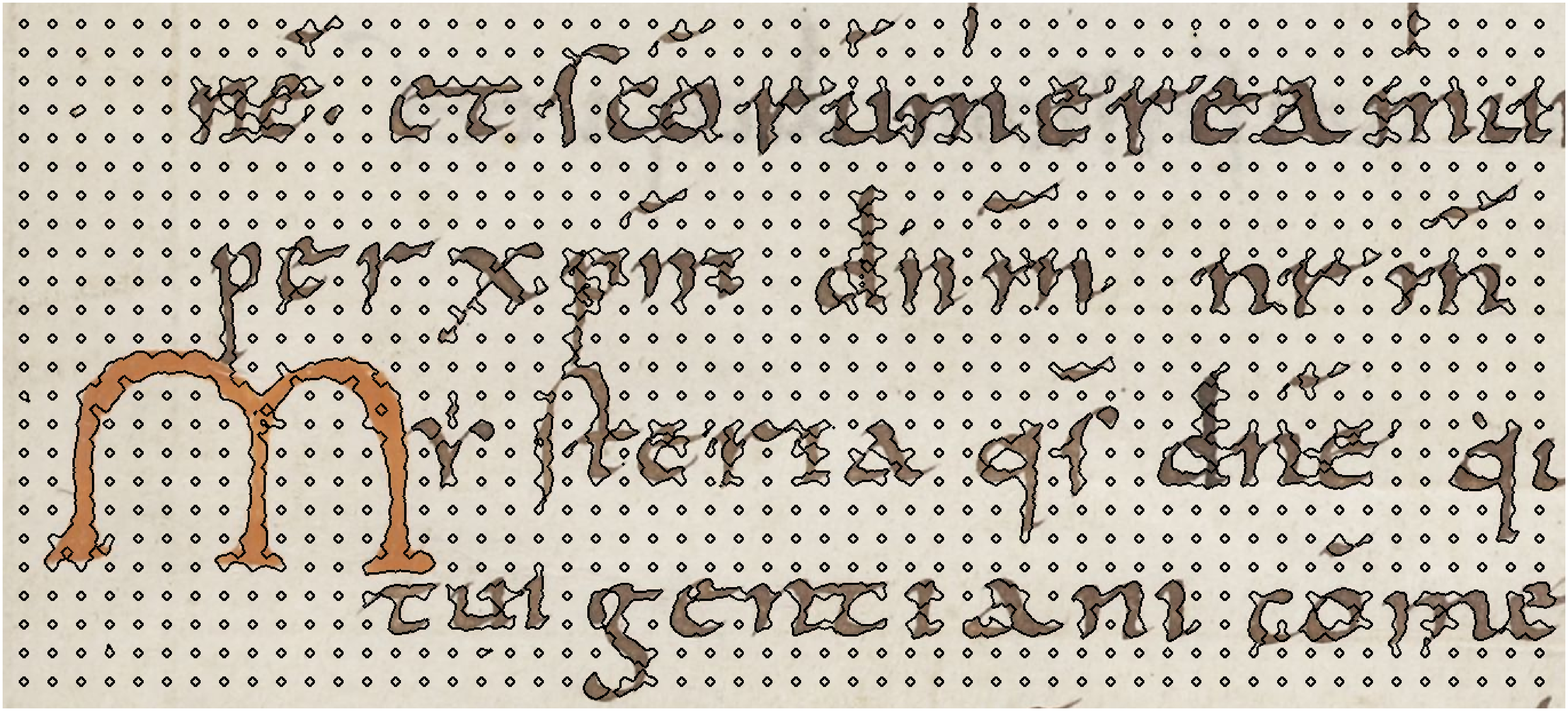} \\
 (a) Initial contour & (b) Iteration 50
\end{tabular}
\begin{tabular}{c}
\includegraphics[width=9cm]{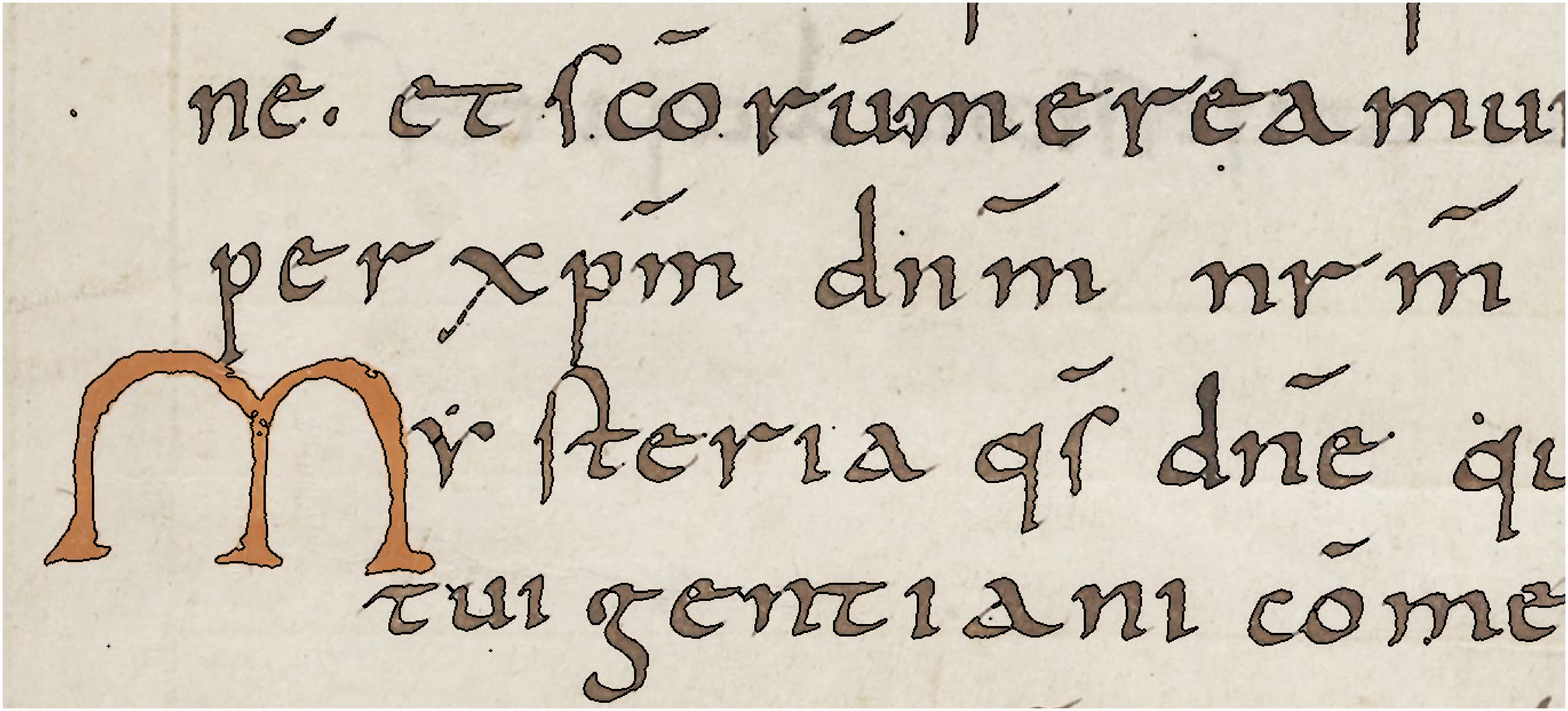} \\
 (c)  Final contour \\
\end{tabular}
\begin{tabular}{c}
\includegraphics[width=9cm]{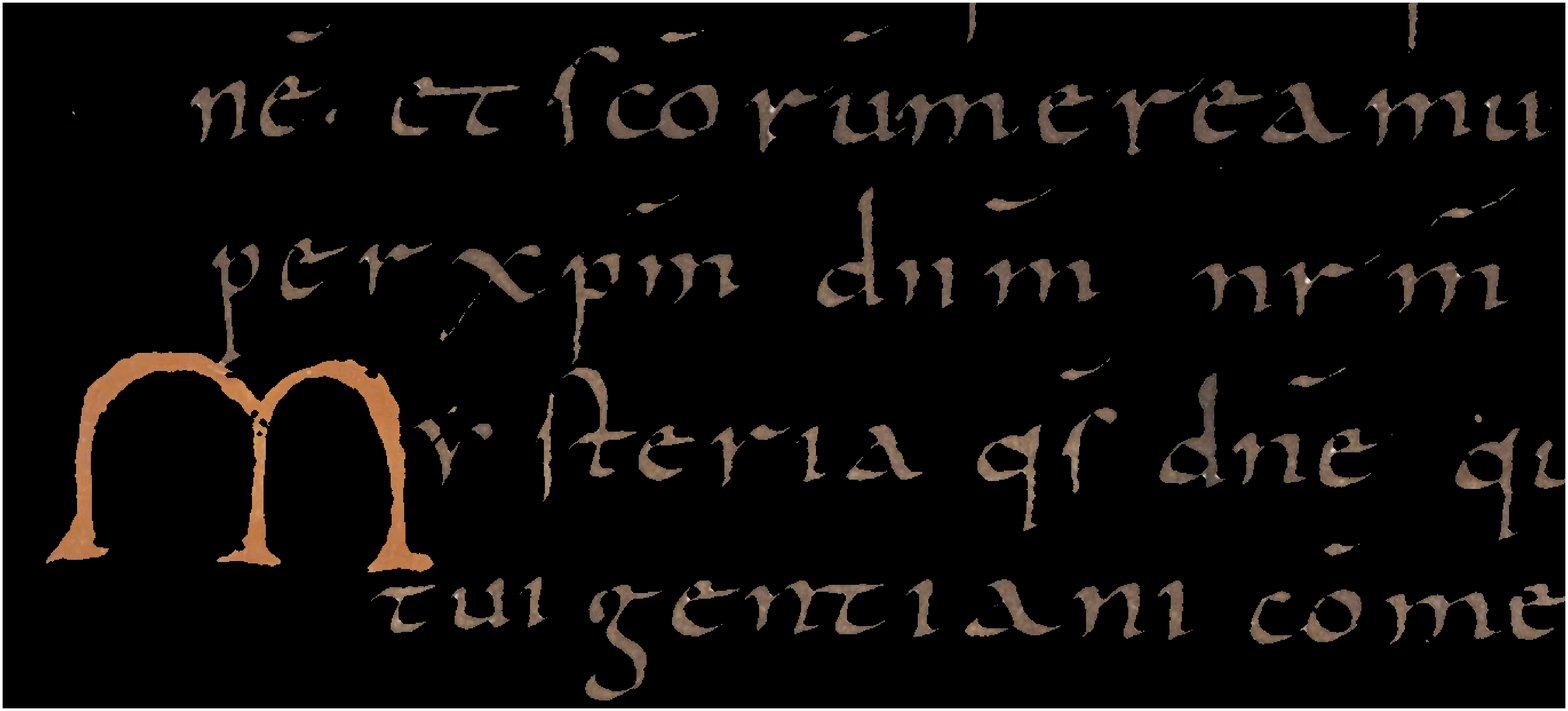} \\
 (d)  Final segmentation \\
\end{tabular}
\caption{Evolution of the active contour (in black) using the mutual shape (evolution equation \eqref{eq:evol_MI_JH} with $\lambda=10$). The active contour is in black. In the first image (a), the initial contour corresponds to a set of bubbles. An Intermediate result obtained from  $50$ iterations is displayed in image (b) and the final contour in (c). The segmented object is displayed in (d) (the background is in black color).}
\label{fig:evol_parchment}
\end{figure}

The mutual shape is then computed using active contours (Fig.\ref{fig:evol_parchment}). The initial contour is chosen as a set of little circles currently named as ``bubbles'' in the framework of active contours. The text is well segmented as displayed in Fig.\ref{fig:evol_parchment}.c showing the potential application of this method to build a consensus segmentation from a set of different simple binarization techniques not necesseraly all well chosen and composed of a set of pixels that is not connected. This example also shows that our algorithm is able to handle a shape composed of different separated components.

\begin{figure}[h]
\center
\begin{tabular}{cc}
 \includegraphics[width=5.5cm]{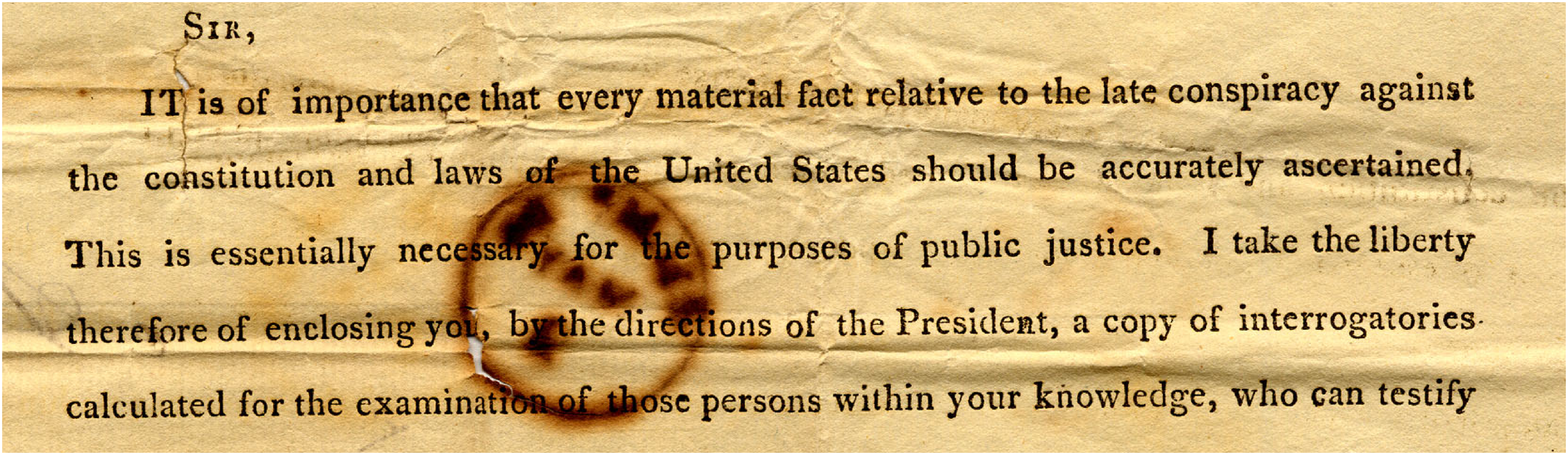}  & \includegraphics[width=5.5cm]{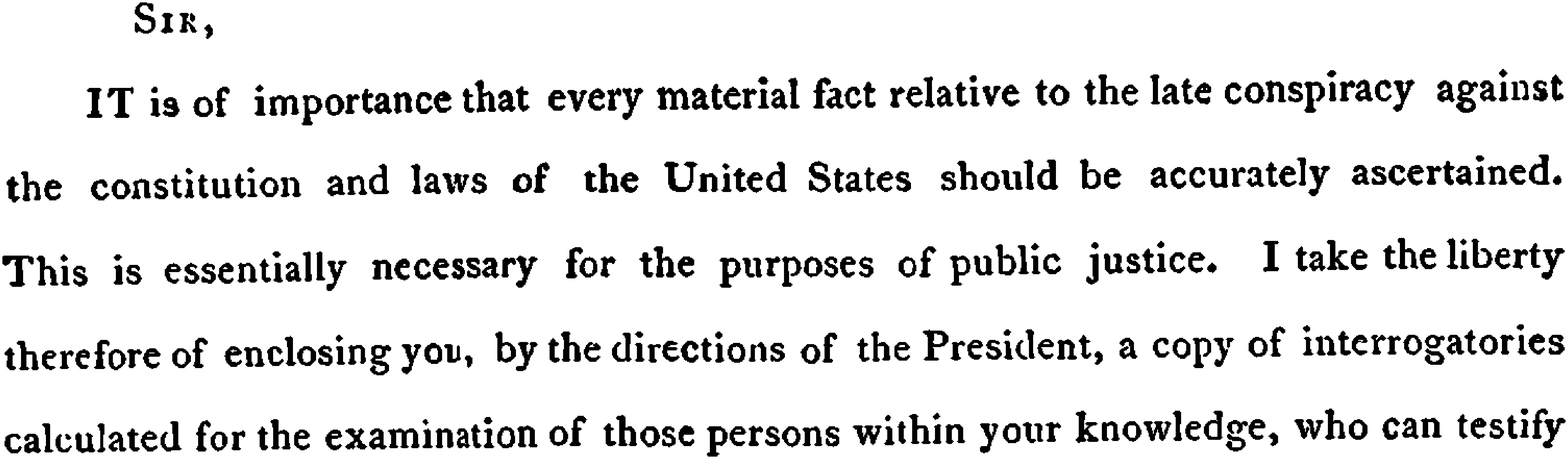} \\
(a) & (b)
\end{tabular}
\caption{An original image from an old manuscript from the DIBCO database (a) and its corresponding reference binarization.}
\label{fig:pr08}
\end{figure}

\begin{figure}[h]
\center
\begin{tabular}{c c}
\includegraphics[width=5cm]{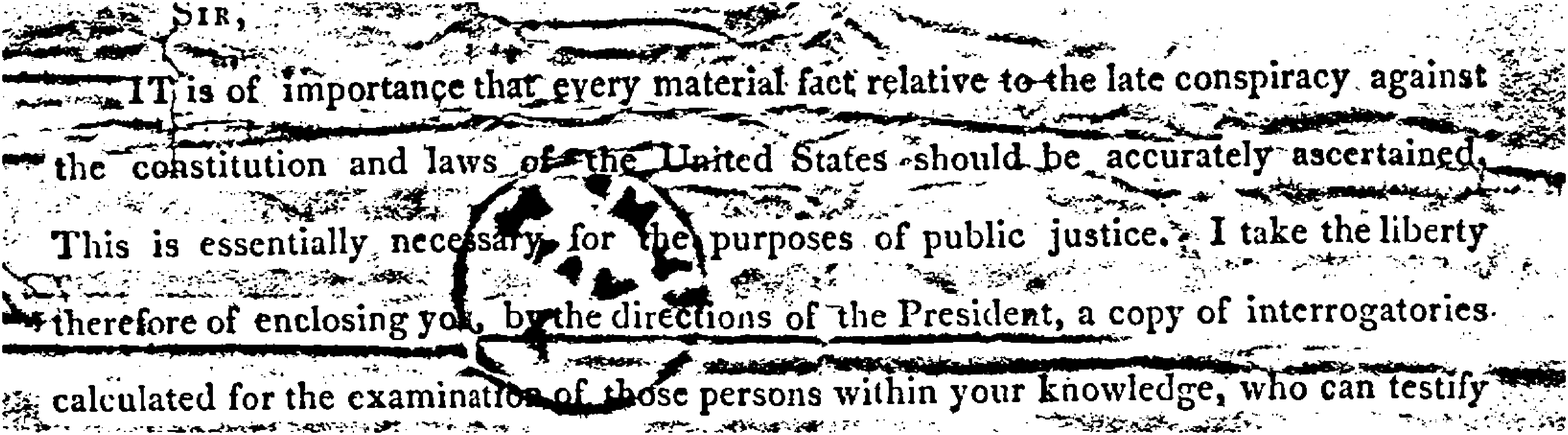} & \includegraphics[width=5cm]{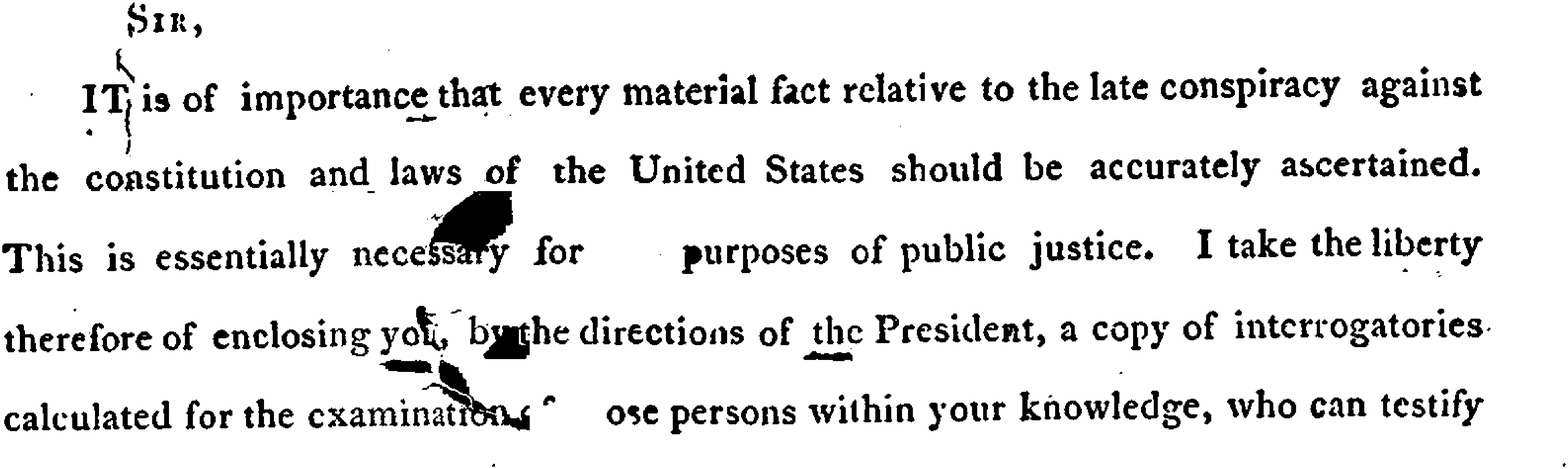} \\
(a) mask 1 & (b) mask 2 \\ \\
\includegraphics[width=5cm]{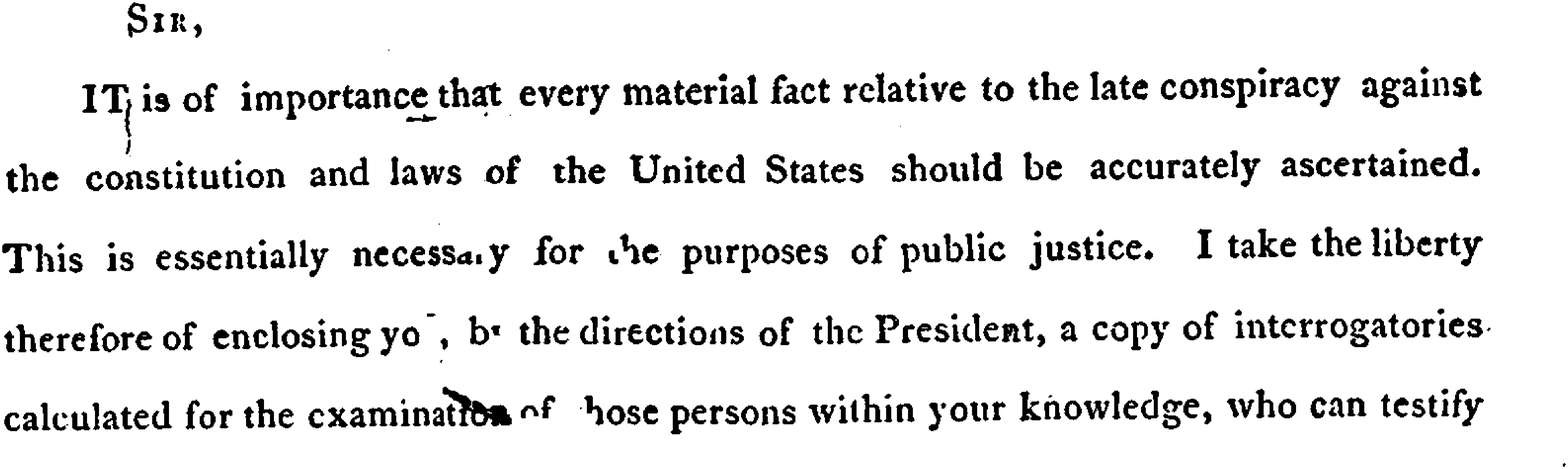} & \includegraphics[width=5cm]{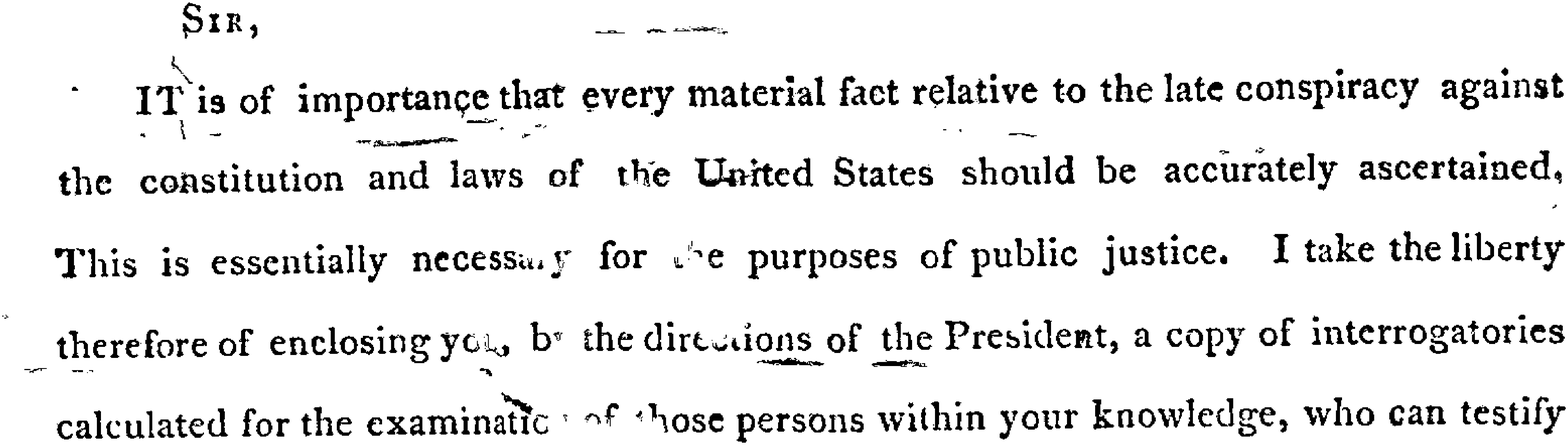} \\
(c) mask 3 & (c) mask 4 \\ \\
\includegraphics[width=5cm]{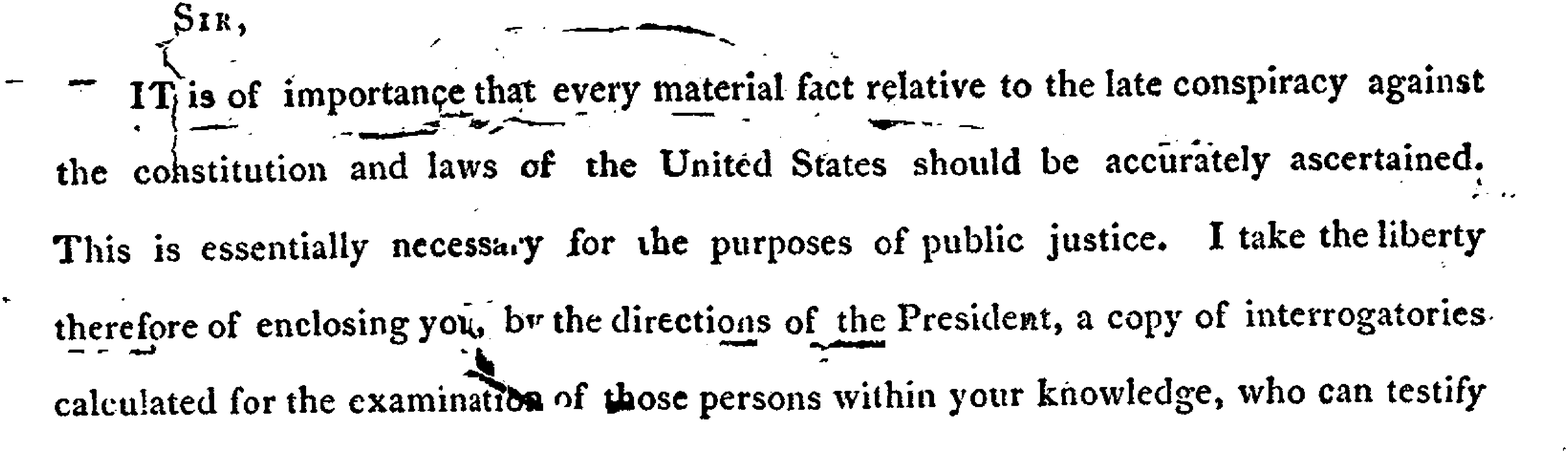} & \includegraphics[width=5cm]{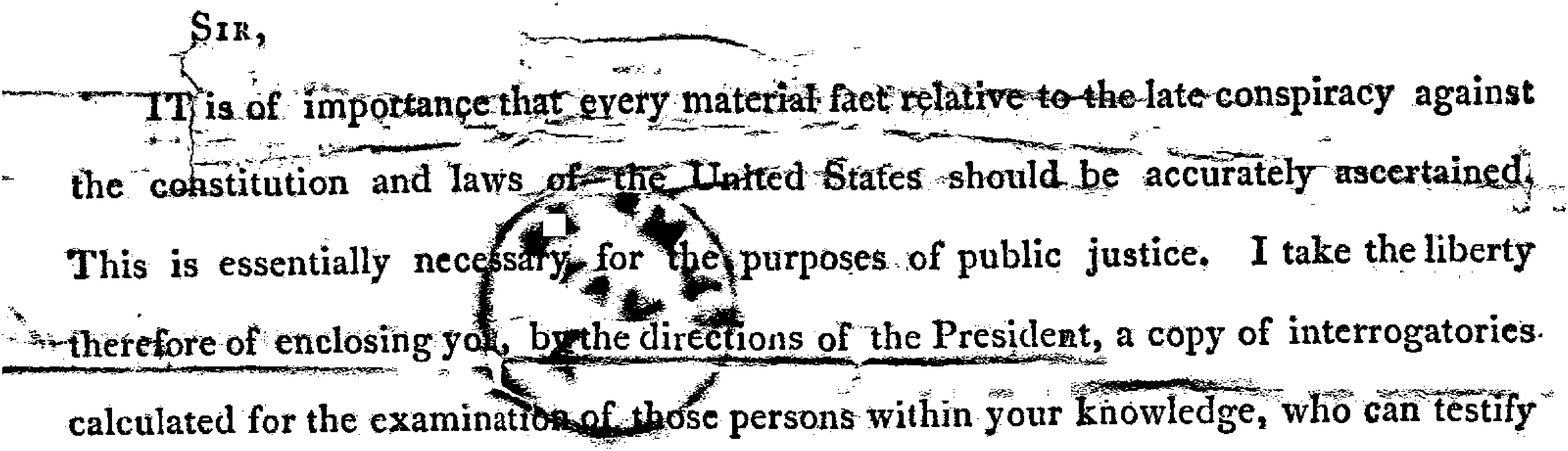} \\
(e) mask 5 & (f) mask 6 \\ \\
\includegraphics[width=5cm]{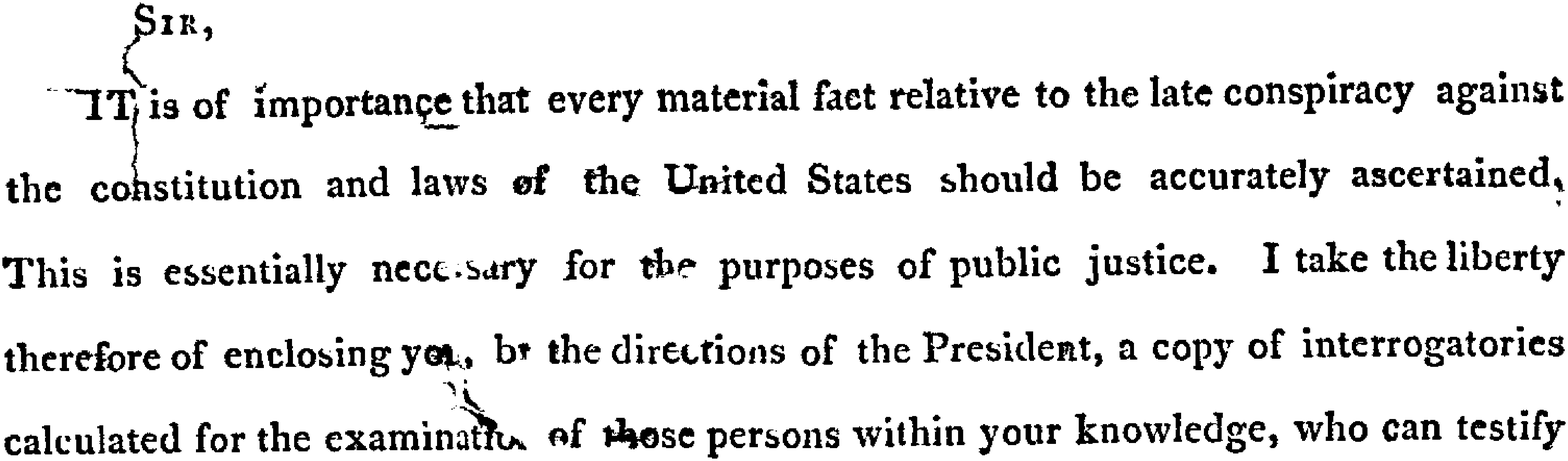} & \\
(g) mask 7 \\ \\
\end{tabular}
\caption{ The different segmentation masks of the text (a,b,c,d,e,f,g) taken from the 2013 DIBCO database.}
\label{fig:pr8mask}
\end{figure}

Let us now take another example of applicability of the mutual shape for segmentation fusion and evaluation of different methods of text binarization taken from the DIBCO database 2013 \cite{DIBCO13}.  Let us consider for example the original image given in Fig.\ref{fig:pr08}.a where the object of interest is the whole text and for which we have the reference segmentation given in  Fig.\ref{fig:pr08}.b. 
The input masks corresponds to different algorithms tested during this challenge and are all available in the database. They are given in Fig.\ref{fig:pr8mask}.  

For all these masks, we can compute the Dice Coefficient with the reference segmentation. The different values are given in Table.\ref{dibco_dc}.
\begin{table}[htb]
 \begin{center}
\begin{tabular} {|c | c | c | c | c | c | c| c|}
  \hline
  i  & $m_1$ & $m_2$ & $m_3$ & $m_4$ & $m_5$ & $m_6$ &$m_7$ \\
 \hline
$DC$  & 0.54 & 0.89 & 0.95 & 0.87 & 0.91 & 0.76 & 0.91 \\
  \hline
\end{tabular}
\caption{Computation of the Dice Coefficient with the reference segmentation (Fig.\ref{fig:pr08}.b) for the segmentations $m_1$ to $m_7$ (Fig.\ref{fig:pr8mask}).}
\label{dibco_dc}
\end{center}
\end{table}

\begin{figure}[h]
\center
\begin{tabular}{c}
\includegraphics[width=8cm]{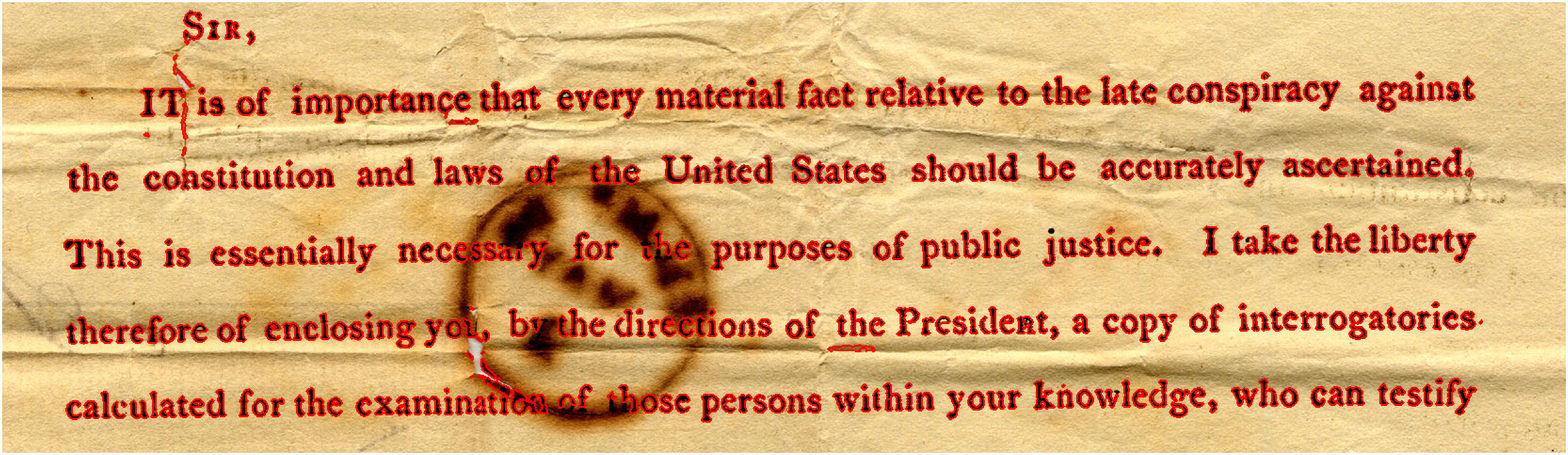} \\
 (a) Final contour \\
 \includegraphics[width=8cm]{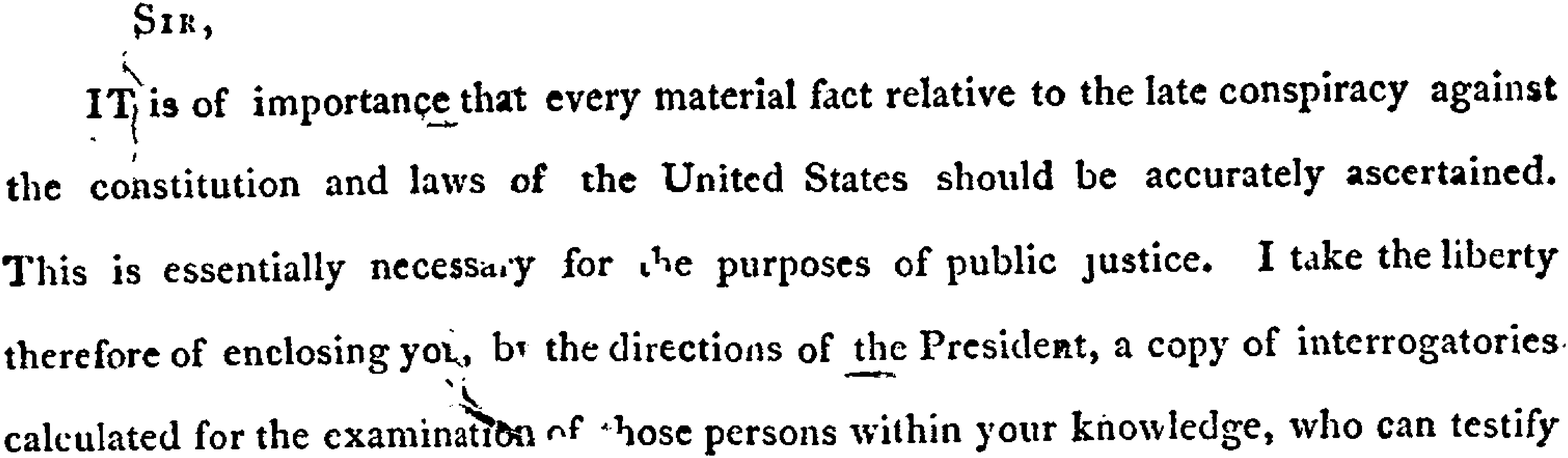} \\
 (b) Resulting mask
\end{tabular}
\caption{The mutual shape (final contour and resulting mask) computed using the masks $m_1$ to $m_7$ of the DIBCO image PR08.}
\label{fig:mutualshape_dibco}
\end{figure}

Let us now compute the mutual shape and compare the quality of the obtained result to the reference segmentation. The obtained  mutual shape (final contour and the associated mask) are given in Fig.\ref{fig:mutualshape_dibco}. For this mutual shape, we find $DC=0.93$ which outperforms the DC coefficient of all the different masks in entry except the mask $m_3$. The resulting mutual shape is then interesting for an intelligent fusion of different segmentation results. Let us now compare the ranking obtained using the mutual shape algorithm through the joint computation of the $p_i$ and $q_i$ coefficients. The different values of $p_i$ and $q_i$ are given in Table.\ref{tab:piqidibco} and allow us to rank the different segmentation methods as follows (from the best one to the worst one according to the sum of $p_i$ and $q_i$) : $m_3$,$m_5$,$m_2$,$m_7$,$m_6$,$m_4$,$m_1$. The ranking obtained using the reference mask and the DC coefficients leads to :$m_3$,$m_7$,$m_5$,$m_2$,$m_4$,$m_6$,$m_1$. We can observe that the ranking is the same for the first and the last mask. There are some difference of ranking between comparable masks such as $m_7$, $m_2$ and $m_5$. The mask $m_4$ corresponds to an under-segmentation and the mask $m_6$ to an over-segmentation which explains their places in the end of the ranking.

\begin{table}[htb]
 \begin{center}
\begin{tabular} {|c | c | c | c | c | c | c|c|}
  \hline
  i  & $m_1$ & $m_2$ & $m_3$ & $m_4$ & $m_5$ & $m_6$ &$m_7$\\
 \hline
$p_i$  & 0.973 & 0.975 & 0.984 & 0.872 & 0.982 & 0.992 & 0.969 \\
$q_i$ & 0.862 & 0.985 & 0.993 & 0.997 & 0.989 & 0.949 & 0.990 \\
  \hline
\end{tabular}
\caption{ Sensitivity and specificity parameters $p_i$  and $q_i$ for the segmentations $m_1$ to $m_7$ displayed in Fig.\ref{fig:pr8mask}.}
\label{tab:piqidibco}
\end{center}
\end{table}

In this last example, we show that the mutual shape leads to an interesting segmentation result by performing an intelligent fusion of different segmentation entries. The obtained ranking is interesting but can be different to the ranking performed using a reference mask and the DC coefficient. 

\subsection{Application to segmentation of cardiac magnetic resonance images}
\label{ssec:medical}

The segmentation of cardiac structures is an active research field in all medical modalities \cite{Fleureau_IRBM09}, where expert performance is still higher than image segmentation algorithms performance. As experts segmentations can vary, it was proposed to use STAPLE algorithm to define a consensus segmentation between different experts \cite{Suinesiaputra_MIA14}. Furthermore, to reduce the drawbacks of each specific image segmentation algorithm, it was proposed to take advantage of the results of different segmentation algorithms and the interest of combining different segmentation results using STAPLE was shown \cite{Lebenberg_PLOS15}. When compared to individual methods, using these combined segmentations provided better estimates of the clinical parameters of interest; this was demonstrated by a supervised approach using experts delineations and a non supervised evaluation approach described in \cite{Lebenberg_TMI12}. In this specific context, the estimation of a mutual shape was  tested for the non supervised evaluation and the fusion of different segmentation methods of the left ventricular cavity from cardiac cine magnetic resonance images \cite{JehanBesson_GRETSI11,jehan_icip14}. For instance, the excellent behavior of mutual shape towards outliers was demonstrated. In this section, we propose a first comparison between the mutual shape approach and STAPLE. At the difference of mutual shape, the STAPLE algorithm does not introduce any regularization term and can thus provide unsmoothed results, which are not relevant on a physiological basis.

\begin{figure}[h]
\center
\begin{tabular}{ccc} 
\includegraphics[width=2cm]{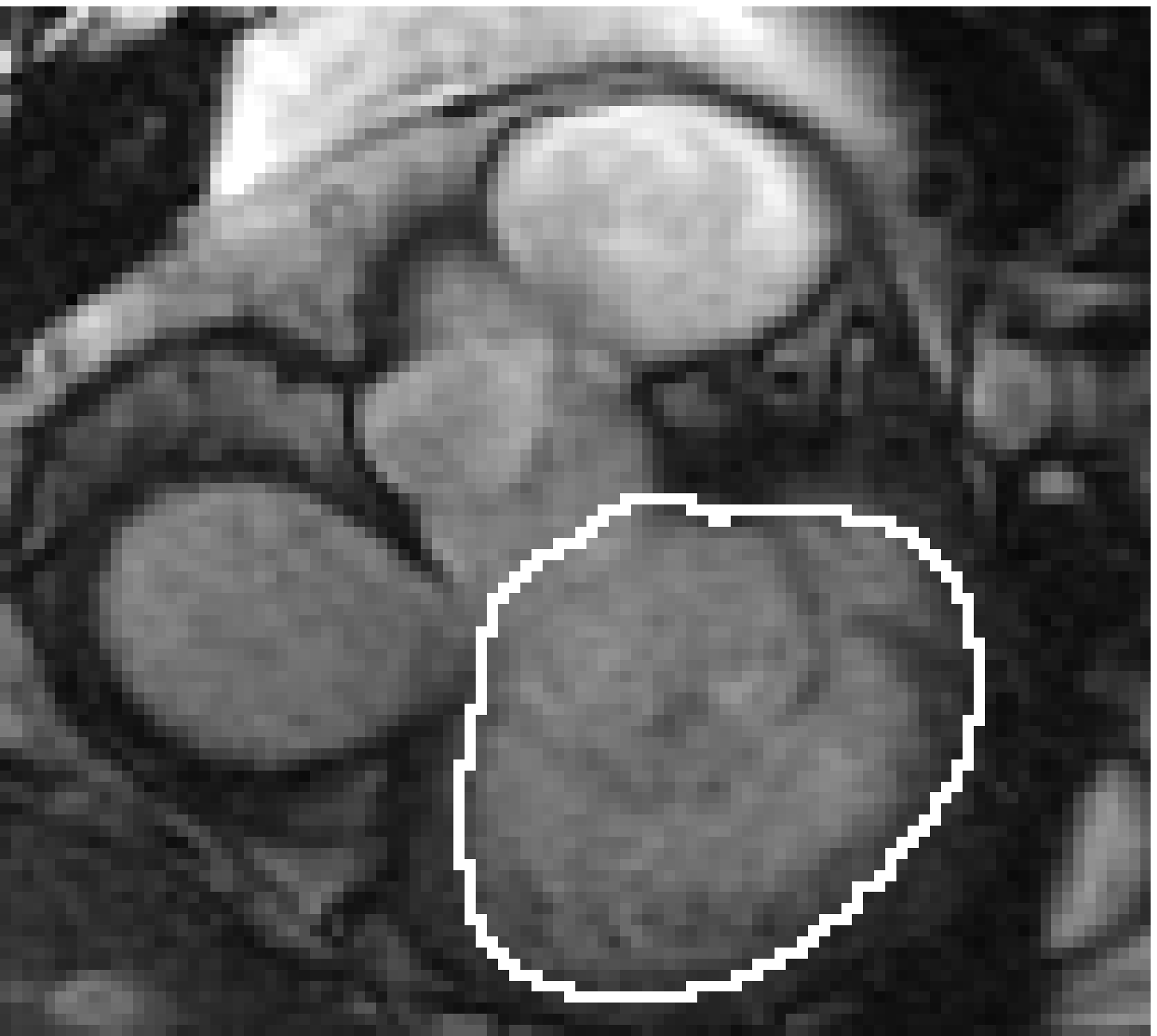} &
\includegraphics[width=2cm]{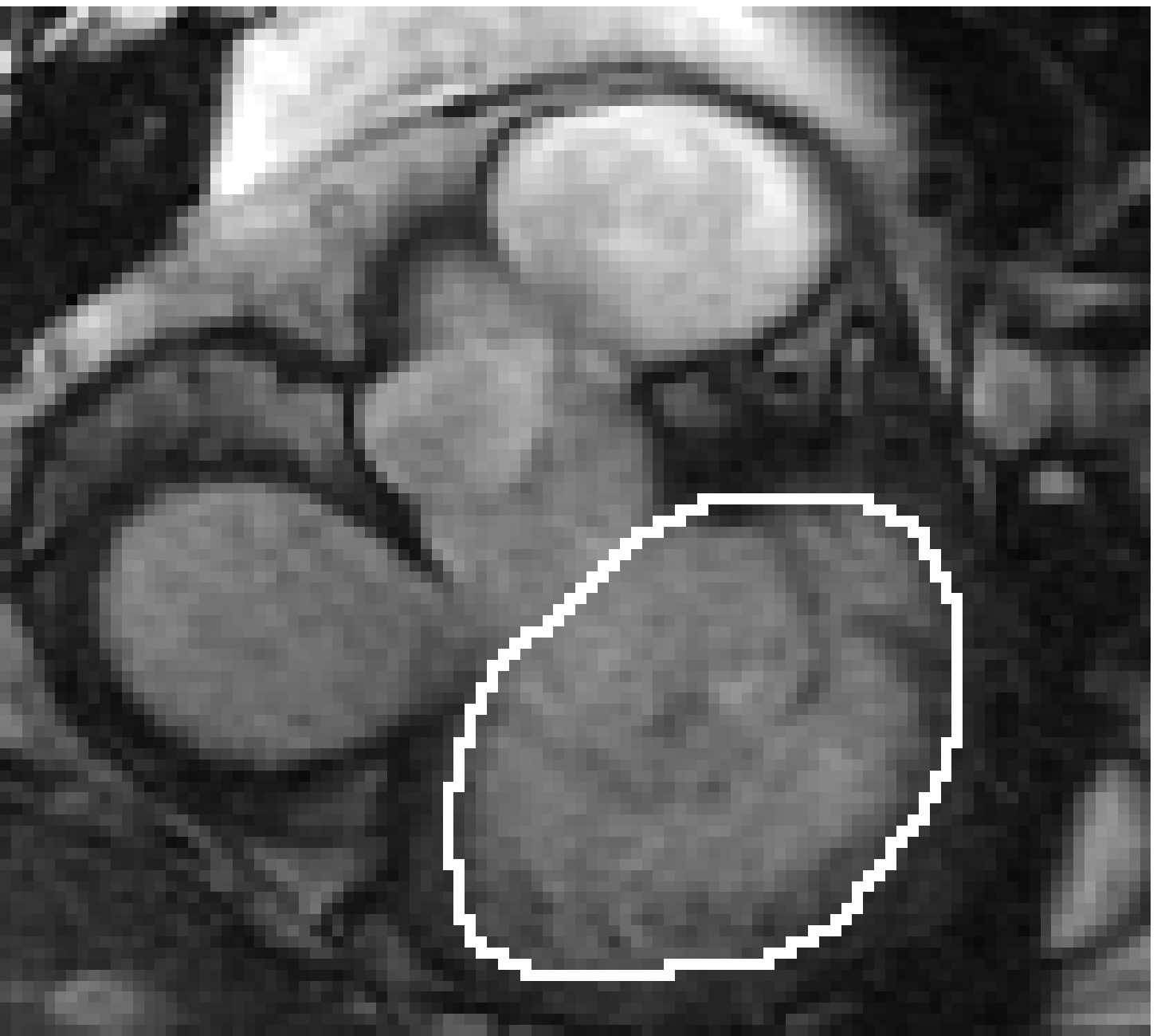} &
\includegraphics[width=2cm]{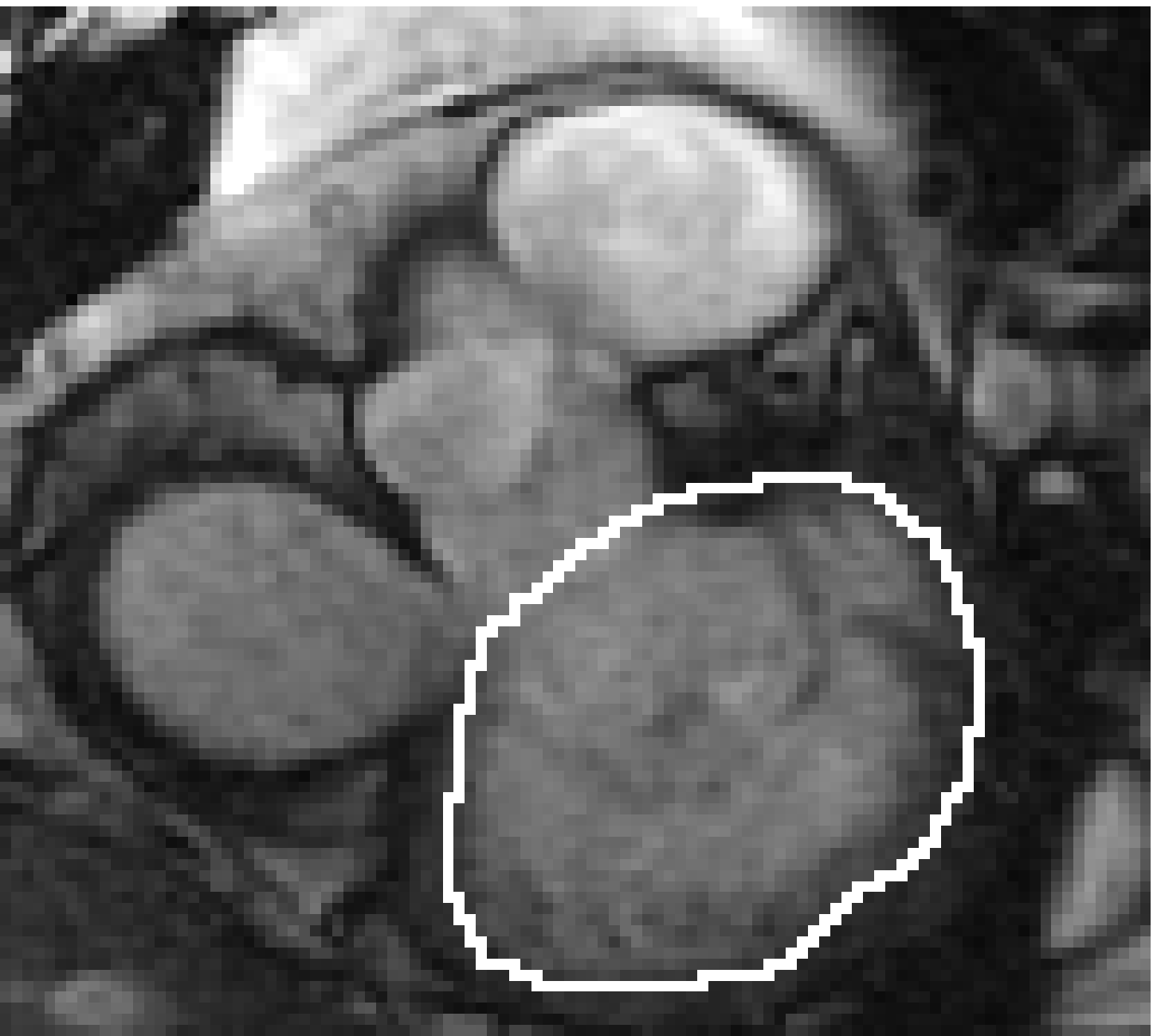}  \\
 $Exp_1$ &  $Exp_2$ & $Exp_3$  \\
\end{tabular}
\caption{Segmentation methods of the left ventricle provided by three experts $Exp_1$ to $Exp_3$. }
\label{fig:endo_exp_SCHFNI12}
\end{figure}

\begin{figure}[h]
\center
\begin{tabular}{ccccc} 
\includegraphics[width=2cm]{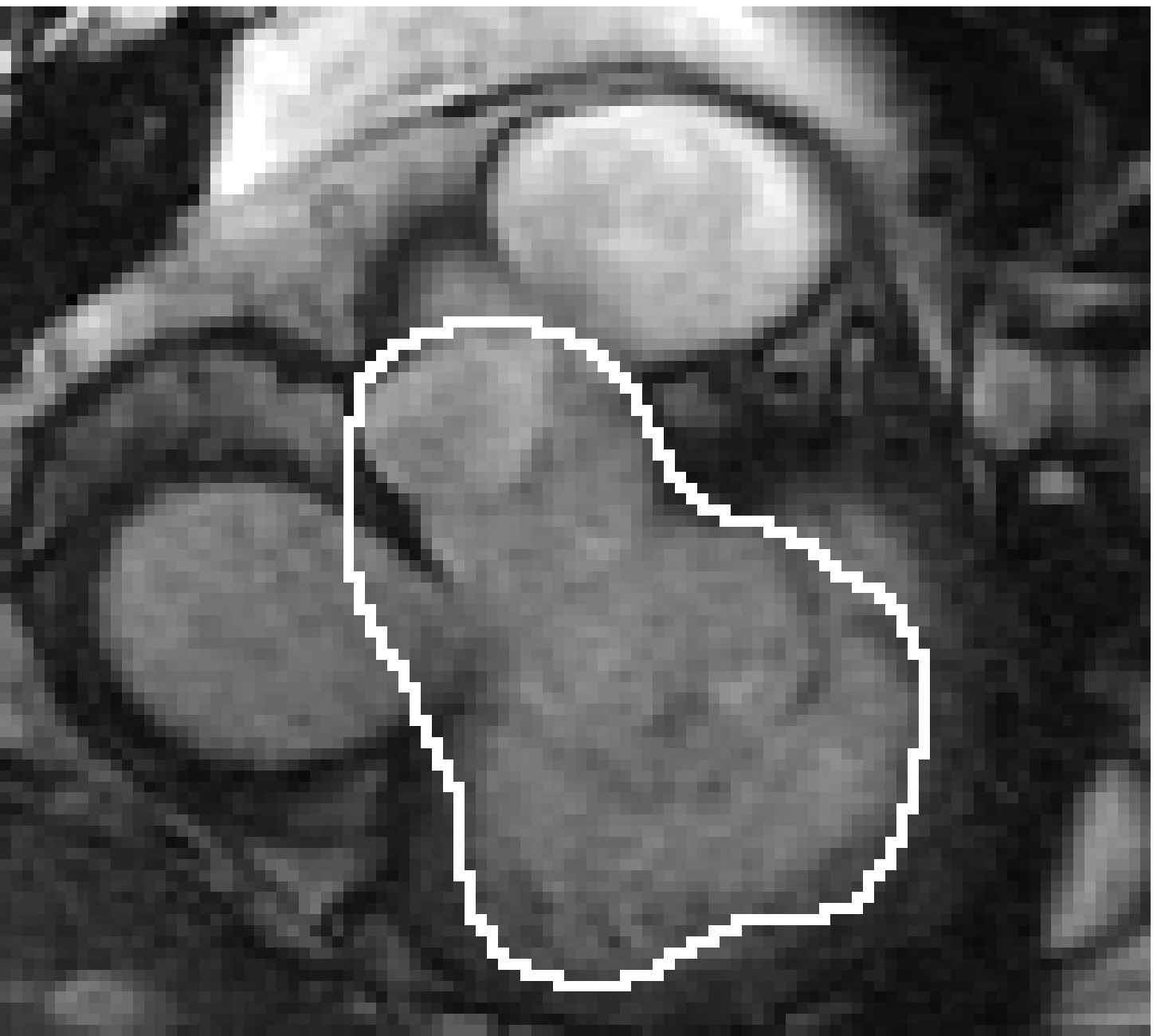} &
\includegraphics[width=2cm]{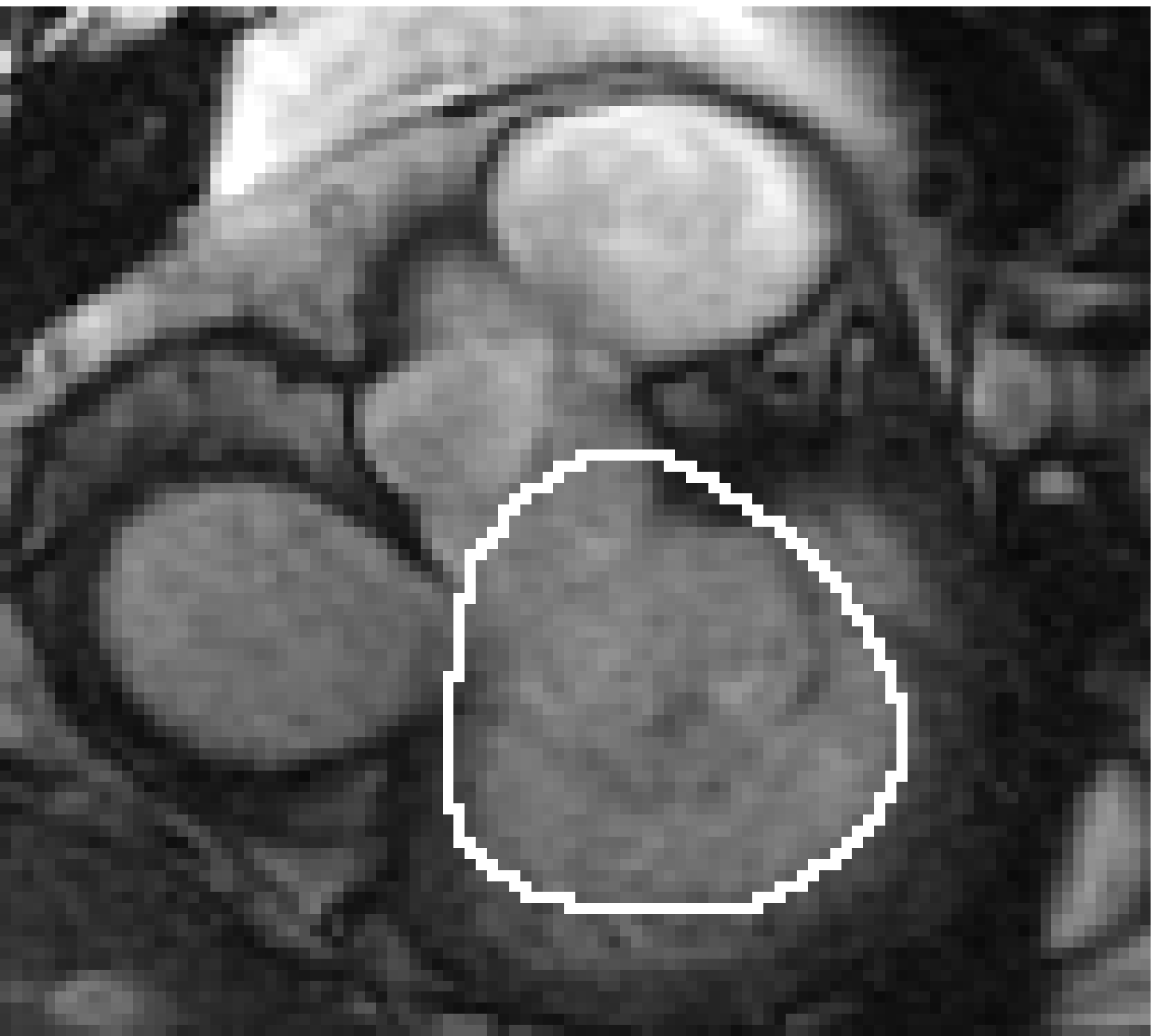} &
\includegraphics[width=2cm]{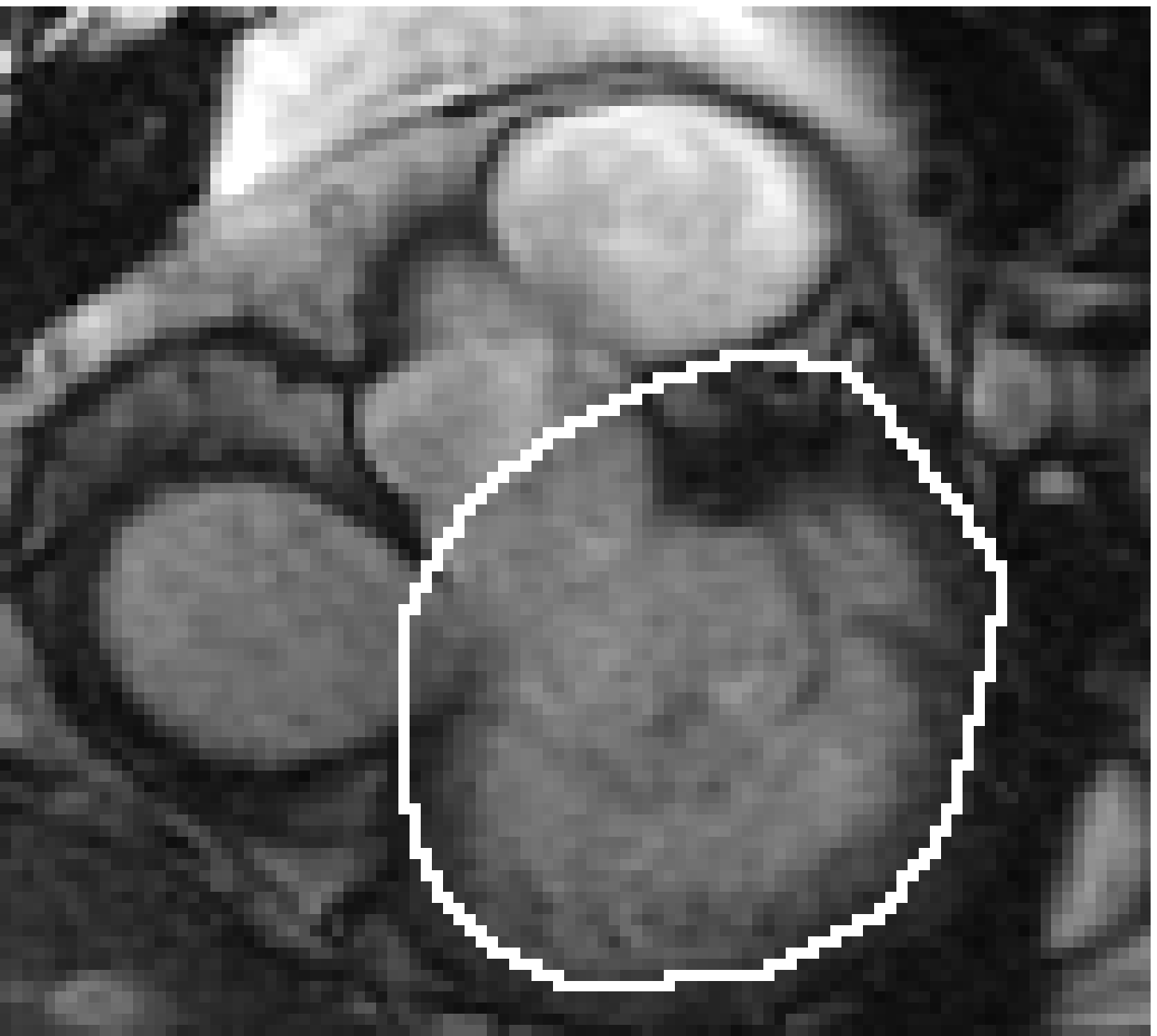} &
\includegraphics[width=2cm]{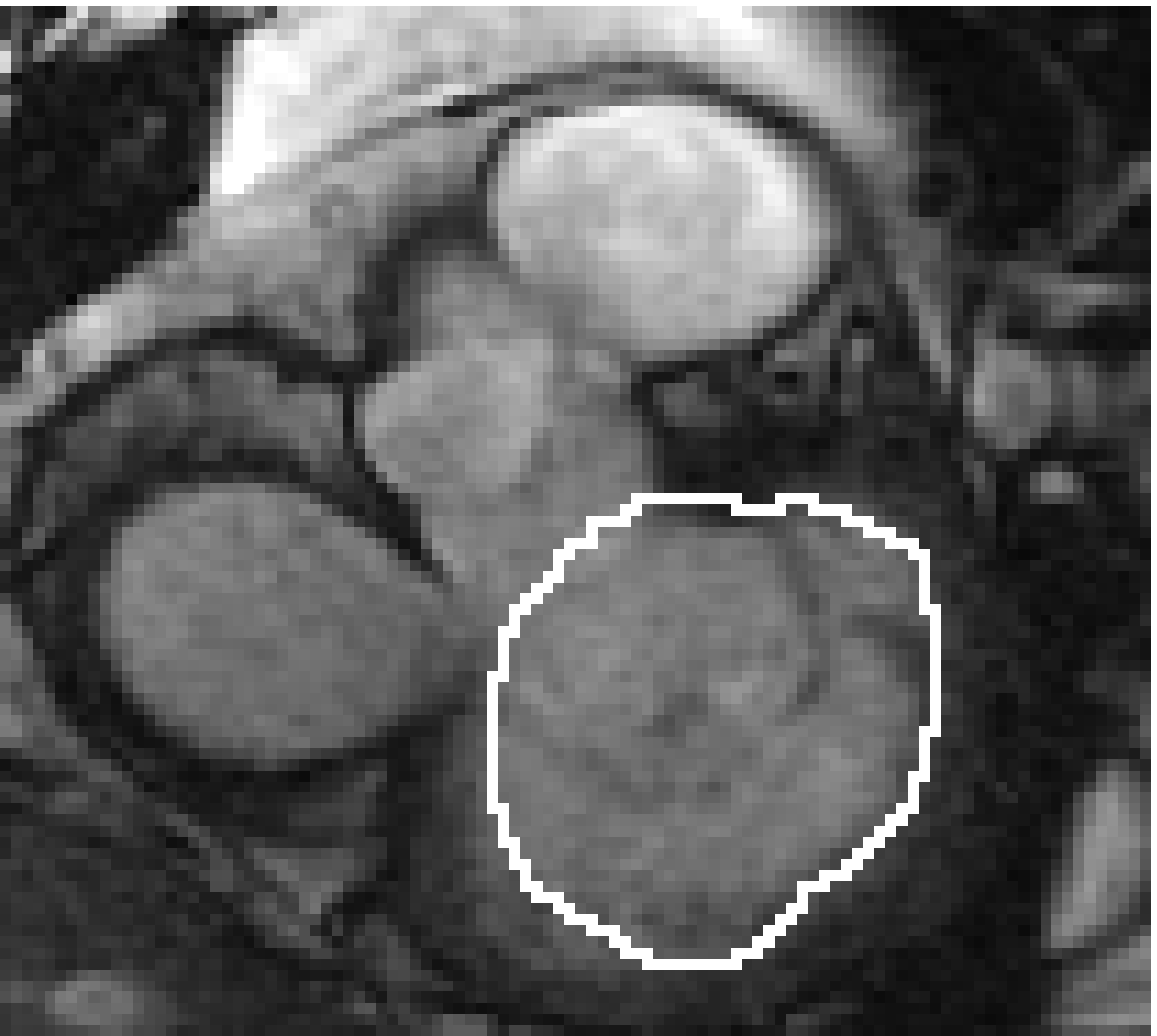} &
\includegraphics[width=2cm]{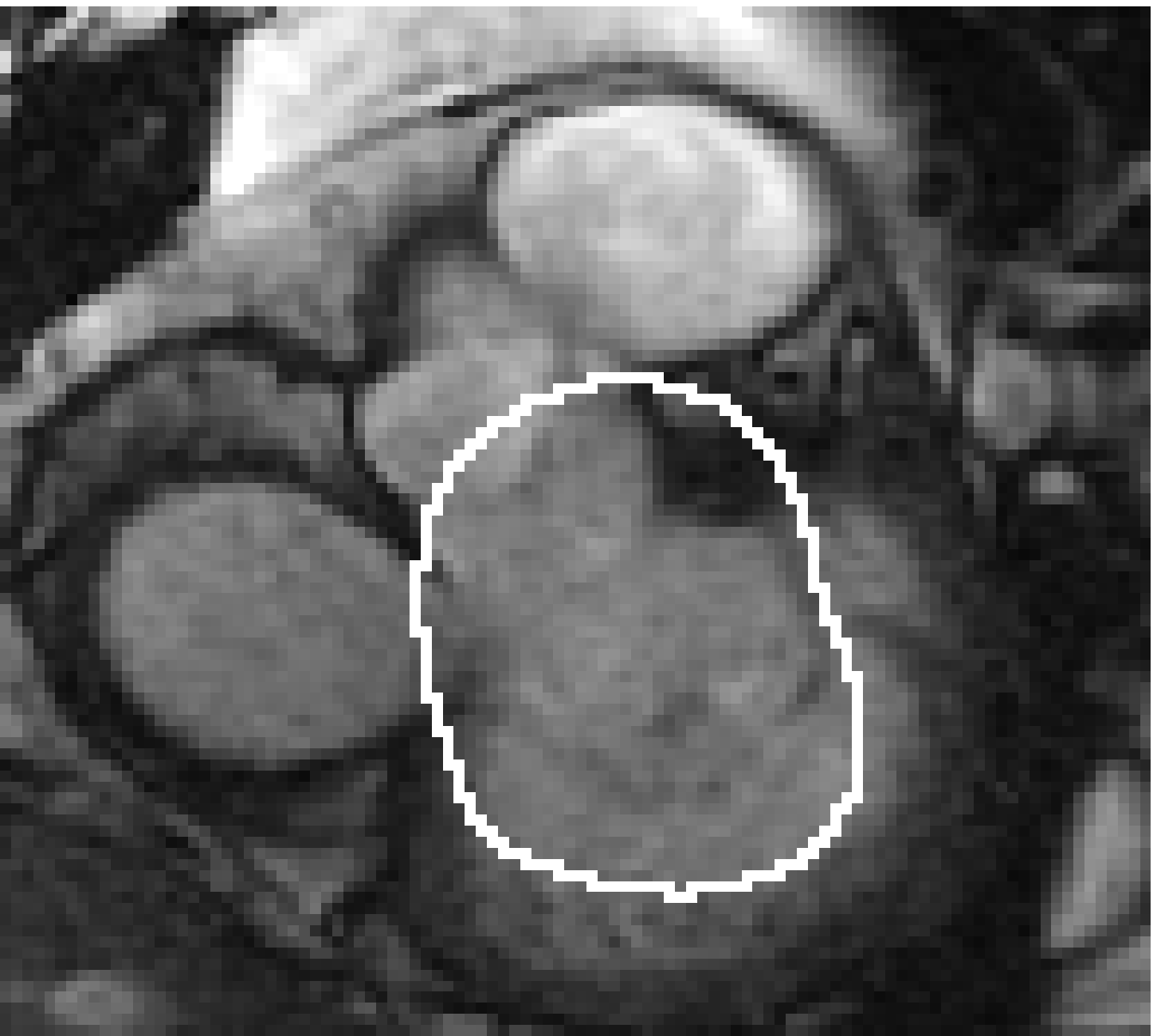} \\
 $m_1$ &  $m_2$ & $m_3$ & $m_4$ & $m_5$ \\
\end{tabular}
\caption{Segmentation methods of the left ventricle provided by $5$ automated algorithms $m_1$ to $m_5$. }
\label{fig:endo_SCHFNI12}
\end{figure}

The segmentation entries are selected inside a database that contains the results obtained by three experts and different algorithms \cite{Constantinides_IRBM_2009,IGMI_CouNajCou2008,Lalande04,Schaerer_MIA_2010,Constantinides_EMBS12}. The corresponding contours are displayed in Fig.\ref{fig:endo_exp_SCHFNI12} and Fig.\ref{fig:endo_SCHFNI12}. In this specific example, the endocardium is not well delimited by the automated algorithms, due to the presence of the aortic root, which leads to very different segmentations.

\begin{figure}[h]
\center
\begin{tabular}{ccccc}
\includegraphics[width=2cm]{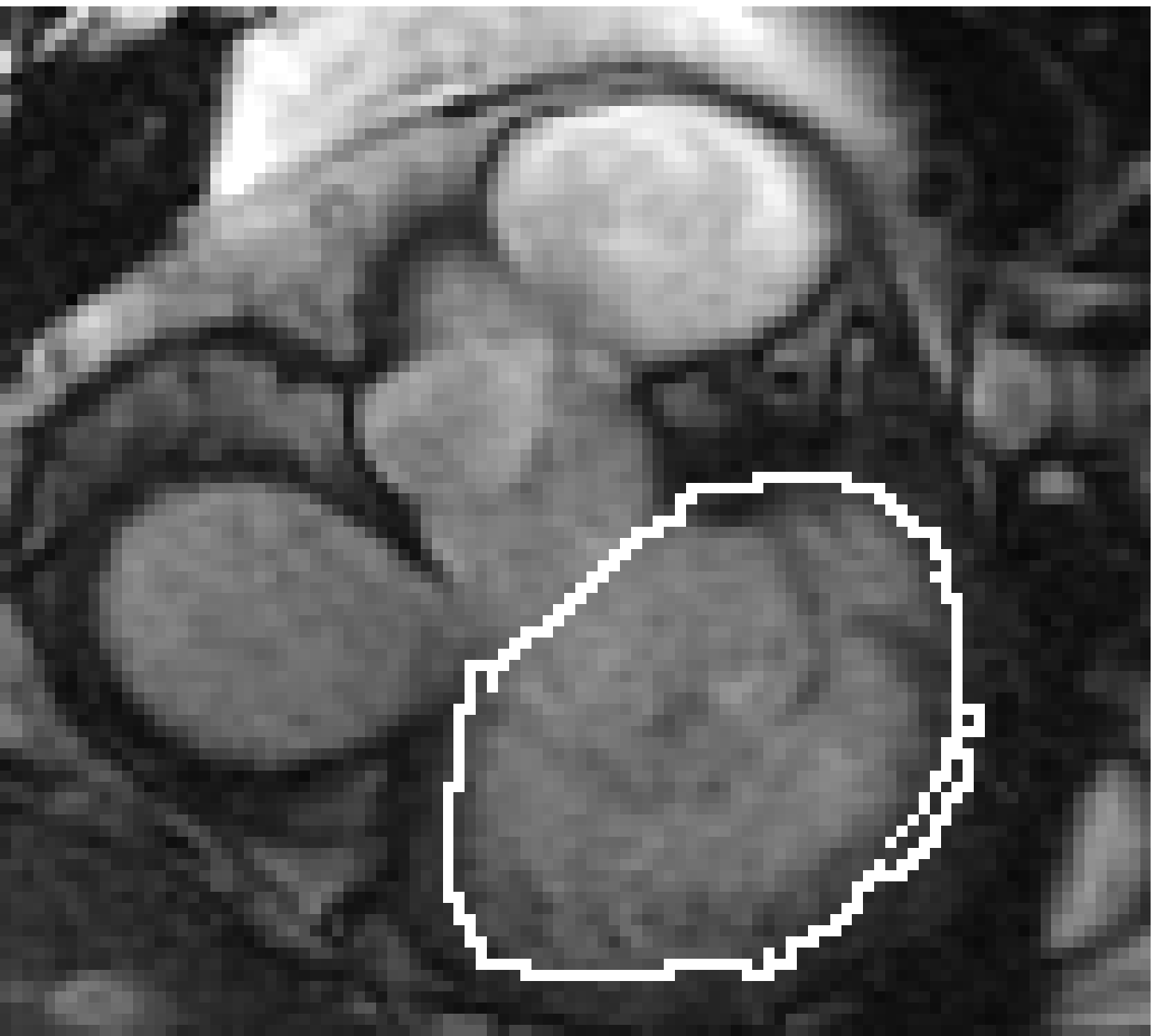} &
\includegraphics[width=2cm]{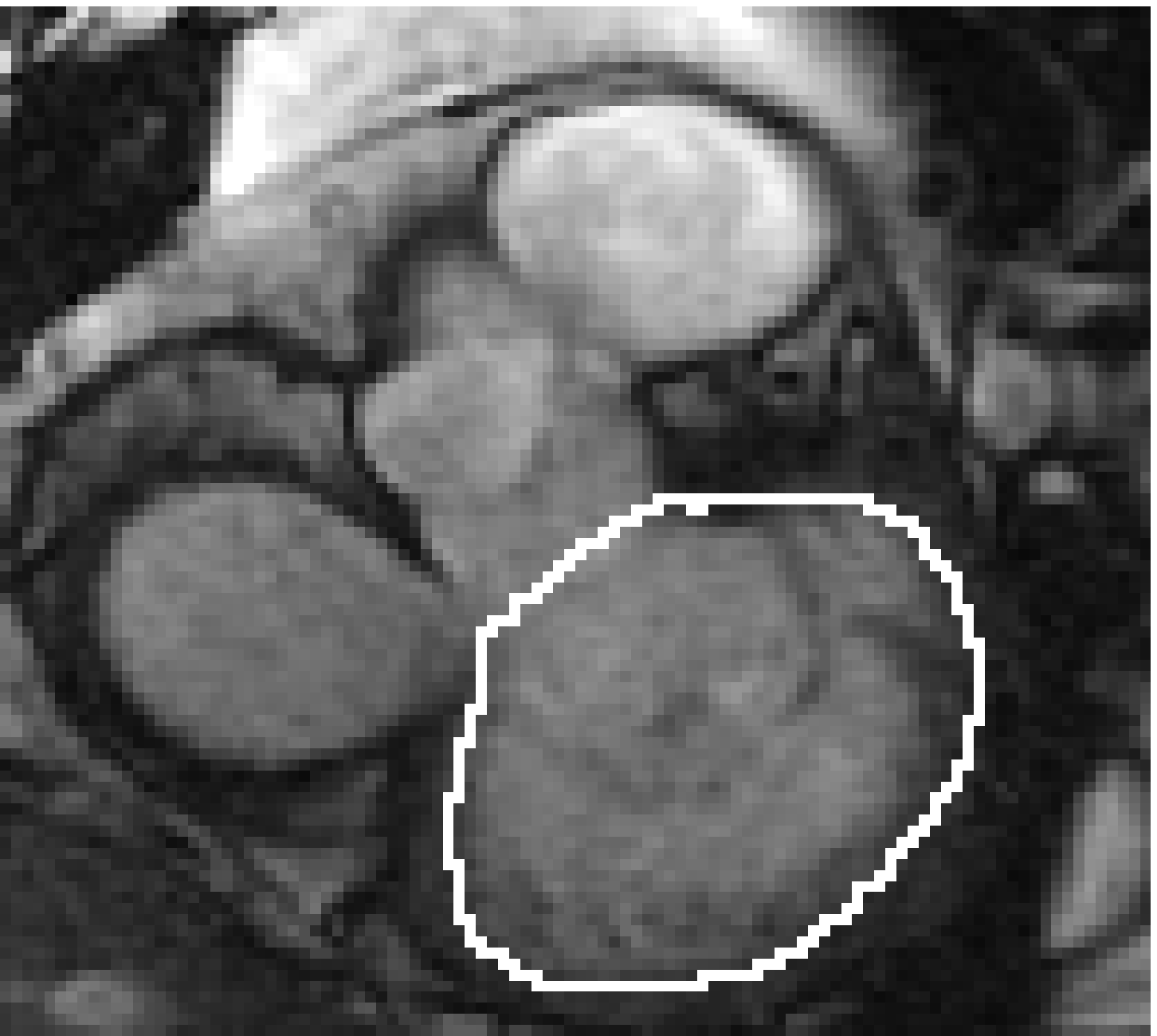} &
\includegraphics[width=2cm]{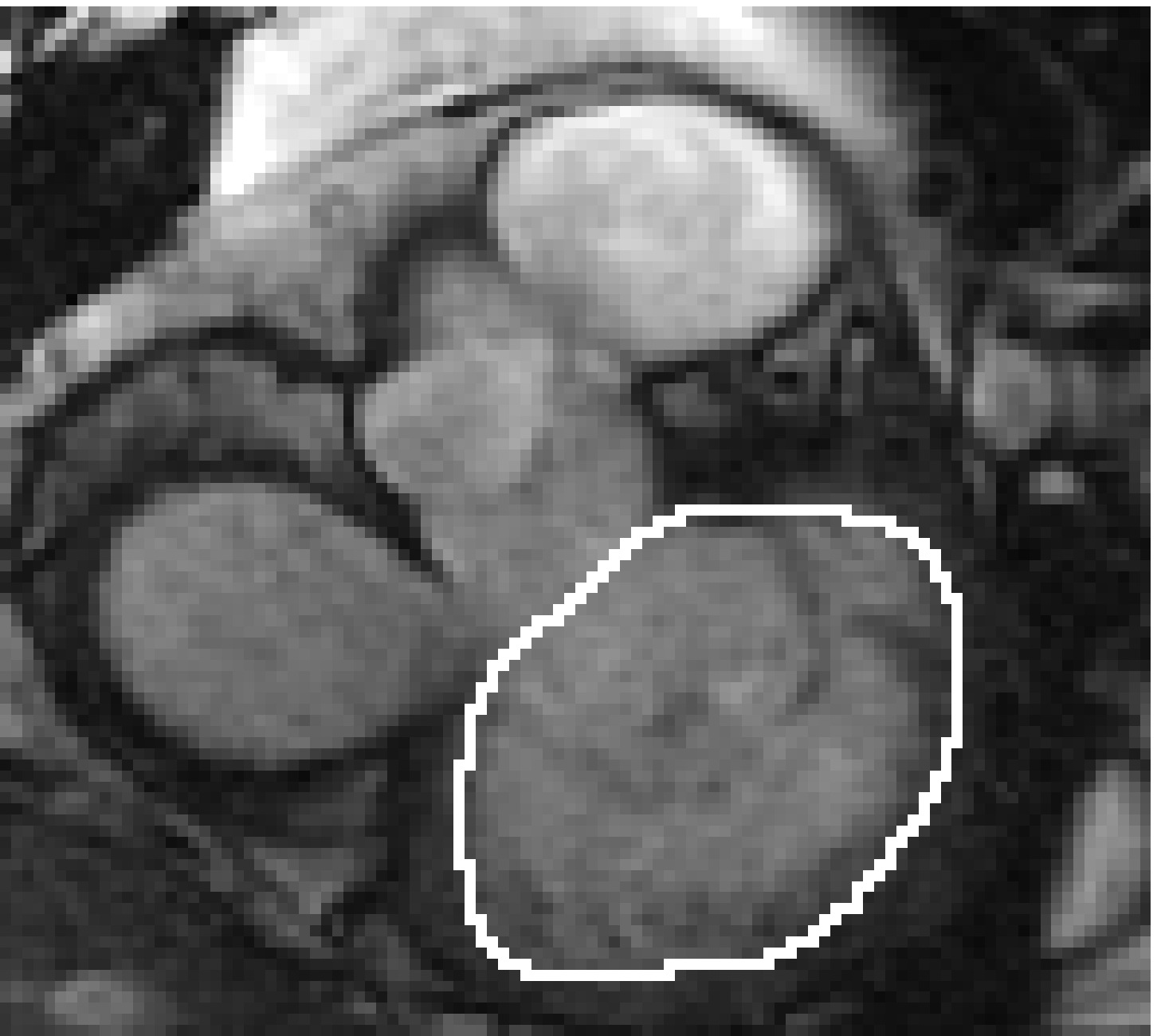} &
\includegraphics[width=2cm]{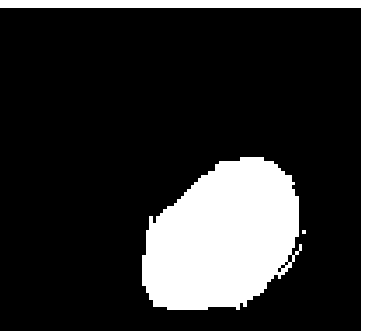} &
\includegraphics[width=2cm]{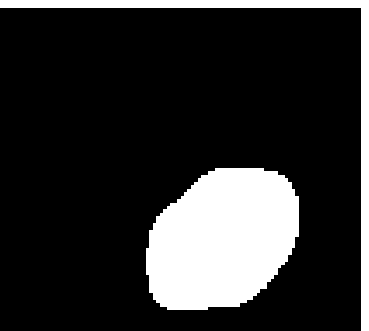} \\
(a)  &  (b)  & (c)  & (d)  & (e) \\
\end{tabular}
\caption{Estimation of different consensus estimates using the contours given by the three experts $Exp_1$ to $Exp_3$ (Fig.\ref{fig:endo_exp_SCHFNI12}) using STAPLE algorithm (a), SD approach (b), and mutual shape (c). Filled masks correspond to STAPLE (d) and mutual shape (e)} 
\label{fig:endo_fusion_exp_SCHFNI12}
\end{figure}

\begin{figure}[h]
\center
\begin{tabular}{cccc}
\includegraphics[width=2cm]{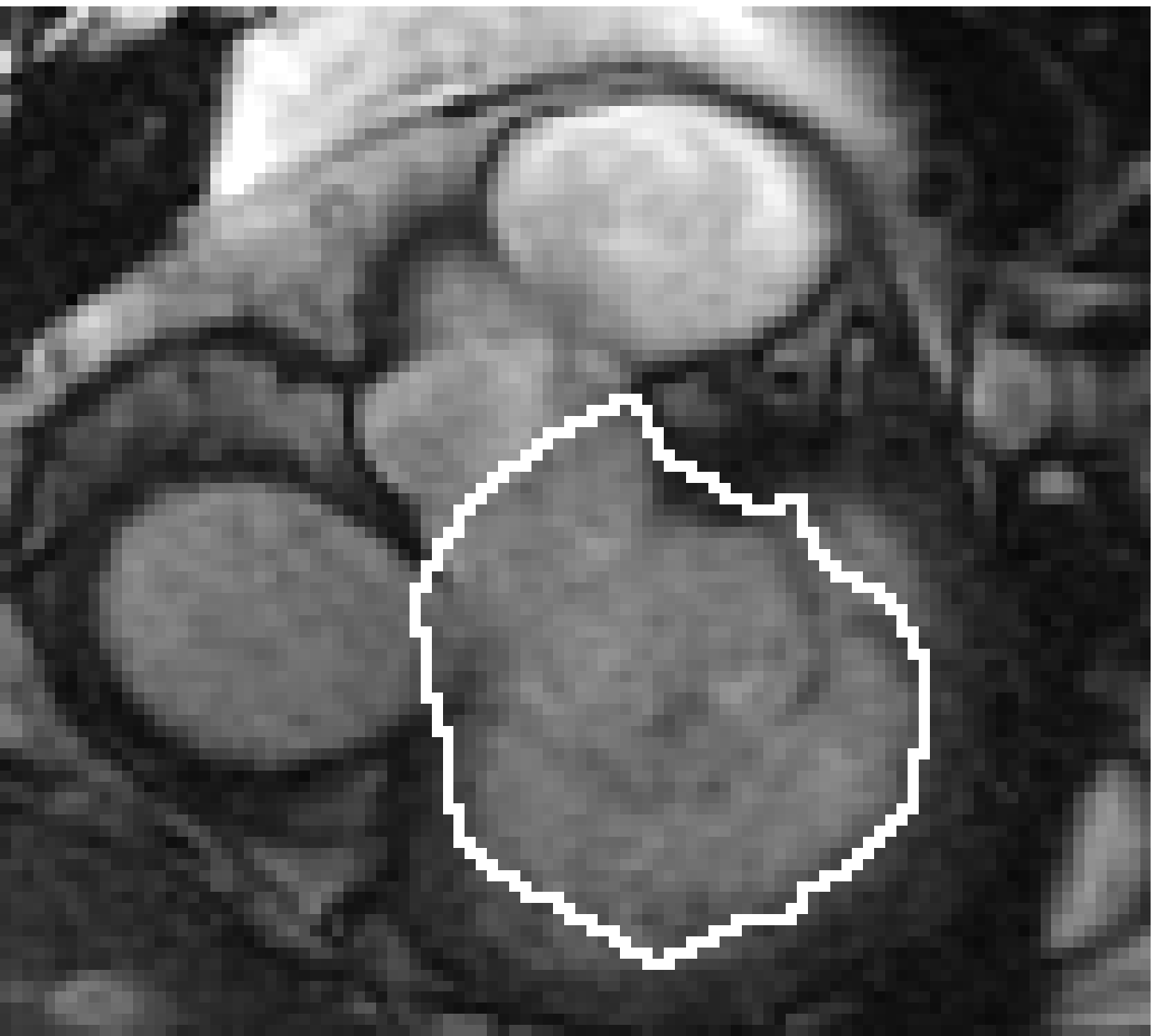} &
\includegraphics[width=2cm]{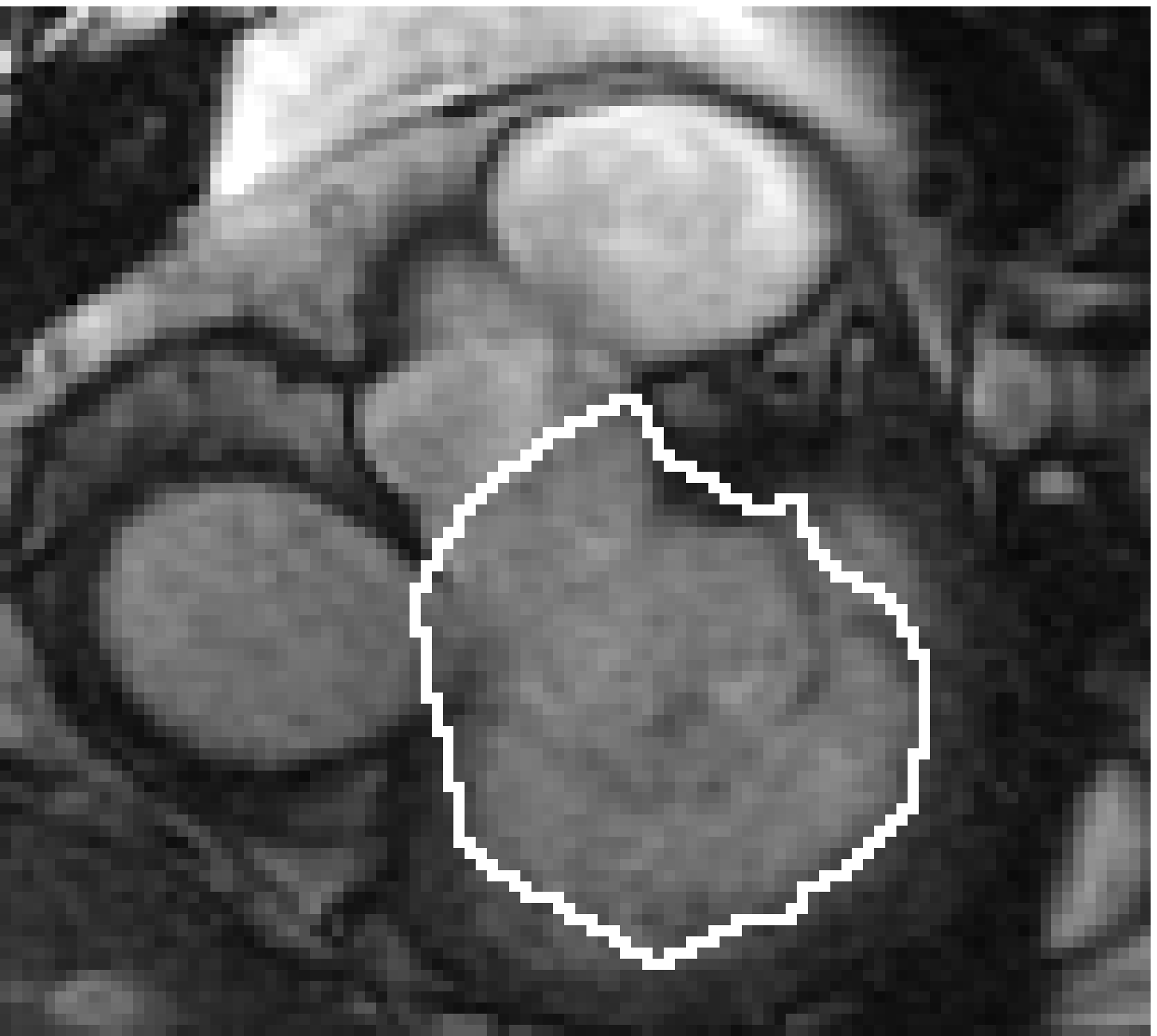} &
\includegraphics[width=2cm]{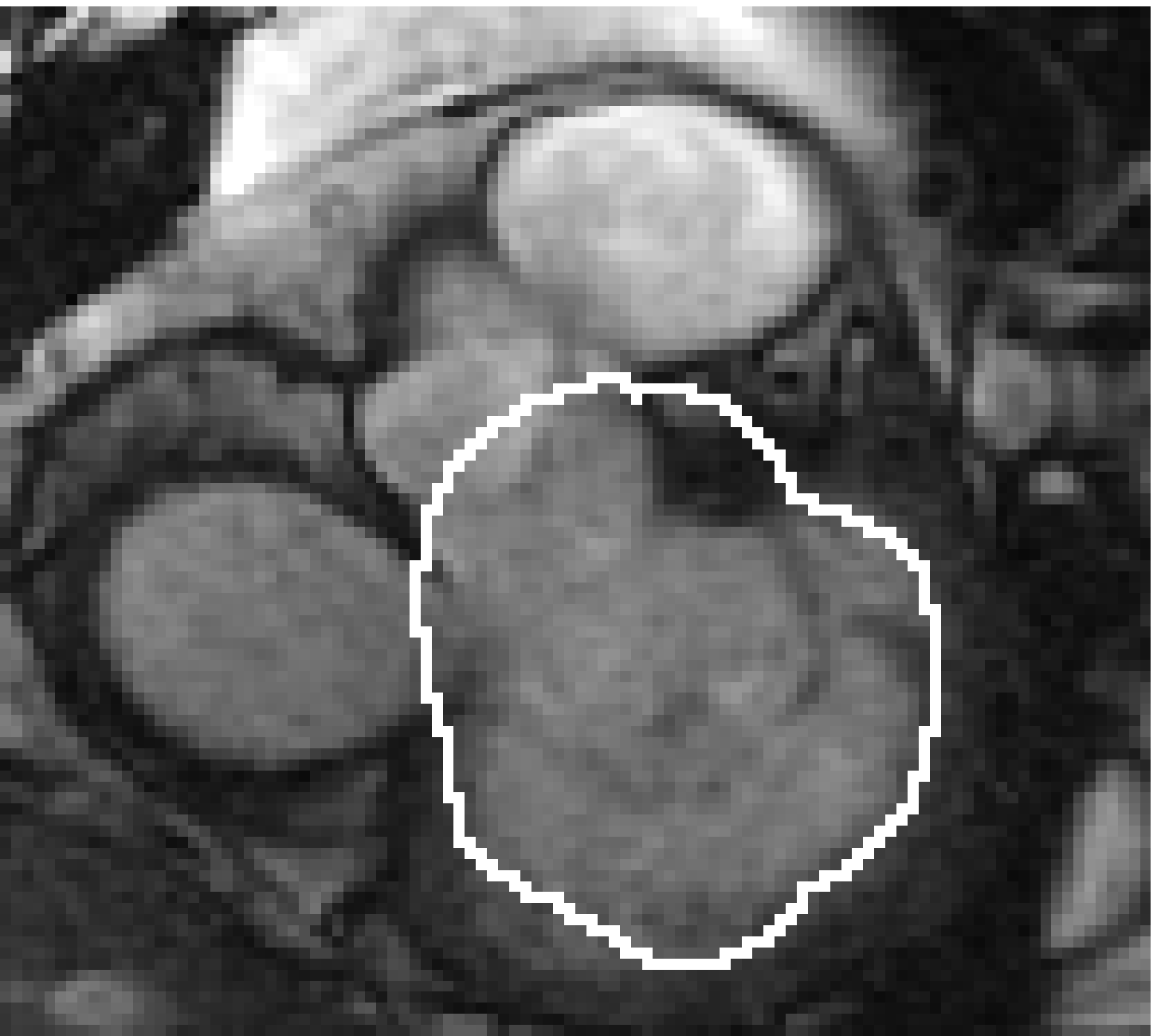} &
\includegraphics[width=2cm]{img/medical_Exp_SC_HFNI12.eps}  \\
(a)  &  (b)  & (c)  & (d)   \\
\end{tabular}
\caption{Estimation of different consensus estimates using the masks $m_1$ to $m_5$ (Fig.\ref{fig:endo_SCHFNI12}) using STAPLE algorithm (a), SD approach (b), and mutual shape (c),  contour delineated by $Exp_2$  (d).} 
\label{fig:endo_fusion_SCHFNI12}
\end{figure}

\begin{table}[h]
    \begin{center}
    \begin{tabular} {|c | c | c | c |}
  \hline
  $i$  & $Exp_1$ & $Exp_2$ & $Exp_3$  \\
 \hline
$p_i $ & 0.999 & 0.999 & 0. 999 \\
$q_i $ & 0.889 & 0.973 & 0.853   \\
  \hline
\end{tabular}
    \end{center}
\vspace{-0.5cm}
\caption{\small  Sensitivity and specificity parameters $p_i$  and $q_i$ for the segmentations $Exp_1$ to $Exp_3$ (Fig.\ref{fig:endo_exp_SCHFNI12}).}
\label{table:classification1}
\end{table}

\begin{table}[h]
    \begin{center}
    \begin{tabular} {|c | c | c | c | c | c |}
  \hline
  $i$  & $m_1$ & $m_2$ & $m_3$ & $m_4$ & $m_5$ \\
 \hline
$p_i $ & 0. 884 & 0.715 & 0.956 & 0.714 & 0.787 \\
$q_i $ & 0.847 & 0.999 & 0.784 & 0.999 & 0.999  \\
  \hline
\end{tabular}
    \end{center}
\vspace{-0.5cm}
\caption{\small  Sensitivity and specificity parameters $p_i$  and $q_i$ for the segmentations $m_1$ to $m_5$ displayed in Fig.\ref{fig:endo_SCHFNI12}.}
\label{table:classification2}
\end{table}

Fig.\ref{fig:endo_fusion_exp_SCHFNI12} shows the consensus segmentations estimated by STAPLE (a), SD (b), and mutual shape (c), using the three experts entries. The regularization parameter $\lambda$ was set equal to $100$. Table \ref{table:classification1} shows that $Exp_2$ provided for this specific case the best contour, but this result (superiority of $Exp_2$ ) was already reported elsewhere \cite{Lebenberg_PLOS15}. Filled masks of STAPLE (Fig.\ref{fig:endo_fusion_exp_SCHFNI12}.d) and mutual shape (Fig.\ref{fig:endo_fusion_exp_SCHFNI12}.e) clearly demonstrate that STAPLE does not necessarily provide smoothed contours at the difference of mutual shape.

Finally, Fig.\ref{fig:endo_fusion_SCHFNI12} shows the consensus segmentations, estimated by STAPLE (a), SD (b), and mutual shape (c), using the five segmentation entries of the automated algorithms ($m_1$ to $m_5$). The regularization parameter $\lambda$ was set equal to $10$.  The parameters $p_i$ and $q_i$ are estimated jointly with the mutual shape (see Table \ref{table:classification2}). At the difference of STAPLE, the contour provided by mutual shape is smooth. However both methods provide results that are quite different from the expert entries. This difficult case shows that mutual shape and STAPLE both depend on the accuracy of the segmentation entries. 

\section{Conclusion}
 
In this work, we search for a mutual shape that minimizes the sum of joint entropies while maximizing the sum of mutual information between each entry shape and the unknown reference shape. We give a geometrical interpretation of this criterion using area measures. The optimization is performed using active contours by computing a shape gradient and the associated evolution equation. Shape derivatives are computed and detailed for the given criterion. Our theoretical formalism is valid for 2D slices or 3D images. The main contribution of this paper lies in the proposition of a theoretical criterion for the estimation of a consensus shape both for segmentation fusion and evaluation without reference.

Some experimental results are provided on both synthetic images and real images for segmentation fusion and evaluation. Indeed, the proposed mutual shape is able to build a consensus shape from a set of different segmentations of the same object and can then be used as an intelligent fusion of different segmentation entries. Moreover, the algorithm estimates jointly the sensibility and specificity parameters (as first proposed by Warfield et al \cite{Warfield_TMI_04}). It then provides an evaluation or ranking of the proposed segmentation methods on the basis of these two parameters.

The experimental results performed on a synthetic image allow to better understand the difference between the mutual shape, a simple union, an average shape and a simple majority voting method. Moreover, the estimated mutual shape is robust to very aberrant outliers thanks to the joint minimization of the joint entropy with the maximization of the mutual information. 

We then propose some tests on real images for different applications. A first test is performed on a real color image and we show that the estimated mutual shape is robust and accurate. The second type of application concerns the segmentation of the text in old manuscripts from a set of simple segmentation entries. The results obtained for this application bring new opportunitues in the field of segmentation by demonstrating that an improved segmentation method may be designed by taking benefit of the fusion of several simple segmentation methods. The intelligent fusion of simple algorithms can probably lead to a new powerful segmentation algorithm. For this part, the choice of the segmentation entries is still an open issue but one may think of a learning phase for an interactive choice of the different segmentation methods chosen to build the consensus. We may call this new kind of process : segmentation by intelligent voting using a consensus shape. The last application, devoted to medical images, also shows that the mutual shape can be useful for the evaluation of different segmentation methods without any reference. However, such a classification may be a little different than the one obtained using the true reference shape. Indeed it corresponds to a classification regarding with the most preponderant shapes in the set of entries and not to an absolute classification with a fixed reference. The classification is clearly dependent on the choice of the different segmentation methods in entry. This unsupervised evaluation process may however be useful to detect abnormal segmentation methods in a set of different segmentation entries. 
 
One perspective of this work may also concern the addition of prior terms inside the variational criterion. For example, a shape prior can be interesting when the objective is to segment the left ventricular cavity. If the objective is different, some other prior shapes may be added (such as the homogeneity of the inside region for example, the gradient, or the target color for color segmentation). Our mathematical framework seems well adapted for this purpose since other information may be easily and rigorously introduced in the criterion to minimize.

\section*{Acknowledgements}

The medical application took place in a larger project named MediEval supported by the GdR 2647 Stic Sant\'e (CNRS-INSERM). We thank all the partners of this project for providing us the different segmentations for the evaluation part. 

\section*{Appendix}
\subsection*{Proof of Theorem \ref{theo:der_IM}}

First of all, we compute the shape derivatives of the probabilities $p_i$ and $q_i$ which depend on the domain. 
By applying the theorem \ref{th:der}, we find :
\begin{eqnarray*}
<p_i'(\mu),\V>=-\frac{1}{|\mu|} \int_{\partial \mu} K(d_i(\x)-1) (\V \cdot \N)  d\mathbf{a} \nonumber \\
+ \frac{\int_{\partial \mu}(\V \cdot \N)  d\mathbf{a}}{|\mu|^2}\int_{\mu} K(d_i(\x)-1) d\x
\end{eqnarray*}
which reduces to :
\begin{eqnarray}
<p_i'(\mu),\V>=\frac{1}{|\mu|} \int_{\partial \mu} (p_i-K(d_i(\x)-1)) (\V \cdot \N)  d\mathbf{a} 
\end{eqnarray}
In the same way, we compute the shape derivative of $q_i$ :
\begin{eqnarray}
<q_i'(\mu),\V>=\frac{1}{|\overline{\mu}|} \int_{\partial \overline{\mu}} (-q_i+K(d_i(\x))) (\V \cdot \N)  d\mathbf{a} 
\end{eqnarray}
We can also compute the shape derivatives of $\mu$ and $\overline{\mu}$:
\begin{eqnarray*}
<|\mu|',\V>= - \int_{\partial \mu}  (\V \cdot \N)  d\mathbf{a} \\
<|\overline{\mu}|',\V>= \int_{\partial \mu}  (\V \cdot \N)  d\mathbf{a}
\end{eqnarray*}
Let us denote $\varphi(p)=p\log(p)+(1-p)\log(1-p)$, the conditional entropy then becomes :
$$
H(D_i/T)=-\Big[ \frac{|\mu|}{|\Omega|} \varphi(p_i) + \frac{|\overline{\mu}|}{|\Omega|} \varphi(q_i)  \Big]
$$
By using chain derivation rules, we find :
\begin{eqnarray*}
<H(D_i/T)',\V>&=& -\frac{1}{|\Omega|} \Big[ |\mu| <p_i',\V> \varphi'(p_i) \\
&+&|\overline{\mu}| <q_i',\V> \varphi'(q_i) +  <|\mu|',\V> \varphi(p_i) \\
&+&  <|\overline{\mu}|',\V> \varphi(q_i) \Big]
\end{eqnarray*}
where $\varphi'(p)=\log(p)-\log(1-p)$.
Replacing the shape derivatives by the previous formulas, we find Theorem \ref{theo:der_IM}.

\subsection*{Proof of Theorem \ref{theo:der_JH} }
First of all, we compute the shape derivatives of the joint probabilities $p(d_i,t)$ which depend on the domain.
Using theorem \ref{th:der}, we find for a=1 or a=0 :
$$
<p(d_i=a,t=0)',\V>= -\frac{1}{|\Omega|}  \int_{\partial \mu}  K(d_i(\x) -a)  (\V \cdot \N)  d\mathbf{a} 
$$
and
$$
<p(d_i=a,t=1)',\V>= \frac{1}{|\Omega|}  \int_{\partial \mu}  K(d_i(\x) -a)  (\V \cdot \N)  d\mathbf{a} 
$$
Let denote $\Psi(p)=p \log p$, the joint entropy then becomes :
\begin{eqnarray*}
H(D_i,T)=-\Psi\left((p(d_i=0,t=0)\right)-\Psi\left(p(d_i=1,t=0)\right) \\
-\Psi\left(p(d_i=0,t=1)\right)-\Psi\left(p(d_i=1,t=1)\right)
\end{eqnarray*}
By using chain derivation rules, we find :
\begin{eqnarray*}
<H(D_i,T)',\V>=-<p(d_i=1,t=1)',\V> \Psi'(p(d_i=1,t=1)) \\
 -<p(d_i=0,t=1)',\V> \Psi'(p(d_i=0,t=1)) \\
-<p(d_i=1,t=0)',\V> \Psi'(p(d_i=1,t=0)) \\ 
-<p(d_i=0,t=0)',\V> \Psi'(p(d_i=0,t=0))
\end{eqnarray*}
where $\Psi'(p)=\log(p) +1$.
Replacing the shape derivatives by the previous formulas, we find Theorem \ref{theo:der_JH}.

\small
\bibliographystyle{plain}
\bibliography{preprint_mutual_shape_mars2017}

\begin{thebibliography}{10}

\bibitem{Aubert03}
G.~Aubert, M.~Barlaud, O.~Faugeras, and S.~Jehan-Besson.
\newblock Image segmentation using active contours : Calculus of variations or
  shape gradients.
\newblock {\em SIAM Journal on Applied Mathematics}, 63(6):2128--2154, 2003.

\bibitem{Berkels_JMIV2010}
B.~Berkels, G.~Linkmann, and M.~Rumpf.
\newblock An \uppercase{SL}(2) invariant shape median.
\newblock {\em Journal of Mathematical Imaging and Vision}, 37(2):85--97, 2010.

\bibitem{Bresson2007}
X.~Bresson, S.~Esedoglu, P.~Vandergheynst, J.~Thiran, and S.~Osher.
\newblock Fast global minimization of the active contour/snake model.
\newblock {\em Journal of Mathematical Imaging and Vision}, 28(2):151--167,
  2007.

\bibitem{Chan01}
T.~F. Chan and L.~A. Vese.
\newblock Active contour without edges.
\newblock {\em IEEE Transactions on Image Processing}, 10(2):266--277, 2001.

\bibitem{Chan00}
T.F. Chan, B.Y. Sandberg, and L.A. Vese.
\newblock Active contours without edges for vector-valued images.
\newblock {\em Journal of Visual Communication and Image Representation},
  11(2):130--141, 2000.

\bibitem{Charpiat_ICIP03}
G.~Charpiat, O.~Faugeras, and R.~Keriven.
\newblock Shape metrics, warping and statistics.
\newblock In {\em International Conference on Image Processing}, volume~2,
  pages 627--630, 2003.

\bibitem{Charpiat_FCM04}
G.~Charpiat, O.~Faugeras, and R.~Keriven.
\newblock Approximations of shape metrics and application to shape warping and
  empirical shape statistics.
\newblock {\em Foundations of Computational Mathematics}, 5(1):1--58, 2005.

\bibitem{Warfield_TMI_12}
O.~Commowick, A.~Akhondi-Asl, and S.K. Warfield.
\newblock Estimating a reference standard segmentation with spatially varying
  performance parameters.
\newblock {\em IEEE Transactions on Medical Imaging}, 31(8):1593--1606, 2012.

\bibitem{Constantinides_IRBM_2009}
C.~Constantinides, R.~El~Berbari, A.~de~Cesare, Y.~Chenoune, E.~Roullot,
  A.~Herment, E.~Mousseaux, and F.~Frouin.
\newblock Development and evaluation of an algorithm for the automated
  segmentation of the left and right ventricles on cine \uppercase{MRI}.
\newblock {\em IRBM}, 30(4):188--191, 2009.

\bibitem{Constantinides_EMBS12}
C.~Constantinides, E.~Roullot, M.~Lefort, and F.~Frouin.
\newblock Fully automated segmentation of the left ventricle applied to cine
  \uppercase{MR} images: description and results on a database of 45 subjects.
\newblock In {\em IEEE Engineering in Medicine and Biology Society}, pages
  3207--3210, 2012.

\bibitem{IGMI_CouNajCou2008}
J.~Cousty, L.~Najman, M.~Couprie, S.~Cl\'ement-Guinaudeau, T.~Goissen, and
  J.~Garot.
\newblock {Segmentation of 4D cardiac MRI: automated method based on
  spatio-temporal watershed cuts}.
\newblock {\em Image and Vision Computing}, 28(8):1229--1243, 2010.

\bibitem{cover-thomas:91}
T.M. Cover and J.A. Thomas.
\newblock {\em Elements of Information Theory}.
\newblock Wiley-Interscience, 1991.

\bibitem{Cremers03b}
D.~Cremers, T.~Kohlberger, and C.~Schn\"orr.
\newblock Shape statistics in kernel space for variational image segmentation.
\newblock {\em Pattern Recognition}, 36(2):1929--1943, 2003.

\bibitem{Zolesio}
M.C. Delfour and J.P. Zol{\'e}sio.
\newblock {\em Shapes and Geometries: Metrics, Analysis, Differential Calculus,
  and Optimization}.
\newblock Advances in design and control. Society for Industrial and Applied
  Mathematics (SIAM), 2001.

\bibitem{duda-hart:73}
R.~Duda and P.~Hart.
\newblock {\em {Pattern Classification and Scene Analysis}}.
\newblock {John Wiley \& Sons, Inc.}, 1973.

\bibitem{Fleureau_IRBM09}
J.~Fleureau, M.~Garreau, D.~Boulmier, C.~Leclercq, and A.~Hernandez.
\newblock 3\uppercase{D} multi-object segmentation of cardiac \uppercase{MSCT}
  imaging by using a multi-agent approach.
\newblock {\em IRBM}, 30(3):104--113, 2009.

\bibitem{Foulonneau_PAMI06}
A.~Foulonneau, P.~Charbonnier, and F.~Heitz.
\newblock Affine-invariant geometric shape priors for region-based active
  contours.
\newblock {\em IEEE Transactions on Pattern Analysis and Machine Intelligence},
  28(8):1352--1357, 2006.

\bibitem{Herbulot_JMIV06}
A.~Herbulot, S.~Jehan-Besson, S.~Duffner, M.~Barlaud, and G.~Aubert.
\newblock Segmentation of vectorial image features using shape gradients and
  information measures.
\newblock {\em Journal of Mathematical Imaging and Vision}, 25(3):365--386,
  2006.

\bibitem{jehan_icip14}
S.~Jehan-Besson, C.~Tilmant, A.~de~Cesare, A.~Lalande, A.~Cochet, J.~Cousty,
  J.~Lebenberg, M.~Lefort, P.~Clarysse, R.~Clouard, L.~Najman, L.~Sarry,
  F.~Frouin, and M.~Garreau.
\newblock A mutual reference shape based on information theory.
\newblock In {\em IEEE International Conference on Image Processing}, pages
  887--891, 2014.

\bibitem{JehanBesson_GRETSI11}
S.~Jehan-Besson, C.~Tilmant, A.~De de~Cesare, F.~Frouin, L.~Najman, A.~Lalande,
  L.~Sarry, C.~Casta, P.~Clarysse, C.~Constantidin{\`e}s, J.~Cousty, M.~Lefort,
  A.~Cochet, and M.~Garreau.
\newblock Estimation d{\textquoteright}une forme mutuelle pour
  l{\textquoteright}{\'e}valuation de la segmentation en imagerie cardiaque.
\newblock In {\em GRETSI - Traitement du Signal et des Images}, 2011.

\bibitem{Kass88}
M.~Kass, A.~Witkin, and D.~Terzopoulos.
\newblock Snakes : Active contour models.
\newblock {\em International Journal of Computer Vision}, 1:321--332, 1988.

\bibitem{Kendall_84}
D.G. Kendall.
\newblock Shape manifolds, procrustean metrics, and complex projective spaces.
\newblock {\em Bulletin of the London Mathematical Society}, 16(2):81--121,
  1984.

\bibitem{Kim_ICIP02}
J.~Kim, J.W. Fisher, A.~Yezzi, M.~Cetin, and A.S. Willsky.
\newblock Nonparametric methods for image segmentation using information theory
  and curve evolution.
\newblock In {\em IEEE International Conference on Image Processing}, pages
  797--800, 2002.

\bibitem{Lalande04}
A.~Lalande, N.~Salve, A.~Comte, M.-C. Jaulent, L.~Legrand, P.M. Walker,
  Y.~Cottin, J.E. Wolf, and F.~Brunotte.
\newblock Left ventricular ejection fraction calculation from automatically
  selected and processed diastolic and systolic frames in short axis
  cine-\uppercase{MRI}.
\newblock {\em Journal of Cardiovascular Magnetic Resonance}, 6(4):817--827,
  2004.

\bibitem{Lebenberg_TMI12}
J.~Lebenberg, I.~Buvat, A.~Lalande, P.~Clarysse, C.~Casta, A.~Cochet,
  C.~Constantinid\`es, J.~Cousty, A.~de~Cesare, S.~Jehan-Besson, M.~Lefort,
  L.~Najman, E.~Roullot, L.~Sarry, C.~Tilmant, M.~Garreau, and F.~Frouin.
\newblock Non supervised ranking of different segmentation approaches:
  application to the estimation of the left ventricular ejection fraction from
  cardiac cine \uppercase{MRI} sequences.
\newblock {\em IEEE Transactions on Medical Imaging}, 31(8):1651--1660, 2012.

\bibitem{Lebenberg_PLOS15}
J.~Lebenberg, A.~Lalande, P.~Clarysse, I.~Buvat, C.~Casta, A.~Cochet,
  C.~Constantinid\`es, J.~Cousty, A.~de~Cesare, S.~Jehan-Besson, M.~Lefort,
  L.~Najman, E.~Roullot, L.~Sarry, C.~Tilmant, F.~Frouin, and M.~Garreau.
\newblock Improved estimation of cardiac function parameters using a
  combination of independent automated segmentation results in cardiovascular
  magnetic resonance imaging.
\newblock {\em PLOS ONE}, 10(8):e0135715, 2015.

\bibitem{Lec_icip06}
F.~Lecellier, S.~Jehan-Besson, J.~Fadili, G.~Aubert, M.~Revenu, and E.~Saloux.
\newblock Region-based active contours with noise and shape priors.
\newblock In {\em IEEE International Conference on Image Processing}, volume~1,
  pages 1649--1652, 2006.

\bibitem{Leventon_CVPR00}
M.~Leventon, E.~Grimson, and O.~Faugeras.
\newblock Statistical shape influence in geodesic active contours.
\newblock In {\em IEEE International Conference on Image Processing}, pages
  316--323, 2000.

\bibitem{MartinFTM01}
D.~Martin, C.~Fowlkes, D.~Tal, and J.~Malik.
\newblock A database of human segmented natural images and its application to
  evaluating segmentation algorithms and measuring ecological statistics.
\newblock In {\em Proceedings Eighth International Conference on Computer
  Vision}, volume~2, pages 416--423, 2001.

\bibitem{Osher_1988}
S.~Osher and J.A. Sethian.
\newblock Fronts propagating with curvature-dependent speed: Algorithms based
  on \uppercase{H}amilton-\uppercase{J}acobi formulations.
\newblock {\em Journal of Computational Physics}, 79(1):12--49, 1988.

\bibitem{Pandore}
{Pandore: A library of image processing operators (Version 6.6). [Software].
  Greyc Laboratory}.
\newblock {https://clouard.users.greyc.fr/Pandore}, 2013.

\bibitem{Paragios_03}
N.~Paragios.
\newblock A level set approach for shape-driven segmentation and tracking of
  the left ventricle.
\newblock {\em IEEE Transactions on Medical Imaging}, 22(6):773--776, 2003.

\bibitem{DIBCO13}
I.~Pratikakis, B.~Gatos, and K.~Ntirogiannis.
\newblock \uppercase{ICDAR} 2013 document image binarization contest
  (\uppercase{DIBCO} 2013).
\newblock In {\em Proceedings of the 2013 12th International Conference on
  Document Analysis and Recognition}, pages 1471--1476, Washington, DC, USA,
  2013. IEEE Computer Society.

\bibitem{Precioso05}
F.~Precioso, M.~Barlaud, T.~Blu, and M.~Unser.
\newblock Robust real-time segmentation of images and videos using a
  smooth-spline snake-based algorithm.
\newblock {\em IEEE Transactions on Image Processing}, 14(7):910--924, 2005.

\bibitem{Reza_book94}
F.~M. Reza.
\newblock {\em An Introduction to Information Theory}, pages 106--108.
\newblock McGraw-Hill, 1994.

\bibitem{Sauvola_PR2000}
J.~Sauvola and M.~Pietikainen.
\newblock Adaptive document image binarization.
\newblock {\em Pattern Recognition}, 33(2):225--236, 2000.

\bibitem{Schaerer_MIA_2010}
J.~Schaerer, C.~Casta, J.~Pousin, and P.~Clarysse.
\newblock A dynamic elastic model for segmentation and tracking of the heart in
  \uppercase{MR} image sequences.
\newblock {\em Medical Image Analysis}, 14(6):738--749, 2010.

\bibitem{Sokolowski_book_92}
J.~Sokolowski and J.P. Zol\'{e}sio.
\newblock {\em Introduction to shape optimization. Shape sensitivity analysis.}
\newblock Springer Series in Computational Mathematics. Springer-Verlag,
  Berlin, 1992.

\bibitem{Suinesiaputra_MIA14}
A.~Suinesiaputra, B.R. Cowan, A.O. Al-Agamy, M.A. Elattar, N.~Ayache, A.S.
  Fahmy, A.M. Khalifa, P.~Medrano-Gracia, M.P. Jolly, A.H. Kadish, D.C. Lee,
  J.~Margeta, S.K. Warfield, and A.A. Young.
\newblock A collaborative resource to build consensus for automated left
  ventricular segmentation of cardiac \uppercase{MR} images.
\newblock {\em Medical Image Analysis}, 18(1):50--62, 2014.

\bibitem{Angulo_ICPR2010}
S.~Velasco-Forero and J.~Angulo.
\newblock Statistical shape modeling using morphological representations.
\newblock In {\em IEEE International Conference on Image Processing}, pages
  3537--3540, 2010.

\bibitem{Warfield_TMI_04}
S.K. Warfield, K.~H. Zou, and W.~M.~Wells III.
\newblock Simultaneous truth and performance level estimation
  (\uppercase{STAPLE}): an algorithm for the validation of image segmentation.
\newblock {\em IEEE Transactions on Medical Imaging}, 23(7):903--921, 2004.

\bibitem{Yeung_IEEEIT_91}
R.~W. Yeung.
\newblock A new outlook on shannon's information measures.
\newblock {\em IEEE Transactions on Information Theory}, 37(3):466--474, 1991.

\bibitem{Soatto_IJCV02}
A.J. Yezzi and S.~Soatto.
\newblock Deformotion: Deforming motion, shape average and the joint
  registration and approximation of structures in images.
\newblock {\em International Journal of Computer Vision}, 53(2):153--167, 2003.

\end{thebibliography}

\end{document}